\documentclass[twocolumn,times]{aastex62_reep}

\pdfoutput=1

\shorttitle{Multi-Thread Modeling of Soft X-Ray Irradiances}
\shortauthors{Reep et al.}

\graphicspath{{}}

\usepackage[multidot]{grffile}
\usepackage{amsmath}

\usepackage{tabularx}
\usepackage[roman]{parnotes}

\usepackage{fancyhdr}   
\fancypagestyle{specialfooter}{%
 \fancyhf{}
 
 \fancyfoot[C]{Distribution Statement A.  Approved for public release.  Distribution unlimited.}  }

\begin{document}


\title{Simulating Solar Flare Irradiance with Multithreaded Models of Flare Arcades}

\author[0000-0003-4739-1152]{Jeffrey W. Reep}
\affil{Space Science Division, Naval Research Laboratory, Washington, DC 20375}

\author[0000-0001-6102-6851]{Harry P. Warren}
\affil{Space Science Division, Naval Research Laboratory, Washington, DC 20375}

\author[0000-0002-4103-6101]{Christopher S. Moore}
\affil{Harvard-Smithsonian Center for Astrophysics, Cambridge, MA 02138}

\author[0000-0001-5243-7659]{Crisel Suarez}
\affil{Harvard-Smithsonian Center for Astrophysics, Cambridge, MA 02138}
\affil{Fisk University, Nashville, TN 32708}
\affil{Vanderbilt University, Nashville, TN 37235}

\author[0000-0002-6835-2390]{Laura A. Hayes}
\affil{NASA Goddard Space Flight Center, Greenbelt, MD 20771}


\begin{abstract}
Understanding how energy is released in flares is one of the central problems of solar and stellar astrophysics.  Observations of high temperature flare plasma hold many potential clues as to the nature of this energy release. It is clear, however, that flares are not composed of a few impulsively heated loops, but are the result of heating on many small-scale threads that are energized over time, making it difficult to compare observations and numerical simulations in detail.  Several previous studies have shown that it is possible to reproduce some aspects of the observed emission by considering the flare as a sequence of independently heated loops, but these studies generally focus on small-scale features while ignoring the global features of the flare.  In this paper, we develop a multithreaded model that encompasses the time-varying geometry and heating rate for a series of successively-heated loops comprising an arcade.  To validate, we compare with spectral observations of five flares made with the MinXSS CubeSat as well as light curves measured with GOES/XRS and SDO/AIA.  We show that this model can successfully reproduce the light curves and quasi-periodic pulsations in GOES/XRS, the soft X-ray spectra seen with MinXSS, and the light curves in various AIA passbands.  The AIA light curves are most consistent with long duration heating, but elemental abundances cannot be constrained with the model.  Finally, we show how this model can be used to extrapolate to spectra of extreme events that can predict irradiance across a wide wavelength range including unobserved wavelengths.
\end{abstract}



\section{Introduction}
\thispagestyle{specialfooter}   

Understanding the solar irradiance and its impact upon the Earth's ionosphere-thermosphere system is vital to understanding the coupling between the Earth-Sun system.  During a flare, the solar irradiance increases significantly as magnetic energy is released, driving brightenings across the electromagnetic spectrum (see the review by \citealt{fletcher2011}).  These brightenings, particularly in the X-rays, can produce a direct reaction in the ionospheric free electron density (\textit{e.g.} \citealt{hayes2017}).  In order to properly predict the change in solar irradiance, detailed models of the energy release and transport must be developed that can then be used to calculate the spectra at various wavelengths.  Since many wavelengths are not routinely monitored, it is necessary to develop a model to calculate the irradiance at these wavelengths.  The spectra, spanning orders of magnitude in wavelength range, can then be used to determine the impact of flare irradiance variability on the ionosphere (\textit{e.g.,} \citealt{hinteregger1981, Meier2002, Huba2005}). 

Solar flares are driven by a magnetic reconnection event that causes the formation of a series of magnetic loop structures \citep{kopp1976}, referred to as a flare arcade (a clear example is in \textit{e.g.} \citealt{sheeley2004}).  Each loop is heated to temperatures exceeding 10 MK, causing brightenings in the extreme ultraviolet (EUV) and soft X-rays (SXRs).  The magnetic energy is converted to kinetic energy through the acceleration of particles \citep{holman2011}, which drives shocks \citep{cargill1983}, as well as driving magnetohydrodynamic (MHD) wave motions \citep{tarr2017}.  The electron beam is generally thought to be the primary driver of heating, causing efficient heating of the solar chromosphere that drives the ablation of material into the corona, heating and filling the loops (\textit{e.g.} \citealt{fisher1985,reep2015}), as well as producing hard X-ray (HXR) footpoint emission from non-thermal bremsstrahlung \citep{brown1971}.  

Modeling the formation and heating of these loops is challenging for a few reasons.  The number of loops and the properties of the heating of each individual loop (energy, duration, electron beam properties, \textit{etc.}) are not well known, nor are they easy to diagnose from observations \citep{reep2018}.  Instruments like the Reuven Ramaty High Energy Solar Spectroscopic Imager (RHESSI) have been able to measure many of the electron beam properties for large areas of the flare (\textit{e.g.} \citealt{holman2003}), and combinations of many instruments can give good estimates of global properties of flares \citep{emslie2005,milligan2014}, but diagnosing the heating or energy on individual loops remains difficult.  It may in fact be that many of these properties are stochastic, with variations from loop to loop as well as from flare to flare, which would necessitate the simulation of many thousands of loops, requiring a non-trivial computational effort.  It is also unlikely that instrumentation will reach the spatial scales needed to diagnose the heating on individual loops in the near future.  This has led to the rise of multi-threaded modeling of solar flare arcades, rather than single loop models, an idea which was first used to explain blue-wing enhancements seen in a Ca XIX line with Yohkoh by \citet{hori1997,hori1998}.  Subsequent papers have shown that compared to single loop models, multi-threaded models improve the consistency with observed temperatures and densities, cooling rates, long duration heating, light curves, and spectral observations across many passbands \citep{aschwanden2000,reeves2002,warren2006,qiu2012,liu2013,reep2016,rubiodacosta2016,rubiodacosta2017,zhu2018,polito2019}. 

Fortunately, there are a few indications that some of these properties can be diagnosed from spatially unresolved observations.  The total duration of a flare is well-correlated with the separation of the foot-points \citep{toriumi2017}, which has been interpreted as being due to continued reconnection of longer and longer loops \citep{reep2017}.  A direct correlation between the footpoint separation has also been found with the period of quasi-periodic pulsations (QPPs, \citealt{pugh2019}) in flaring emission, suggesting that on-going reconnection of new loops could also be related to the presence of QPPs.  Indeed, QPPs can persist well into the gradual phase of flares \citep{hayes2019}, consistent with the idea that bursty reconnection can continue past the impulsive phase.  This is also supported by observations showing that the energy release in the gradual phase can exceed that in the impulsive phase by more than an order of magnitude \citep{kuhar2017}.  These considerations are critical to construct an accurate model of a flare encompassing a large number of loops that are energized over the whole duration of the flare.  
    
In this paper, we therefore build upon the model of \citet{reep2017} to construct a global flare model where the number of loops, the geometry of the arcade, and the heating rates on each loop are all constrained by basic observations.  We follow the basic premise that the magnetic reconnection event drives a series of successively heated loops, whose heating rate and volume can vary with time.  We test this model on a set of five well-observed flares, comparing both spectra and light curves at EUV and SXR wavelengths.  We constrain these parameters with observations by NOAA's GOES X-ray Sensors (XRS), following the methodology of \citet{warren2005} and \citet{warren2006}, and then compare the resultant spectra with EUV light curves from the Atmospheric Imaging Assembly (AIA, \citealt{lemen2012}) as well as SXR irradiance measurements from the first Miniature X-ray Solar Spectrometer (MinXSS-1) CubeSat \citep{mason2016,moore2018,woods2017}.  We show that this model can often reproduce observations from all three of these instruments, and show what properties can be constrained.  Finally, we show how this model can be extrapolated to larger or smaller flares so that predictions of extreme solar flares can be made.  
    \def\arraystretch{1.15}
    \begin{table*}[t]
        \centering
        \begin{tabularx}{\textwidth}{ c c c c c c c c c }
            Flare Time &  Class & Duration\parnote{FWHM of GOES/XRS 1--8\,\AA\ channel.} & GOES $T_{\text{max}}$\parnote{Uncertainties for GOES temperature and EM are the 1-sigma uncertainties calculated with the TEBBS algorithm \citep{ryan2012}.} & GOES EM$_{\text{max}}$ & MinXSS-1 $T_{\text{max}}$\parnote{High and low temperature components derived from a two-temperature fit, respectively.} & MinXSS-1 EM$_{\text{max}}$ & FIP bias 
             & QPP \parnote{Average period during the impulsive phase, measured in GOES 1--8\,\AA\ channel.} \\ 
            (UT) & & (min) & (MK) & ($10^{49}$ cm$^{-3}$) & (MK) & ($10^{49}$ cm$^{-3}$) & & (s) \\ \hline
            \rowcolor{cyan!20}
            21 Jul 2016 01:30--05:00 & M1.0 & 30.4 & $12.96 \pm 0.32$ & $0.90 \pm 0.064$ & $23.40 \pm 0.70$ & $1.704 \pm 0.031$ & 2.530 & 20 \\
            \rowcolor{cyan!20}
            &  &  &  &  & $10.05 \pm 1.15$ & $0.073 \pm 0.013$ & &  \\
            \rowcolor{red!20}
            23 Jul 2016 01:36--03:30 & M5.0 & 16.8 & $23.93 \pm 0.11$ & $3.20 \pm 0.023$ & $20.03 \pm 0.05$ & $11.84 \pm 0.324$ & 2.192 & 26 \\ 
            \rowcolor{red!20}
            &  &  &  &  & $10.11 \pm 0.04$ & $4.523 \pm 0.074$ & &  \\
            \rowcolor{cyan!20}
            23 Jul 2016 04:50--07:25 & M7.7 & 10.9 & $23.25 \pm 0.10$ & $4.19 \pm 0.024$ & $43.12 \pm 1.04$ & $14.67 \pm 0.558$ & 2.248 & 20 \\ 
            \rowcolor{cyan!20}
            &  &  &  &  & $9.82 \pm 0.23$ & $6.352 \pm 0.076$ & &  \\
            \rowcolor{red!20}
            29 Nov 2016 07:00--07:40 & C7.8 & 3.3 & $20.19 \pm 0.07$ & $0.49 \pm 0.004$ & $17.88 \pm 0.14$ & $0.730 \pm 0.055$ & 3.043  & 12 \\ 
            \rowcolor{red!20}
            &  &  &  &  & $9.68 \pm 0.11$ & $0.269 \pm 0.005$ & &  \\
            \rowcolor{cyan!20}
            01 Apr 2017 20:50--23:00 & M4.4 & 19.7 & $20.82 \pm 0.05$ & $2.54 \pm 0.003$ & $20.50 \pm 0.15$ & $3.84 \pm 0.180$  & 2.706  & 25 \\
            \rowcolor{cyan!20}
            &  &  &  &  & $9.94 \pm 0.05$ & $0.343 \pm 0.073$ & & \\
        \end{tabularx}
        \caption{Summary of observations and the parameters of the five flares that are the subject of this paper.  \label{tab:obs_summary} }
        \parnotes
    \end{table*}

\section{Observations}
\label{sec:observations}

In this paper, we examine five well-observed flares which have good coverage in GOES-15, SDO/AIA, and MinXSS-1.  The flares range in size from C7 to M7, and range in total duration from onset to finish of around 20 minutes to a few hours (or about 3 to 30 minutes FWHM).  Table \ref{tab:obs_summary} summarizes the key parameters of these flares found in this section.  The flares occurred between July 2016 and April 2017, during the period when MinXSS-1 was taking observations.  These flares were chosen for study partly because they were the best observed flares with MinXSS-1, whereas many other flares have significant data gaps during eclipse periods.

\begin{figure*}[t]
    \centering
    \includegraphics[width=0.95\linewidth]{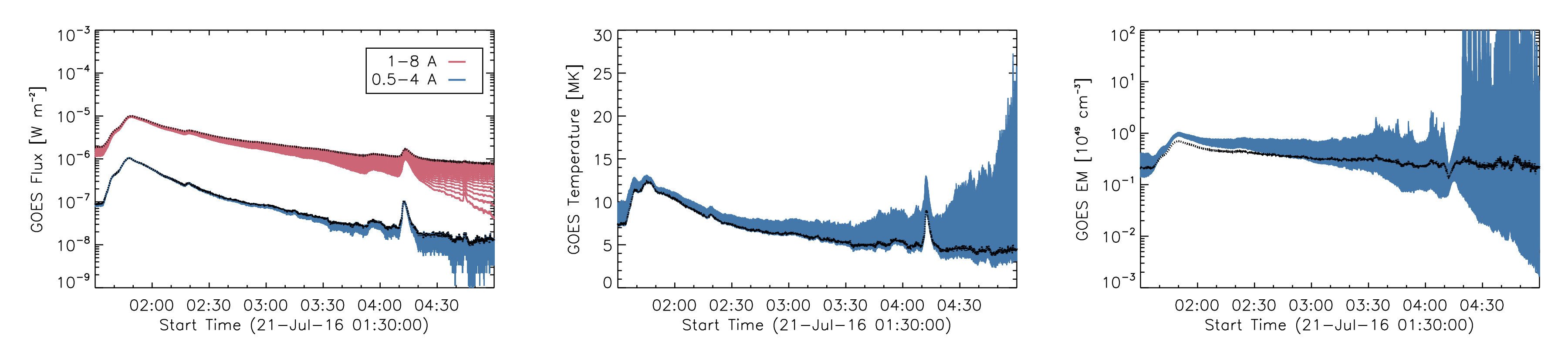}
    \includegraphics[width=0.95\linewidth]{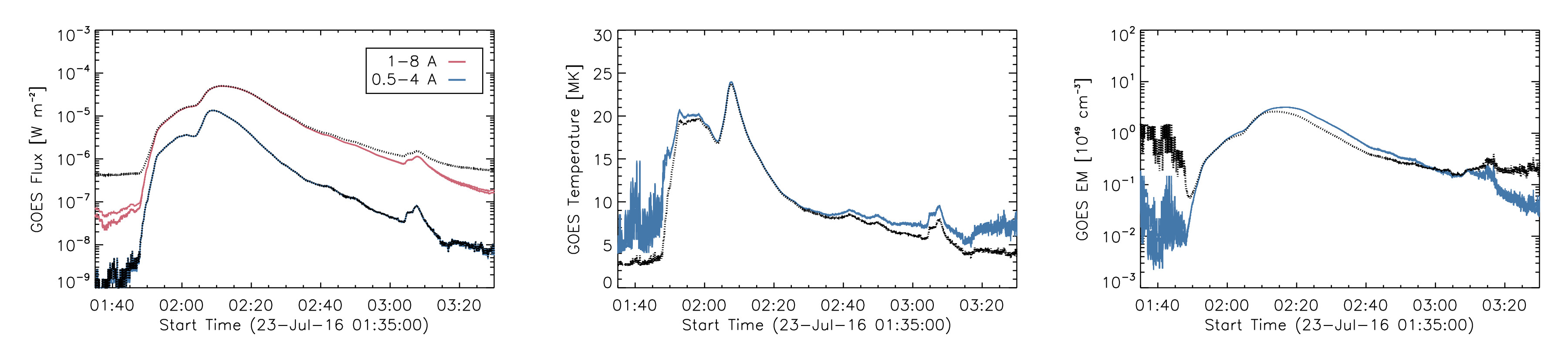}
    \includegraphics[width=0.95\linewidth]{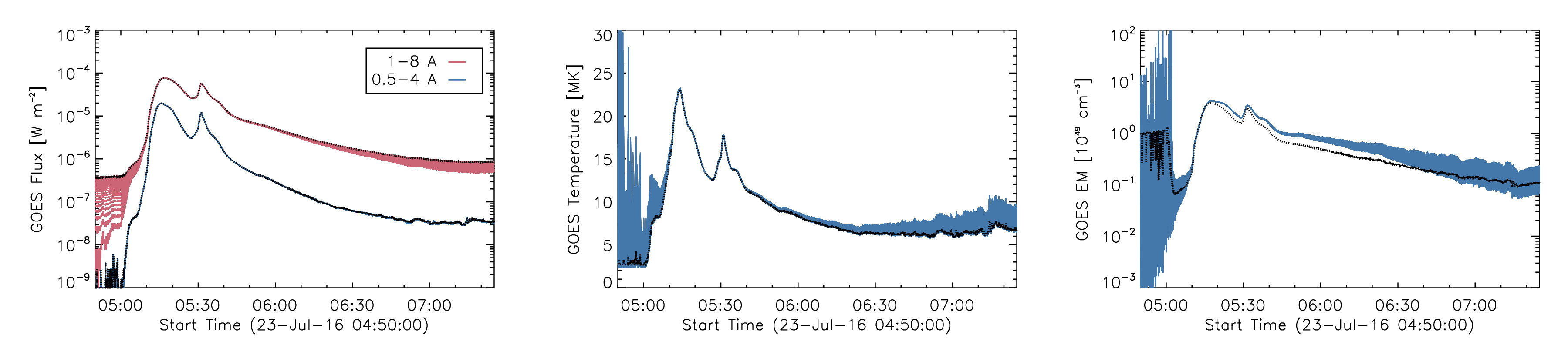}
    \includegraphics[width=0.95\linewidth]{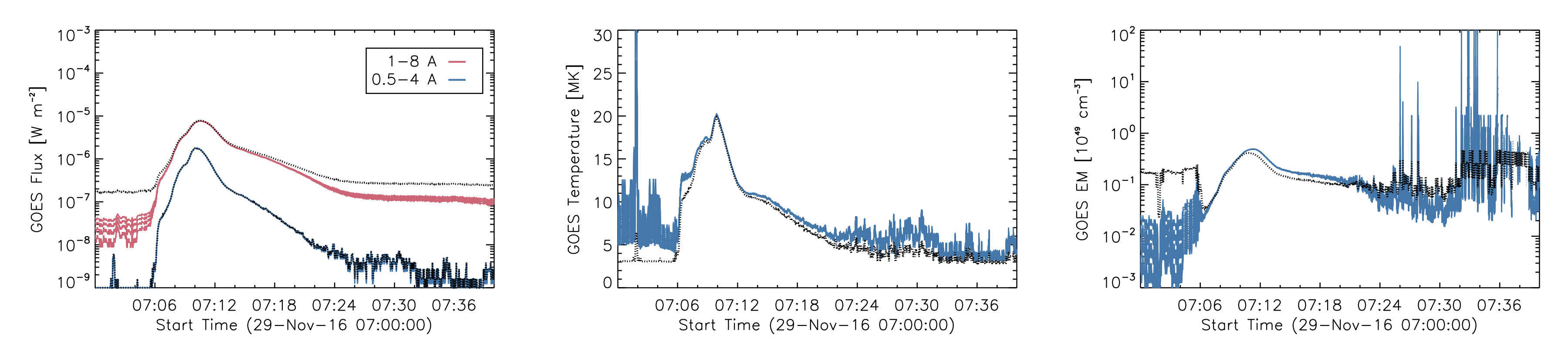}
    \includegraphics[width=0.95\linewidth]{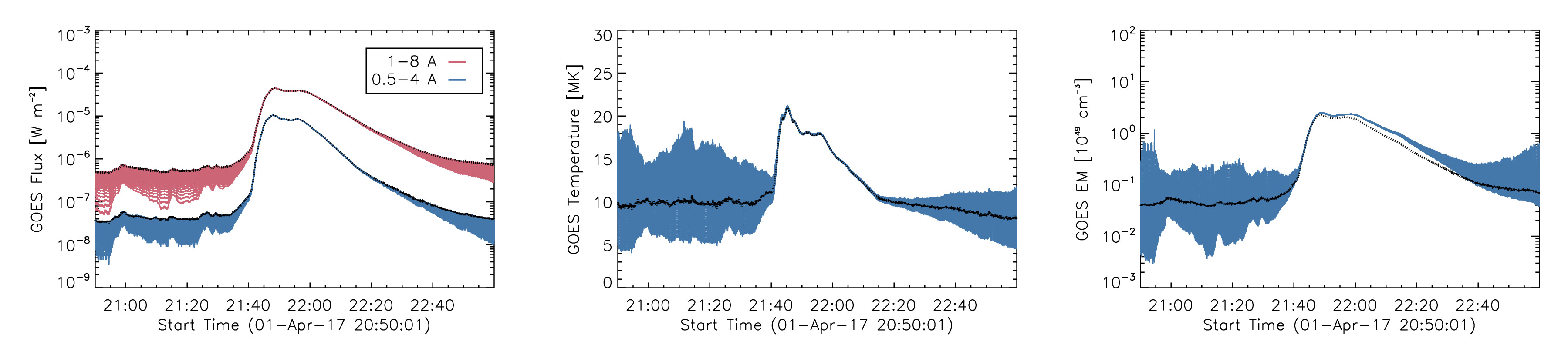}
    \caption{GOES light curves, temperatures, and EMs for the five flares under study.  The plots in the left column show the light curves in both channels, the middle column the GOES-derived temperatures (MK), and the right column the GOES-derived EMs (10$^{49}$ cm$^{-3}$).  The black lines show the values without background subtraction, while the colors are the values found from background subtraction with the TEBBS algorithm \citep{ryan2012}, showing the range of values consistent with valid estimates of the background flux.}
    \label{fig:obs_goes}
\end{figure*}

\begin{figure*}
  \centerline{\includegraphics[width=0.98\linewidth]{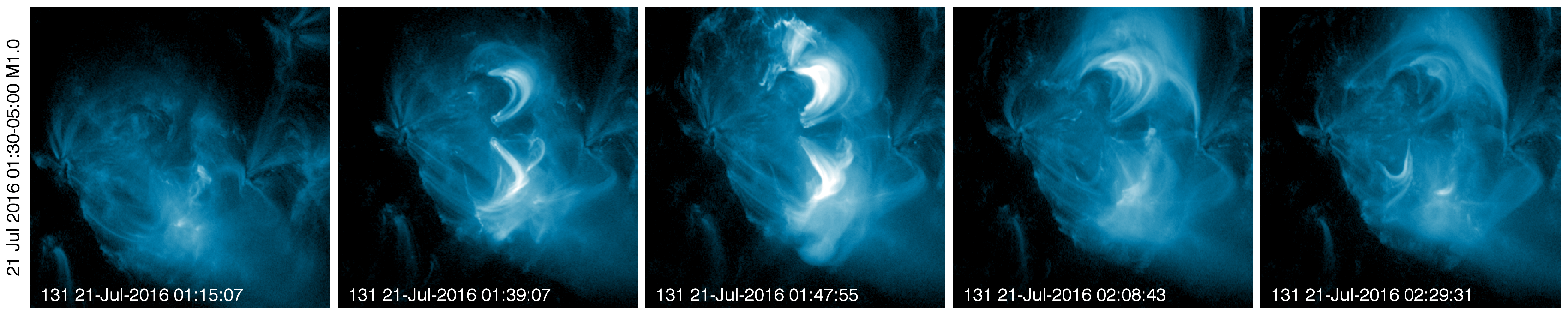}}
  \centerline{\includegraphics[width=0.98\linewidth]{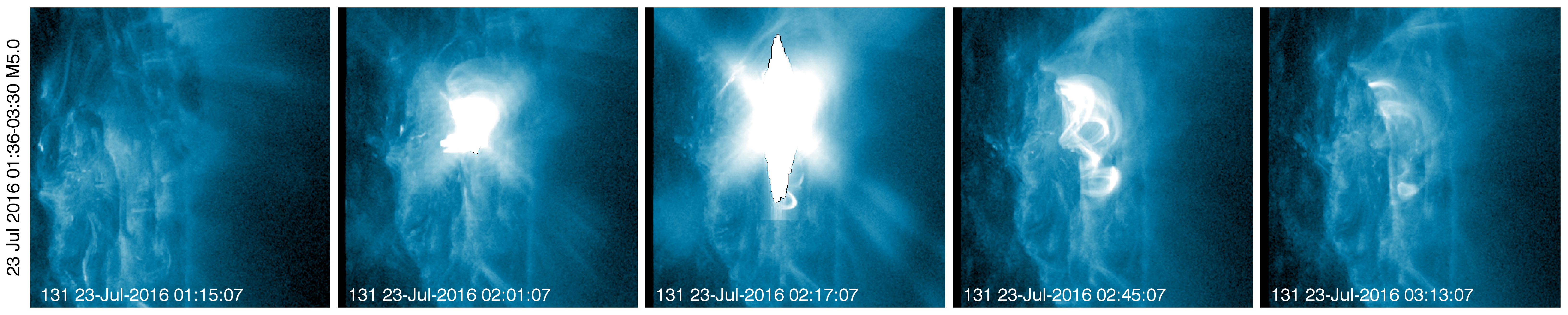}}
  \centerline{\includegraphics[width=0.98\linewidth]{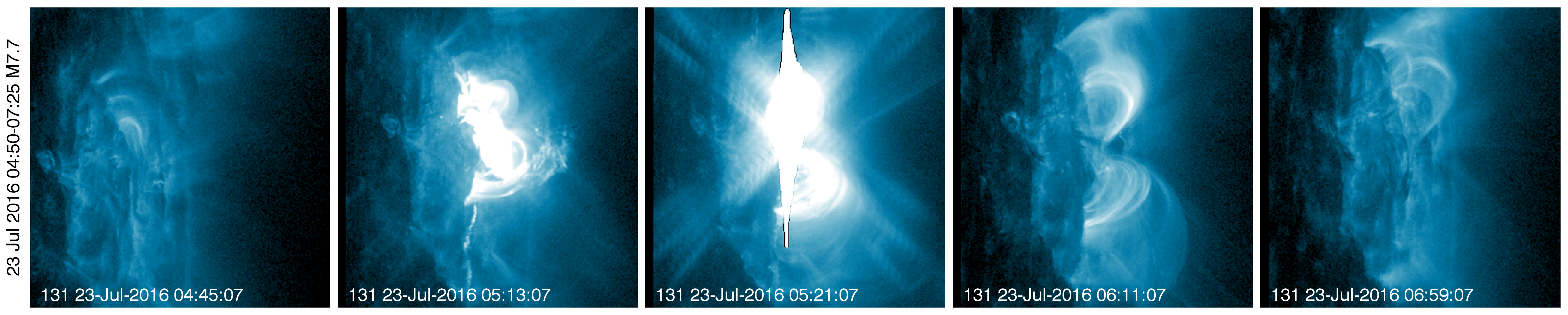}}
  \centerline{\includegraphics[width=0.98\linewidth]{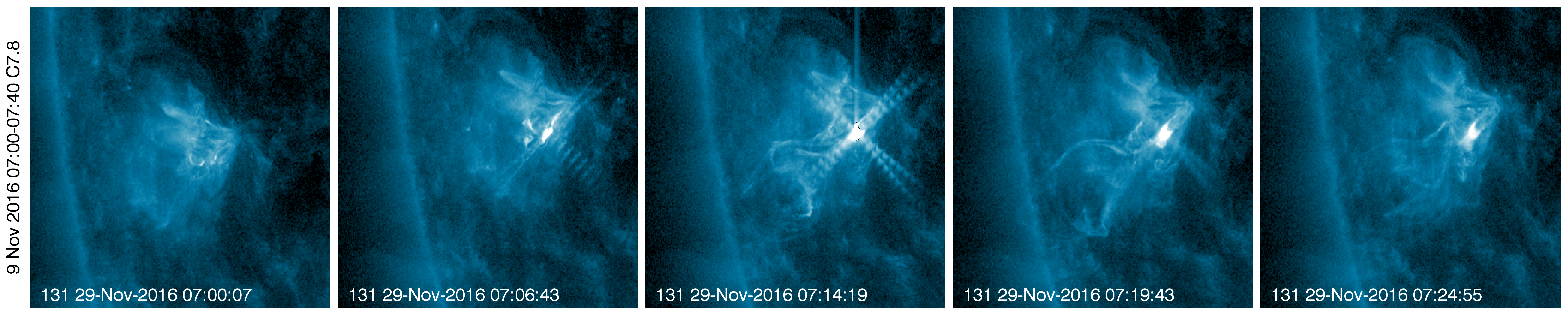}}
  \centerline{\includegraphics[width=0.98\linewidth]{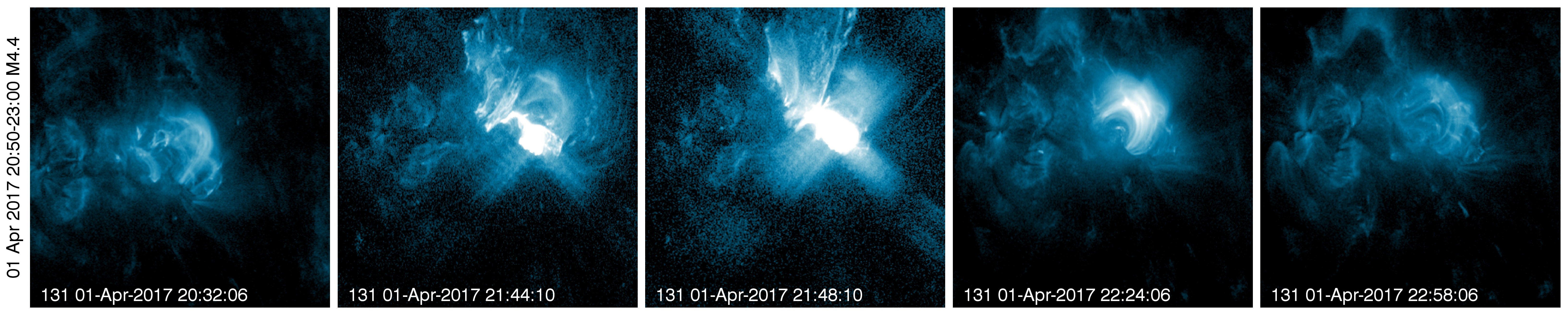}}
  \caption{SDO/AIA 131 context images for the five flares under study (each row). The times shown correspond approximately to pre-flare, rise, peak AIA flux, decay, and event end, respectively. The field of view for each image is $240\arcsec\times240\arcsec$. The intensity scaling for each row is the same.}
  \label{fig:obs_aia}
\end{figure*}

In Figure \ref{fig:obs_goes}, we show the GOES-15 XRS light curves in both channels, GOES-derived temperatures, and GOES-derived emission measures (EMs).  The left  column shows the GOES light curves in 1--8\,\AA\ (red) and 0.5--4\,\AA\ (blue) for each of the five flares, while the middle and right columns show the temperatures (MK) and EMs (10$^{49}$ cm$^{-3}$).  The dotted black lines mark the values of each without background subtraction.  The color ranges show the possible values derived from background subtraction with the TEBBS algorithm \citep{ryan2012}, which provides a better estimate of the temperature and EM, while also giving an estimate of the uncertainty.  In Figure \ref{fig:obs_aia}, we show context images of each of the five flares are shown in SDO/AIA 131.  In flares, the hot plasma in the 131 channel is primarily Fe XXI emission at $\approx 12$\,MK, while the cool emission is dominated by Fe VIII \citep{odwyer2010}.  Respectively in each row, we show the pre-flare, rise, peak in AIA, decay phase, and the end of the event for each case.  The field of view is $240\arcsec\times240\arcsec$, and the intensity scaling is the same for all of the flares.

\begin{figure*}[t]
\centering
\includegraphics[width=0.48\textwidth]{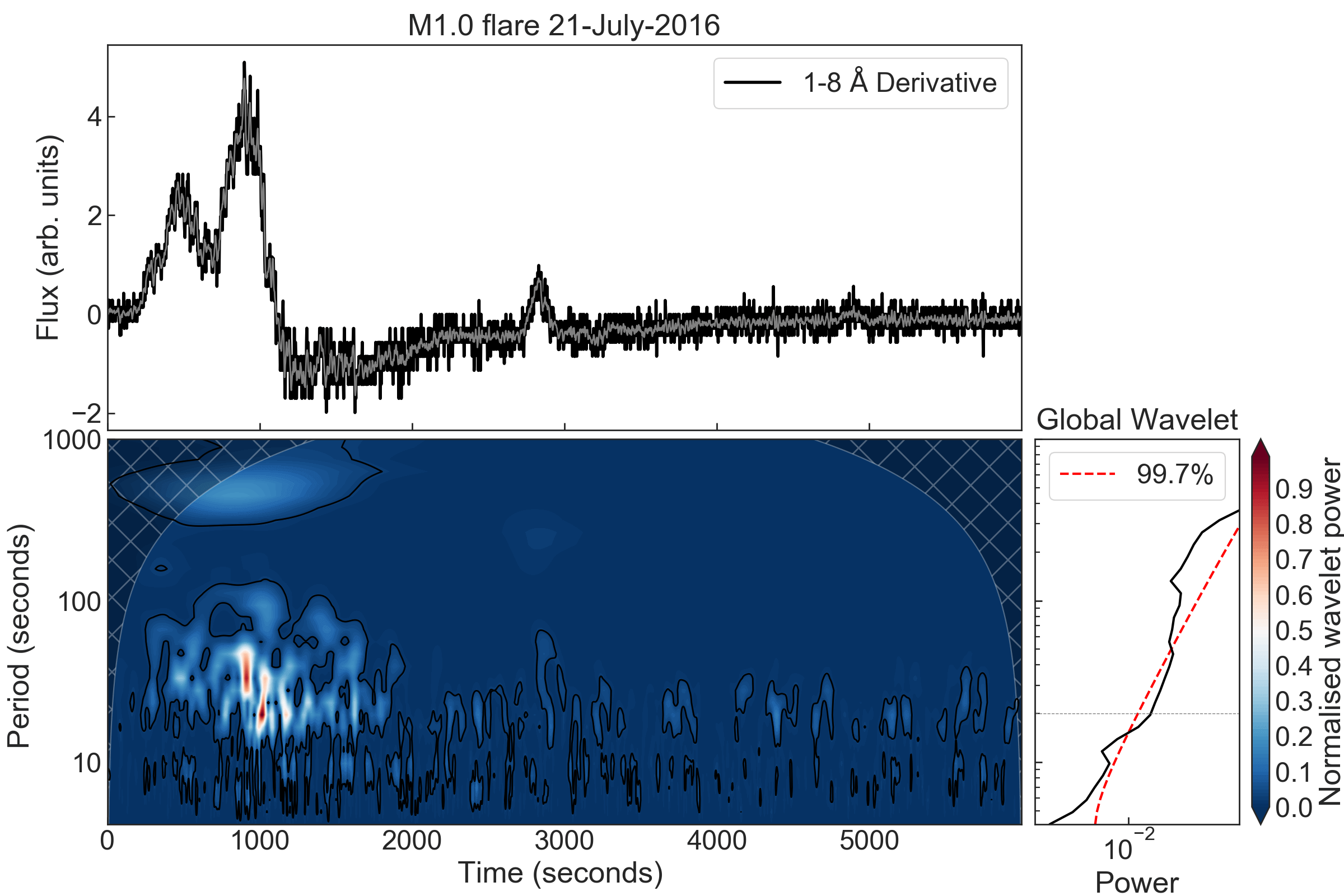}
\includegraphics[width=0.48\textwidth]{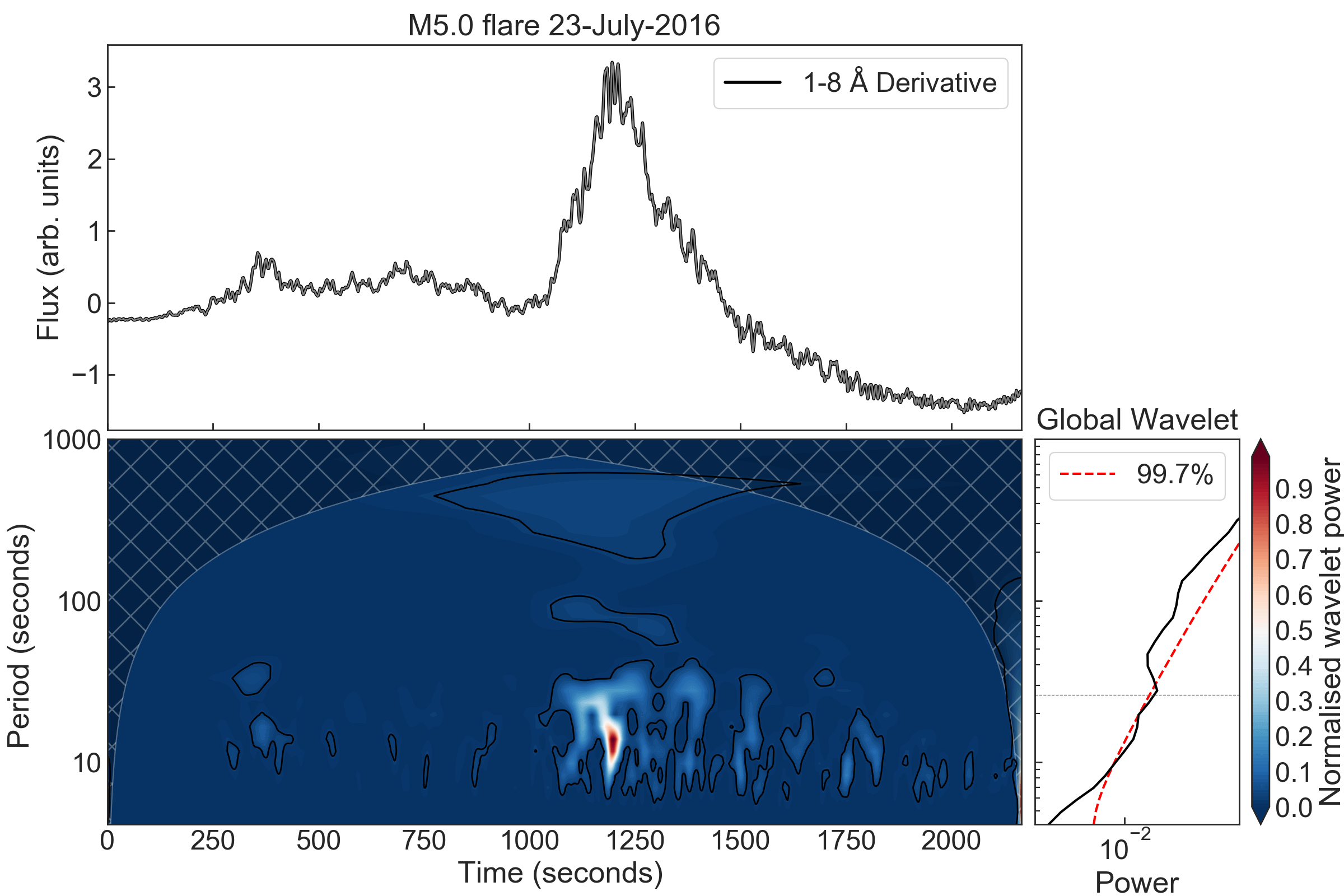}
\includegraphics[width=0.48\textwidth]{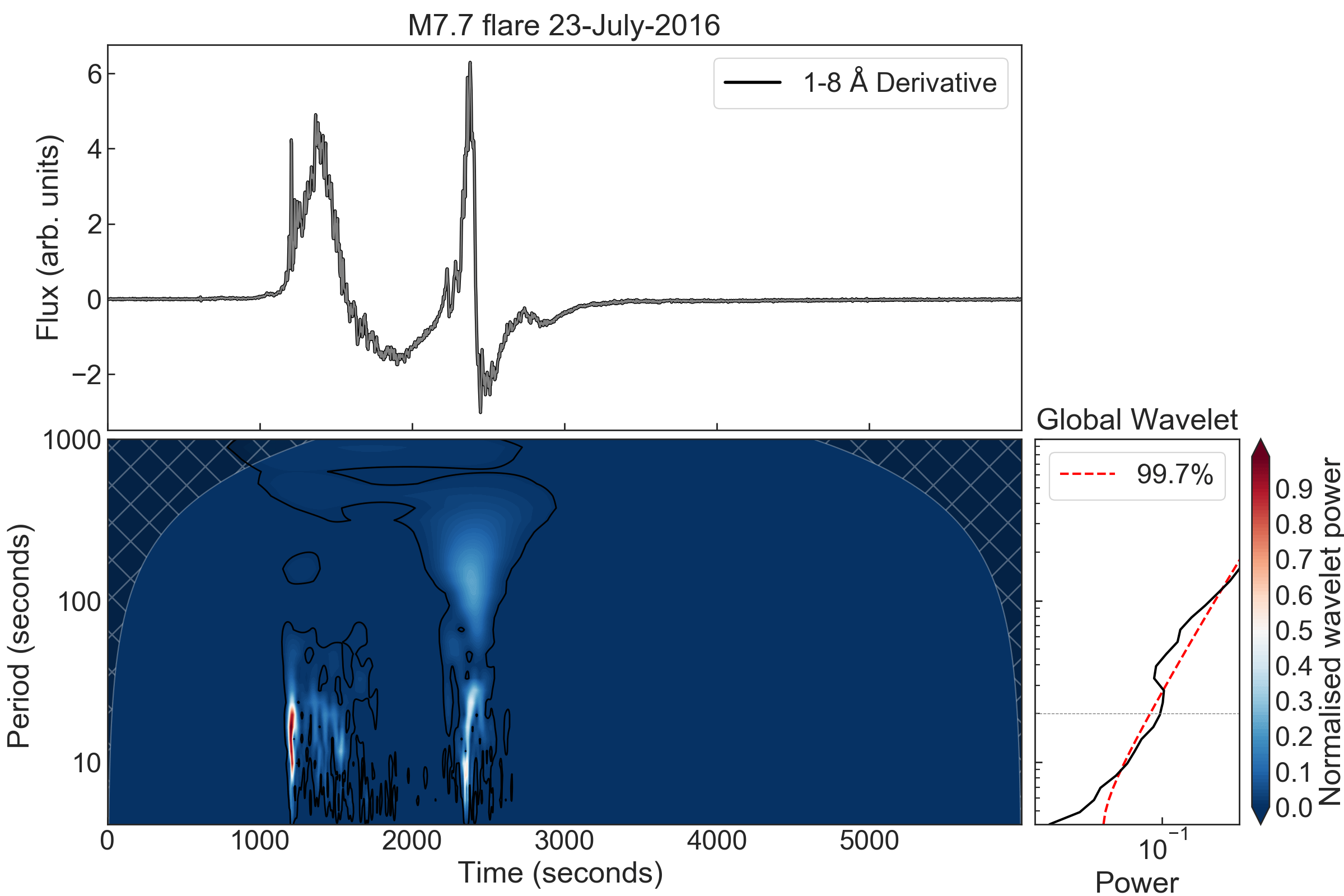}
\includegraphics[width=0.48\textwidth]{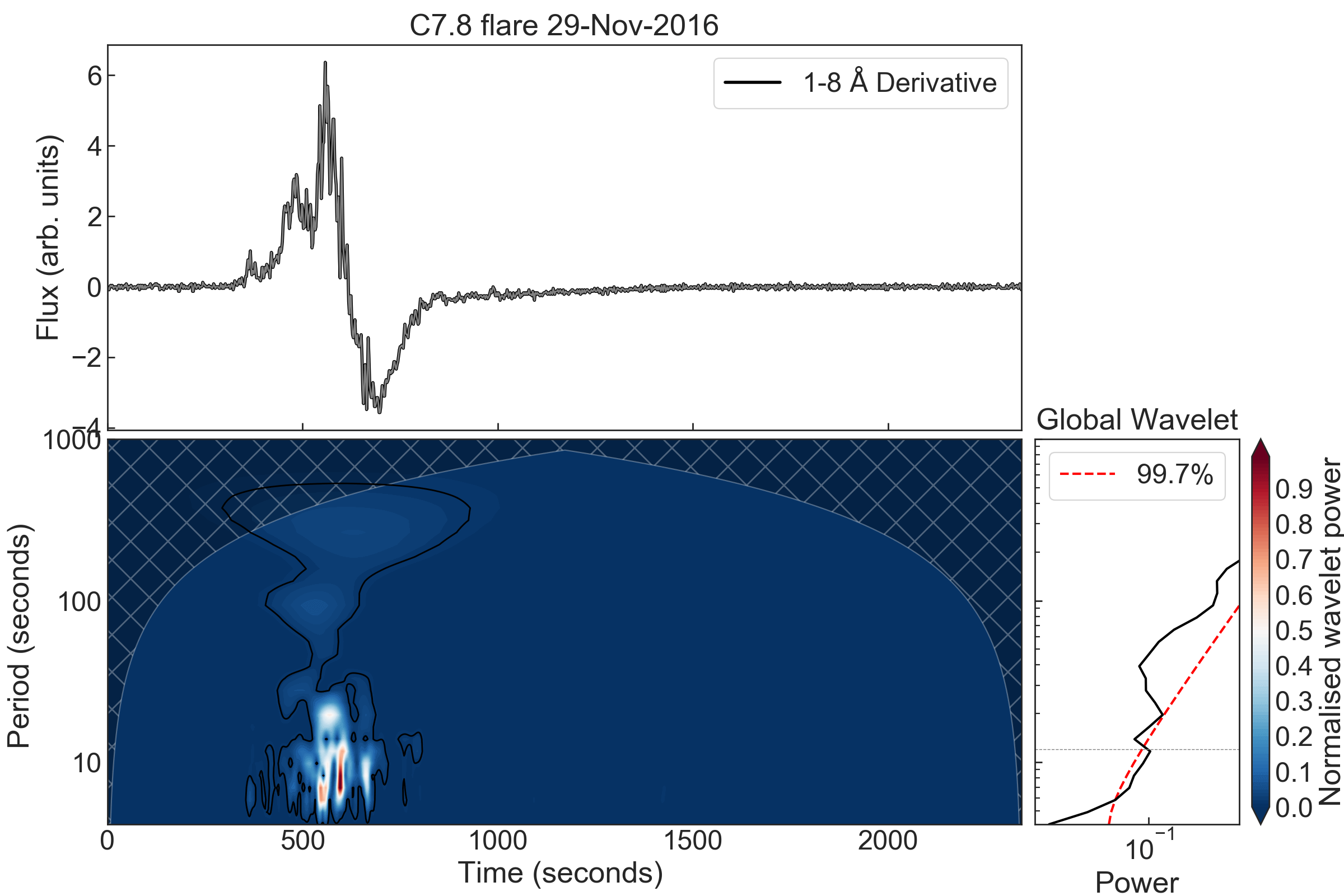}
\includegraphics[width=0.48\textwidth]{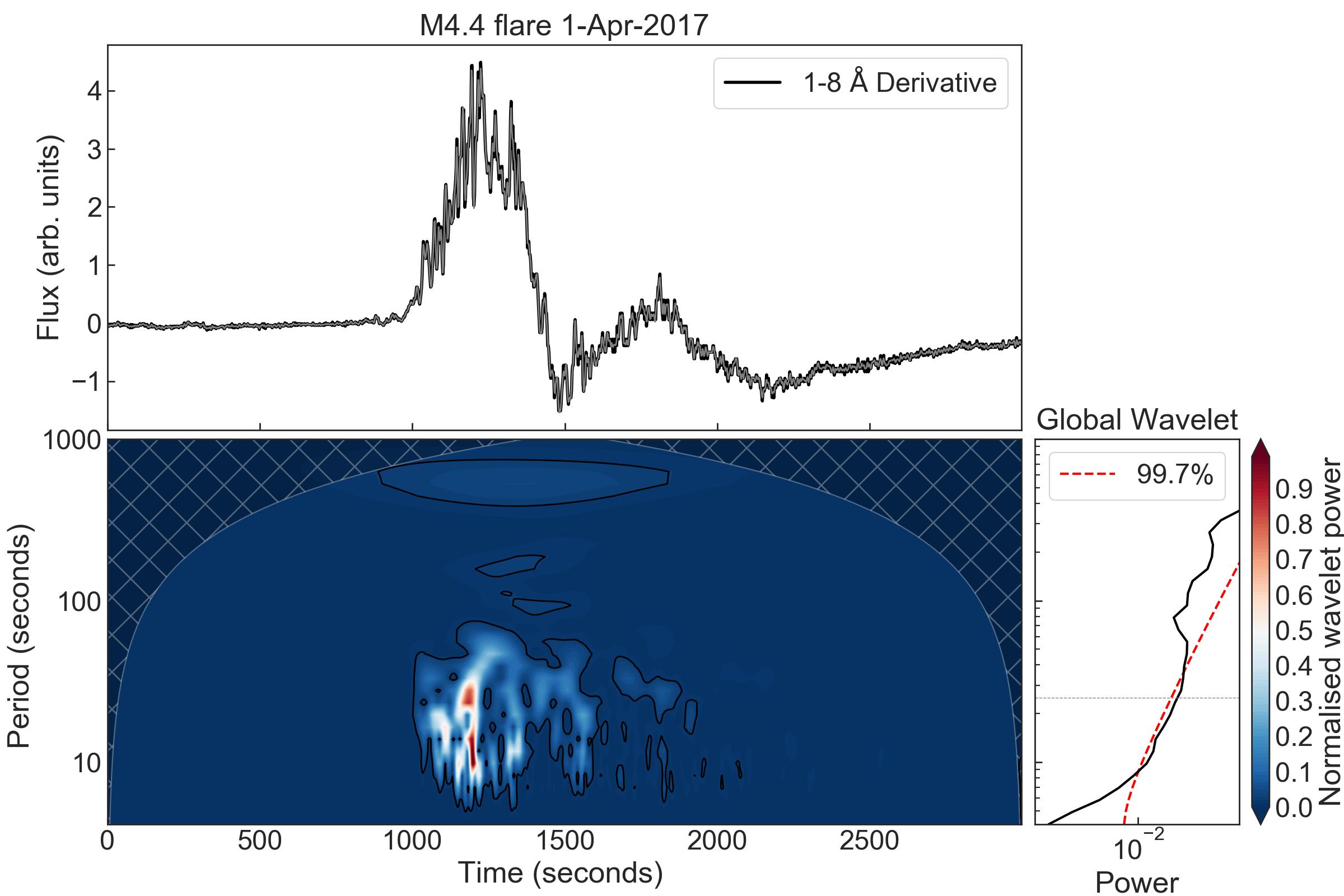}
\caption{Wavelet analysis of the GOES/XRS derivative light curves for the five observed flares (in chronological order, from top left).  For each flare, we show the time derivative (top), the wavelet power spectrum (bottom left), and global wavelet spectrum (bottom right).  There are peaks of significant wavelet power at 20, 26, 20, 12, and 25 s periods, respectively, which can be seen in the global wavelet spectra where the wavelet power exceeds the 99.7\% significance level, marked with a red dashed line. \label{fig:obs_qpp}}
\end{figure*}

A common feature of the soft X-ray and EUV flaring emission is the presence of small amplitude QPPs often identified in the time derivative or detrended form of flaring lightcurves.  Many recent studies have observed such QPPs in GOES XRS observations \citep[e.g.][]{dolla2012,simoes2015, inglis2016, dennis2017}, as well as in other X-ray passbands \citep{chowdhury2015}.  For the flares of interest we search for the presence of QPPs in the GOES/XRS light curves and characterize the dominant pulsation timescales using wavelet analysis.  The time derivative of the lightcurves was used in this analysis to highlight the QPPs, and the standard \cite{torrence_compo} wavelet analysis was performed.  The significance of wavelet power was tested against a power-law (`red-noise') background model.  The results of this is shown in Figure \ref{fig:obs_qpp} for each of the five flares in this paper, in chronological order from top left. In each case, the top plot shows the time derivative of the lightcurve, and the bottom plots show the corresponding wavelet power spectrum and global wavelet spectrum.  For all five flares, there are peaks above the 99.7\% significance level in the global wavelet power with periods (averaged over the impulsive phase) of 20 s, 26 s, 20 s, 12 s, and 25 s, respectively.  Although the underpinning mechanism causing QPPs is not fully understood at present, as discussed further in Section \ref{subsec:qpp}, one possibility is that they are due to the reconnection and energization of new loops (see \citealt{nakariakov2009, mclaughlin2018}).  We will use this interpretation in the modeling work to show that it can consistently reproduce the QPPs in the GOES/XRS light curves.

\begin{figure*}[t]
    \centering
    \includegraphics[width=0.95\linewidth]{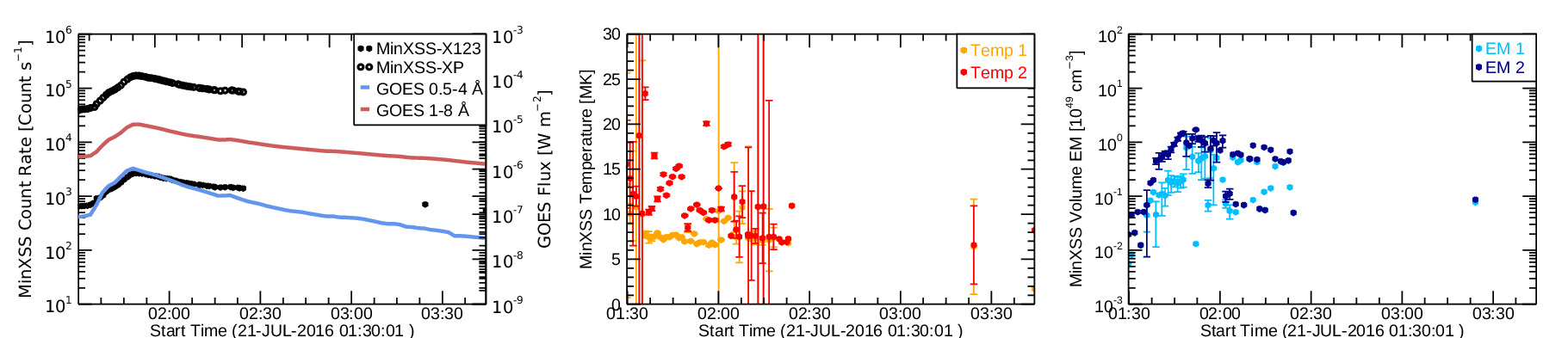}
    \includegraphics[width=0.95\linewidth]{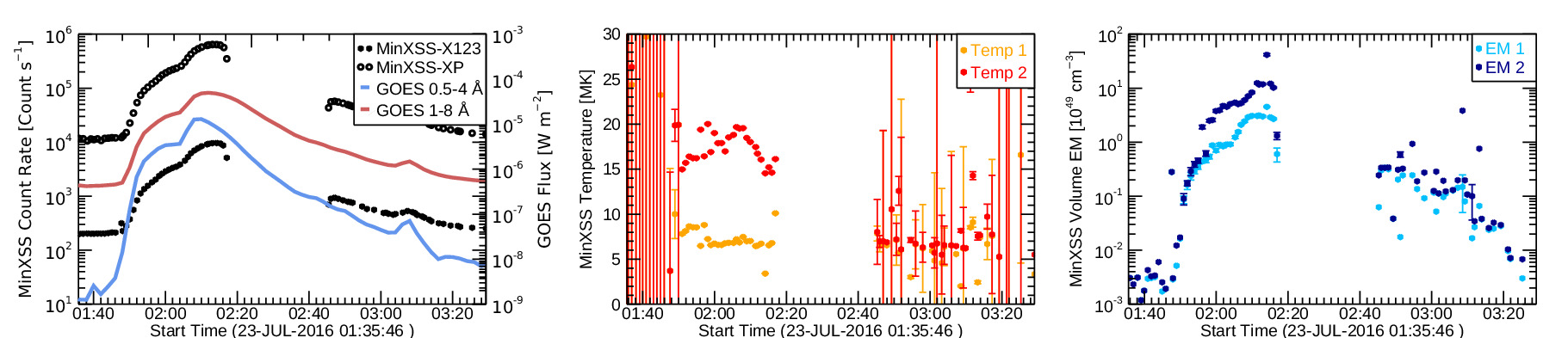}
    \includegraphics[width=0.95\linewidth]{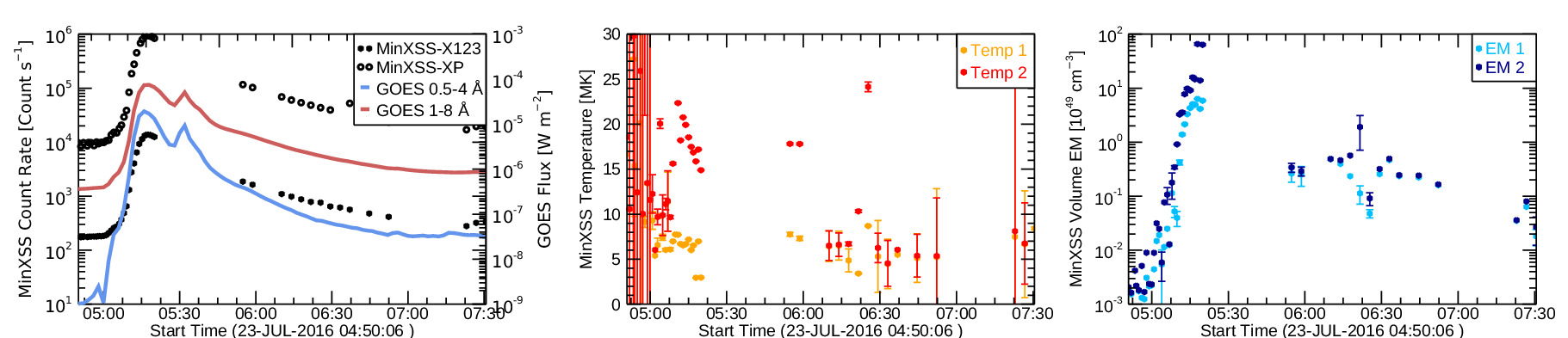}
    \includegraphics[width=0.95\linewidth]{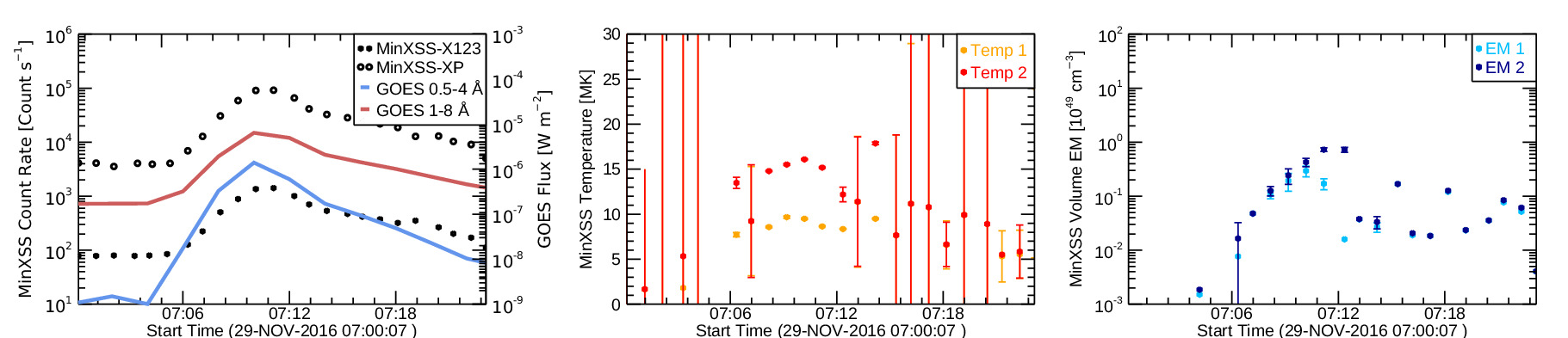}
    \includegraphics[width=0.95\linewidth]{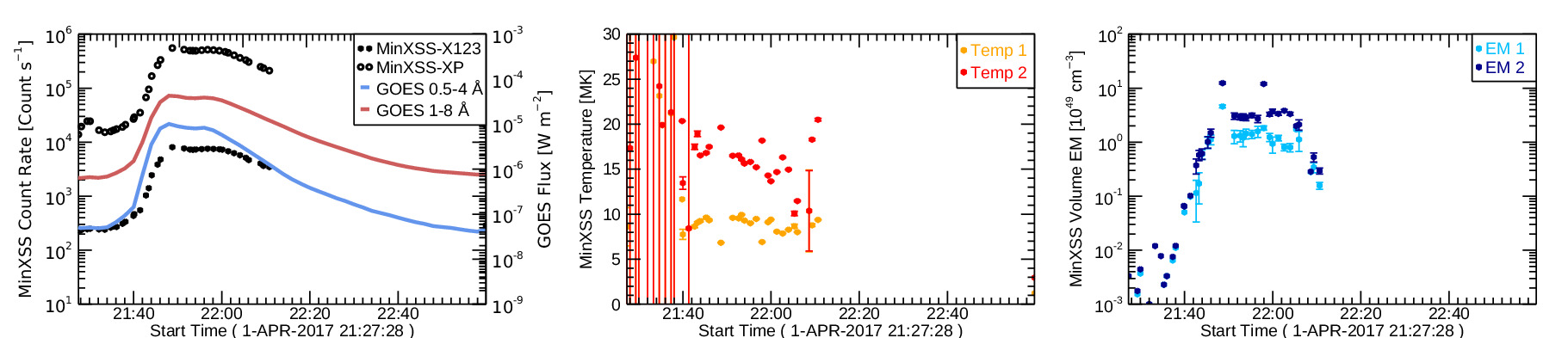}
    \caption{MinXSS-1 light curves, temperatures, and EMs for the five flares under study.  The plots in the left column show MinXSS-1 1-minute average count rate compared to the GOES light curves in both channels, while the middle and right column show the temperature and EM derived with a two-temperature fit to the MinXSS-X123 spectra.}
    \label{fig:obs_minxss}
\end{figure*}

Finally, these flares were also observed by MinXSS-1, and are perhaps the best observed flares in the MinXSS-1 catalogue.  The MinXSS-1 CubeSat \citep{mason2016}, which flew from 16 May 2016 until 6 May 2017, took spatially integrated X-ray spectra with good time coverage for the duration of each of these flares.\footnote{The MinXSS data can be downloaded from \url{http://lasp.colorado.edu/home/minxss/data/}.}  MinXSS-1 had an X-ray spectrometer (X123) with solar soft X-ray sensitivity from 0.7--20 keV with 0.15 keV spectral resolution and an X-ray photometer (XP) spanning 0.5--30 keV \citep{moore2018}.  In Figure \ref{fig:obs_minxss}, we show the light curves observed with MinXSS-X123 and MinXSS-XP for each of the five flares, along with the temperatures and EMs derived from a two-temperature fit to the X123 spectra.  The left column shows the MinXSS-1 light curves using a 1-minute average, as well as the observation times, overlaid on the GOES light curves.  The middle column shows the fitted temperature for each temperature component, with the high temperature component in red and low temperature component in yellow.  The right column shows the fitted EM for the two temperature components, in light and dark blue, as labeled.  We have also fit the enhancement of low first ionization potential (FIP) elements above photospheric values in these flares, shown in Table \ref{tab:obs_summary} (where FIP bias $f = 1.0$ corresponds to photospheric).  For all five flares, the EM of the low-temperature component is comparable to that found with GOES, while the high-temperature component has a much higher EM.  The fitted temperatures, however, show that the GOES temperature is close to the high-temperature component.  We note that the fits to the MinXSS spectra are noisy during the early and late phases, when the signal-to-noise ratio is not large after background subtraction.  This causes the fits to derived quantities such as the temperature to have large uncertainties.

\section{Modeling}

We wish to construct a model that is capable of recreating GOES light curves for a series of loops energized by a reconnection event.  In order to reconstruct the GOES light curves accurately, we explain a method to estimate the lengths of newly formed loops, the energy released on each loop, and the volume of each loop.  We then can constrain the model with some combination of spectra, light curves, and images from other data sets in order to better understand the physics of the flares.  We define a ``ribbon'' as the area of the chromospheric footpoints of the arcade structure, while we use the terms ``loop'' and ``thread'' interchangeably to refer to each individual flux tube that forms and is energized in succession in that arcade, rooted on the ribbon.

In this section, we show how we construct this flare model, and then in Section \ref{sec:simulations}, we test it for each of the five flares presented in Section \ref{sec:observations}, comparing synthetic MinXSS-1 spectra and AIA lightcurves to those observed.  We focus on multiple flares to understand what properties can improve the consistency with observations for all of them, rather than simply tweaking the parameters of a single flare for a better fit. 

\subsection{Loop Length Expansion Model}
\label{subsec:length}

We derive a model that we use to calculate the lengths of loops with time for a flare arcade as its ribbons spread apart, assuming that each flare has a simplistic two-ribbon geometry (see \textit{e.g.} Section 3 of \citealt{fletcher2011}).  In principle, we could determine the loop lengths directly from AIA 1600 or 1700\,\AA\ observations of the flare ribbons.  These images, however, usually saturate in large flares, and additionally would limit the model to flares observed since SDO was launched.  Ribbon dynamics are also difficult to measure in events close to or at the limb.  For these reasons we employ a model based on the ribbon properties derived from a large number of well-observed events.

\begin{figure}
\centering
\includegraphics[width=\linewidth]{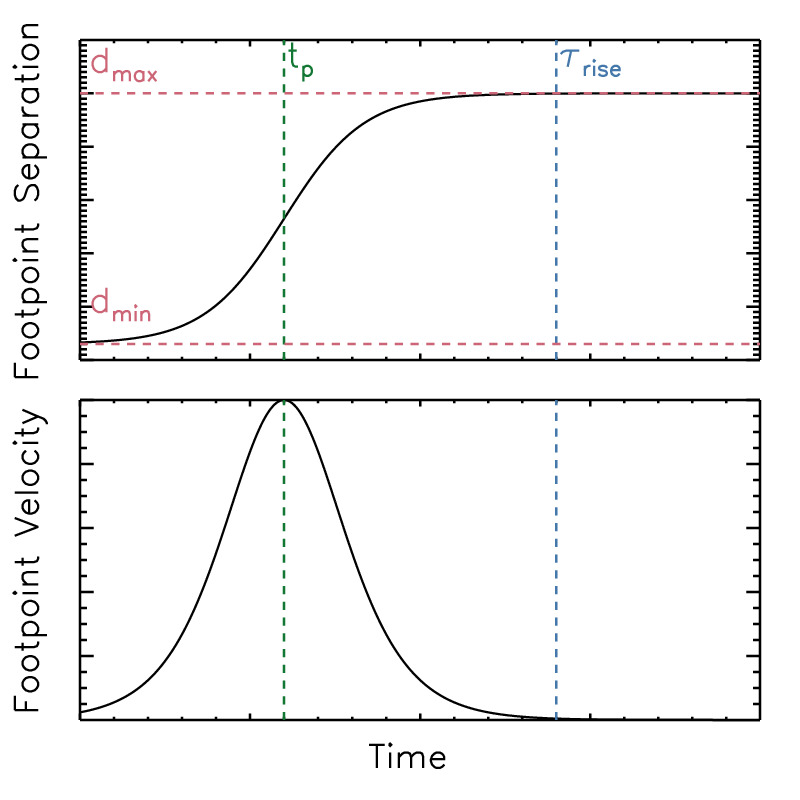}
\caption{The functional form of Equations \ref{eqn:fpsep} and \ref{eqn:fpvel}.  The footpoint separation $d$ increases from the minimum distance $d_{\text{min}}$ to the maximum $d_{\text{max}}$ in time $\tau_{\text{rise}}$.  The speed quickly rises to its maximum value in time $t_{p}$, then falling towards 0 in time $\tau_{\text{rise}}$.}
\label{fig:distance}
\end{figure}

We begin by assuming that the two ribbons are close together initially, such that the separation of loop footpoints is $d_{\text{min}}$, which then expands to some maximum separation $d_{\text{max}}$.  An appropriate function, chosen \textit{ad hoc}, is the hyperbolic tangent function.  That is, for the footpoint separation of a loop $d(t)$ we employ a function of the form $d(t) \propto \tanh{t}$.  The time-scale for it to reach its maximum value is roughly the rising time $\tau_{\text{rise}}$ of the GOES light curve.  For example, see Figure 2 of \citet{toriumi2017}, where the ribbon expansion stops near the peak of the flare.  We therefore write
\begin{align}
    d(t) \propto \tanh{\Bigg(\frac{\omega (t - t_{p})}{\tau_{\text{rise}}}\Bigg)}
\end{align}
\noindent where we derive $\omega$ below and $t_{p}$ is the time of peak velocity ($< \tau_{\text{rise}}$).  Next, to ensure that it reaches the appropriate maximum and minimum values, we scale the function as
\begin{align}
    d(t) = \frac{d_{\text{max}}-d_{\text{min}}}{2} \tanh{\Big(\frac{\omega (t - t_{p})}{\tau_{\text{rise}}}\Big)} + \frac{d_{\text{max}} + d_{\text{min}}}{2}
    \label{eqn:fpsep}
\end{align}
\noindent At early times (much less than $t_{p}$), this gives a value of $d_{\text{min}}$, and for times greater than $\tau_{\text{rise}}$ this tends to $d_{\text{max}}$.  To check that the velocities are reasonable, we additionally calculate the time derivative of the separation:
\begin{align}
    v(t) = \frac{\omega}{\tau_{\text{rise}}} \frac{d_{\text{max}} - d_{\text{min}}}{2}\Bigg[1 - \tanh^{2}{\Big(\frac{\omega (t - t_{p})}{\tau_{\text{rise}}}}\Big)\Bigg]
    \label{eqn:fpvel}
\end{align}
\noindent Figure \ref{fig:distance} shows the functional forms.  In time $\tau_{\text{rise}}$, the distance increases from $d_{\text{min}}$ to $d_{\text{max}}$, while the velocity rises from 0 to its maximum value $v_{p} = \frac{\omega}{\tau_{\text{rise}}}\frac{d_{\text{max}}-d_{\text{min}}}{2}$, and then back towards 0.  This then gives a definition for $\omega$: it determines the peak velocity of the ribbon separation, and can either be calibrated against observations or chosen with a reasonable value.  Compare this model to either \citet{asai2004} or \citet{hinterreiter2018}, who both find similar trends in the ribbon separation and velocities.  Of course, many flares have rather different topologies that would not fit this idealized model, so caution is warranted.

However, we want a model for the loop lengths, not only footpoint separations.  Assuming that the loops that form are semi-circular, we can write that the total loop length $(2L) = \frac{\pi d(t)}{2}$.  Therefore, we have
\begin{align}
    (2L)(t) = \frac{\pi}{4} \Bigg[(d_{\text{max}}-d_{\text{min}}) \tanh{\Big(\frac{\omega (t - t_{p})}{\tau_{\text{rise}}}\Big)} + d_{\text{min}} + d_{\text{max}}\Bigg]
\end{align}

Finally, we wish to eliminate $d_{\text{max}}$, which we first replace with $d_{\text{ribbon}}$, defined in \citet{toriumi2017} as the centroid separation of the two ribbons.  By definition, $d_{\text{max}} = 2 d_{\text{ribbon}} - d_{\text{min}}$.  Further, from that same paper, we know that the ribbon separation is related to the FWHM of the GOES light curves $d_{\text{ribbon}} \propto \tau_{\text{FWHM}}$.  Using that relation, we can rewrite $d_{\text{max}} = 2 d_{\text{ribbon}} - d_{\text{min}} = 2 a \tau_{\text{FWHM}} - d_{\text{min}}$, where $a$ is the fitted coefficient from the FWHM-$d_{\text{ribbon}}$ relation \citep{toriumi2017,reep2017}.  Therefore, we finally have the function:
\begin{align}
    (2L)(t) = \frac{\pi}{2} \Bigg[ (a \tau_{\text{FWHM}} - d_{\text{min}}) \tanh{\Bigg(\frac{\omega (t - t_{p})}{\tau_{\text{rise}}}\Bigg)} + a \tau_{\text{FWHM}}\Bigg]
\end{align}
\noindent where the two time-scales are easily measured from GOES data for any flare.  $d_{\text{min}}$ can either be measured directly from AIA/1600 data or assumed (for example, \citealt{reep2017} assume $d_{\text{min}} = 3$\,Mm).  

We then have a relatively simple functional form to describe the expanding loop lengths.  We use this model for all the flares in this paper.

\subsection{Heating Rate \& Volume}
\label{subsec:heating}

We use GOES light curves to determine the heating rate and volume for a succession of small arcades being energized over the course of the flare.  With the model of ribbon expansion, we can estimate an appropriate length for a loop at any given time, but we still need to determine the total heating rate and area to run an appropriate simulation and compute the expected emission in physical units.  In this section, we explain the method for doing so, which closely follows \citet{warren2005} and \citet{warren2006}.

Conceptually, we infer the appropriate heating rate and volume using the ratio and magnitude of the observed GOES fluxes at fixed intervals in the observations.  \citet{warren2004} derived scaling laws that can estimate the GOES intensities from the energy release:
\begin{align}
F_{\text{1--8\,\AA}} &\approx 3.7 \times 10^{-35} \Bigg( \frac{E L}{V}\Bigg)^{7/2} \frac{V}{L^{2}} \nonumber \\ 
 F_{\text{0.5--4\,\AA}}   &\approx 4.4 \times 10^{-42} \Bigg( \frac{E L}{V}\Bigg)^{9/2} \frac{V}{L^{2}}
\end{align}
\noindent where $E$ is the energy release, $V$ the total volume, and $L$ the loop length.  These scalings can be inverted to give an estimate of the volume and energy release given the flux measurements in both GOES channels.  For simplicity, let the constants be written $C = 3.7 \times 10^{-35}$ and $D = 4.4 \times 10^{-42}$.  Then, we can solve for $V$ and $E$ to find:
\begin{align}
    V &= \Bigg(\frac{F_{\text{1--8\,\AA}}}{C} \Bigg)^{9/2} \Bigg(\frac{D}{F_\text{{0.5--4\,\AA}}} \Bigg)^{7/2} L^{2} \nonumber \\
    E &= \Bigg(\frac{F_{\text{1--8\,\AA}}}{C} \Bigg)^{5/2} \Bigg(\frac{D}{F_{\text{0.5--4\,\AA}}} \Bigg)^{3/2} L
\end{align}
\noindent These inversions show that for a given loop length $L$, the energy release $E$ and volume $V$ that reproduce the GOES fluxes and their ratio can be determined.  This allows us to construct a solution that reproduces GOES light curves with a succession of arcades being energized.  In practice, however, we run a series of hydrodynamic simulations, pre-calculate the GOES light curves from each simulation, and then interpolate to find the best-fit heating rate and volume. The reasoning is the same as with the scaling laws, but the results are more accurate. 

We use this information to determine the heating rate and volume for a series of arcades that can reproduce the GOES light curves from an observed flare.  Starting at the beginning of the flare and the first energized loop, we first calculate its individual light curves using the best fit for the heating rate and volume, and then subtract it off of the total GOES emission, creating a residual light curve.  We then fit the second loop parameters to this residual light curve, \textit{etc.}, iterating over the entire flare duration.  One subtlety is that the observed GOES fluxes are not taken from the current time in the simulation, but a ``look-ahead'' time that reflects the time it takes for plasma to evaporate and fill the arcade. This look-ahead time is the heating duration for the first simulation and the time of the previous peak flux for subsequent simulations.

\section{Simulations}
\label{sec:simulations}

\begin{figure*}
    \centering
    \includegraphics[width=0.28\linewidth]{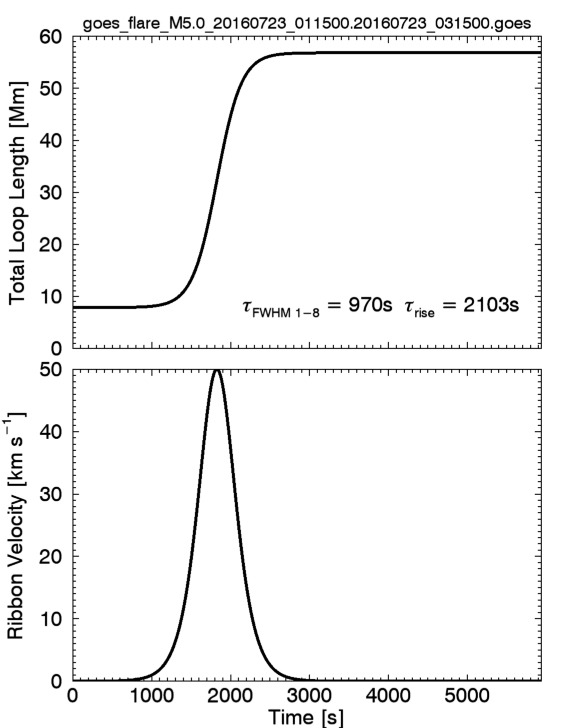}
    \includegraphics[width=0.69\linewidth]{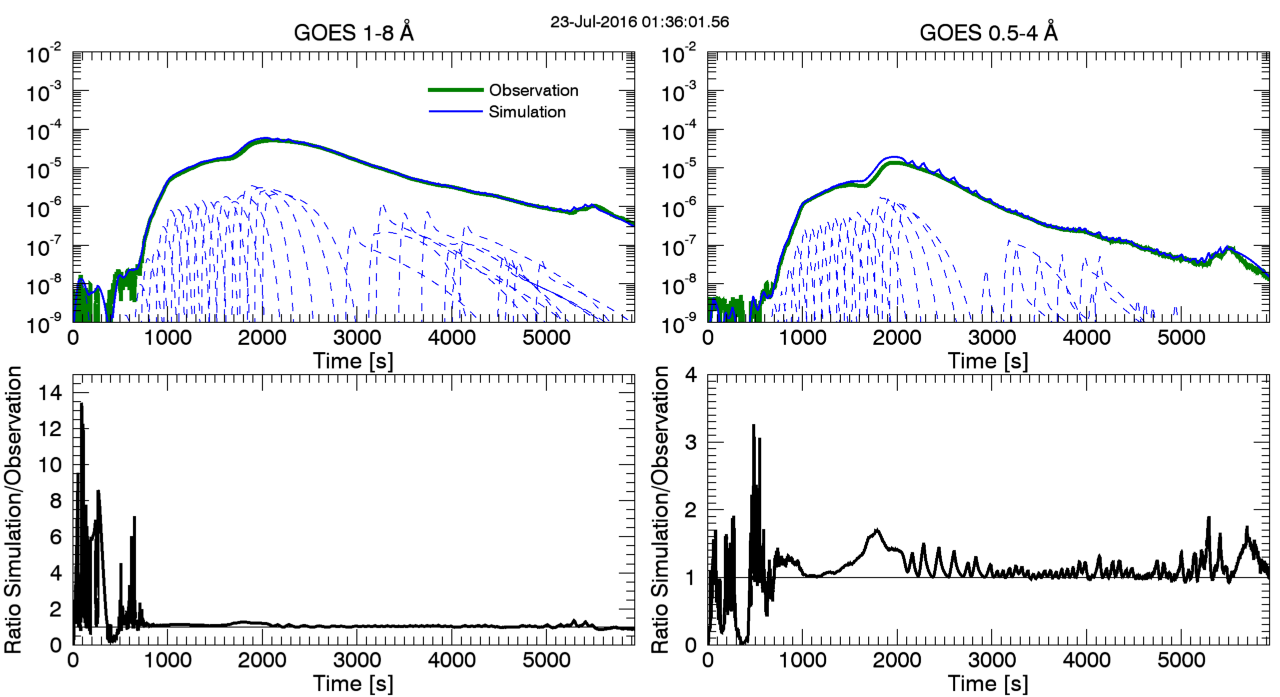}
    \caption{Synthesized and observed GOES light curves for the 2016 July 23 M5.0 flare.  The left plots show the loop expansion and velocity model, where a peak velocity of 50 km s$^{-1}$ has been assumed.  At right, the top plots show each GOES channel (1--8\,\AA\ left, 0.5--4\,\AA\ right), while the bottom plots show the ratio of the simulated vs observed fluxes.  The solid green lines show the observed light curves, while the solid blue lines show the total simulated flux.  Each dashed blue line shows the contribution of a few selected threads to the total light curve, where each thread is heated for 200 seconds.  The agreement is excellent in both channels, which is by construction.  A video of the construction of the light curves is available in the online journal.  \label{fig:modelM5t60}}
    \end{figure*}

To simulate a flare comprised of many hundreds or thousands of loops, all with varying loop lengths and heating rates, we employ the 0D ebtel++ code \citep{barnes2016}\footnote{\url{https://github.com/rice-solar-physics/ebtelPlusPlus}}, based on the older IDL-based EBTEL \citep{klimchuk2008}.  The code solves the coronally-averaged hydrodynamic equations to determine density, ion temperature, and electron temperature as a function of time for a loop subjected to heating by any arbitrary heating function.  The code is quick and robust, capable of running thousands of simulations within minutes on a desktop computer.  We have set up a grid of ebtel++ simulations with loops of lengths ranging from 1 to 200 Mm, heated by impulsive bursts of energy ranging from $10^{-2}$ to $10^{3}$ erg s$^{-1}$ cm$^{-3}$, for durations lasting 20, 60, and 200 seconds (flat heating rate), as well as two cases of 60 and 200 seconds flat heating with a 60 and 200 second decay phase.  We have also simulated these with FIP enhancement factor of 1.0 through 4.0, using the data set of \citet{asplund2009} for the photospheric case, and enhancing the low FIP elements appropriately -- which is done in the radiative losses in the ebtel++ simulations, the synthetic GOES light curve calculations (see appendix), and in the post-processing calculation of MinXSS-1 spectra.\footnote{We modify the abundance set from \citet{asplund2009}.  A FIP enhancement factor of 4.0 corresponds to coronal abundances for low FIP elements.  For high FIP elements, we do not modify the abundance values.  This leads to a small inconsistency with the commonly used coronal abundance set by \citet{feldman1992}: the high FIP elements in the Feldman abundance set have a different abundance than those in the Asplund set.  Presumably this discrepancy is because there were differences in the methodologies of the two studies, rather than there being a real variation in the high FIP abundances.}  We have run a total of 1,750,000 simulations (250 heating rates, 200 loop lengths, 5 heating durations, and 7 abundance sets), hence the choice of ebtel++ for this task rather than a higher dimensional model.

Combining the expanding loop length model described in Section \ref{subsec:length} with the heating model described in \ref{subsec:heating}, we can then construct a full simulation of a flare arcade for any observed flare.  We construct such a model for each of the five flares described in Section \ref{sec:observations}, from which we synthesize emission from GOES, SDO/AIA, and MinXSS-1 that we can use to better constrain the parameters of the flares.   

\subsection{Example Simulation}
\label{subsec:example}

We first focus on one such simulation of the well-studied 2016 July 23 M5.0 flare (\textit{e.g.} \citealt{woods2017,moore2018}).  We begin by assuming that each successive loop is heated for 60 seconds, with a new loop forming every 10 seconds.  These values are assumptions that we will examine in the following sections.  Figure \ref{fig:modelM5t60} shows the model of the ribbon expansion (left), the synthesized GOES light curves in each channel (top right) and the ratio of the modeled to observed GOES fluxes (bottom right).  The 1--8\,\AA\ channel is shown on the left, the 0.5--4\,\AA\ channel on the right.  The solid green curves show the observed light curves, while the solid blue curves show the total synthesized light curves.  Each dashed line shows the contribution of one thread to the light curve (not all threads are shown).  The ribbon expansion velocity is assumed to reach a maximum of 50 km s$^{-1}$, while the minimum and maximum ribbon separations are calculated to be around 8 and 57 Mm.  Finally, below each light curve, the plots show the ratio of the modeled to observed fluxes, which find values close to 1 during the bulk of the flare, excepting the pre-impulsive phase which is noisy due to imperfect background subtraction.  By construction, the observed and modeled GOES light curves are in good agreement during the course of the flare.  A video of the construction of these light curves is available in the online journal.

\begin{figure}
\centering
\includegraphics[width=\linewidth]{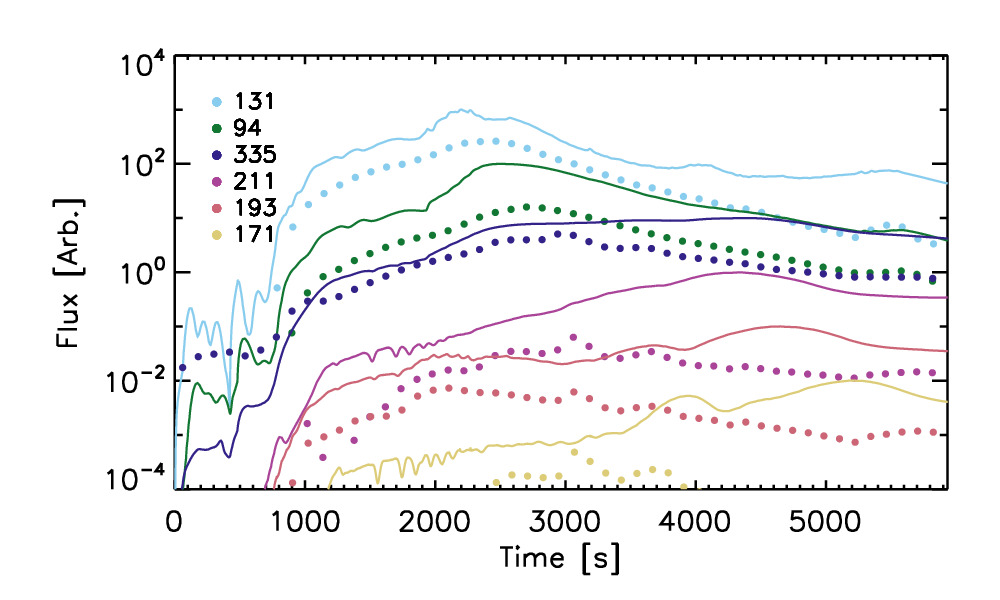}
\includegraphics[width=\linewidth]{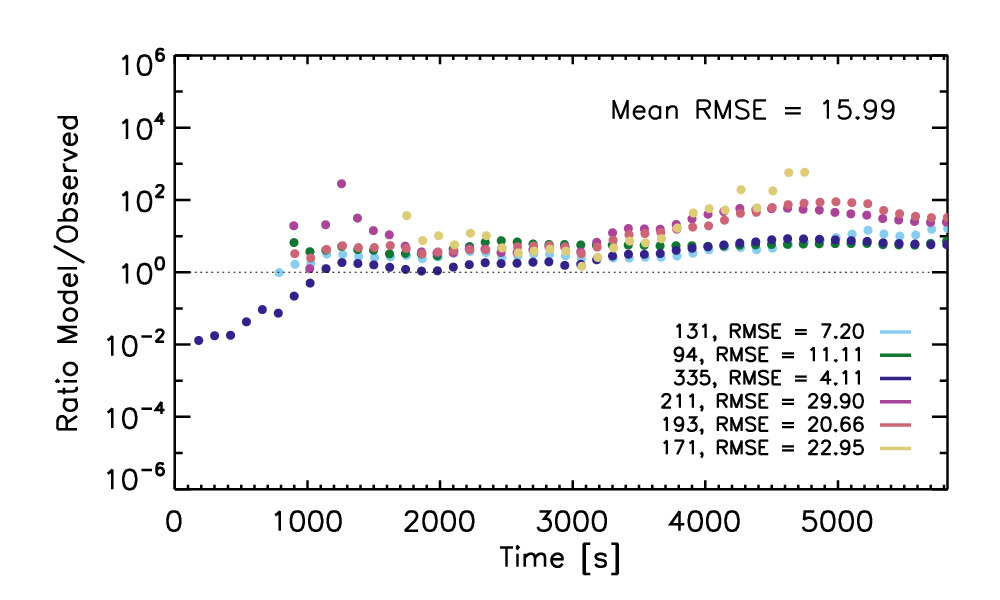}
\caption{The synthesized AIA light curves (top) and the ratio of model-to-observed values (bottom) for the 2016 July 23 M5.0 flare as presented in Section \ref{subsec:example}, corresponding to the GOES light curves shown in Figure \ref{fig:modelM5t60}.  The solid lines shown the AIA light curves in each channel as marked (normalized to powers of 10), while the dots show the observed intensities at the corresponding times.  The agreement is good in the hotter AIA channels, particularly during the impulsive phase, while the cooler channels and gradual phase are worse.  The RMSE between model and observations is indicated.  \label{fig:modelM5aiaminxss}}
\end{figure}

Since the GOES light curves are constructed to be in good agreement, the real test is whether emission in other instruments can be faithfully reproduced.  In Figure \ref{fig:modelM5aiaminxss}, we show synthesized AIA light curves contrasted with the observed values.  These light curves were synthesized using version 9 \citep{dere2019} of the CHIANTI atomic database \citep{dere1997}.  The solid lines show the synthesized AIA light curves, with each color denoting a different channel (top to bottom: 131, 94, 335, 211, 193, and 171).  The intensities have been normalized to integer powers of 10.  The dots show the observed intensities at those same times for each case.  The bottom plot shows the ratio of the model to observed values for each channel, as well as the average root-mean-squared-error (RMSE) for all the channels, 12.09 (defined and discussed in Section \ref{subsec:duration}).  In this case, the emission is over-estimated at essentially all times, in particular in the cooler channels (211, 193, and 171).  It is our goal to determine whether these and similar estimates can be improved by improving upon some of the basic assumptions of the model, and checking whether this works for all five flares presented in Section \ref{sec:observations}.  

\subsection{QPP analysis}
\label{subsec:qpp}

In the example in Section \ref{subsec:example}, it was assumed that a new loop forms every 10 seconds consistently.  This timing is an assumed number, and it is not known what value may be correct.  However, the energization of each individual loop produces small peaks in the GOES light curves as the temperature and density peaks.  This suggests that there should be a small but consistent periodic behavior in the synthesized light curves.  In a previous multithreaded flare model, \citet{rubiodacosta2016} similarly assumed that each peak in the time derivative of the GOES light curves corresponds to the energization of a single loop.  In this work, we attempt to constrain this timing by examining the QPP behavior found in the GOES light curves, and test the characteristic periods determined from the wavelet analysis as loop injection rates. It was found in Section \ref{sec:observations} that the flares [M1.0, M5.0, M7.7, C7.8, M4.4] had QPPs peaking at periods of [20, 26, 20, 12, 25] seconds.  

\begin{figure}
\centering
\includegraphics[width=0.48\textwidth]{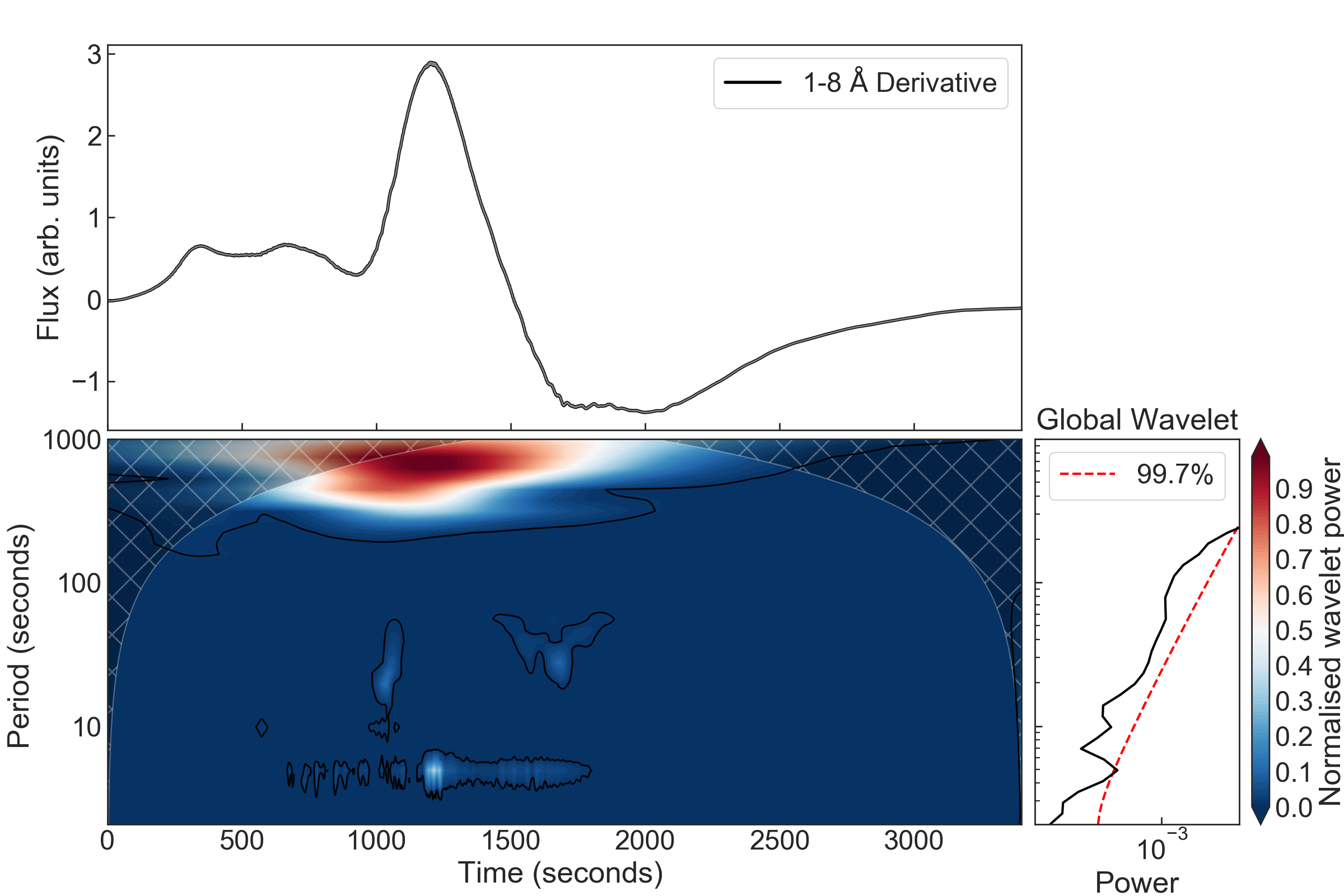}
\includegraphics[width=0.48\textwidth]{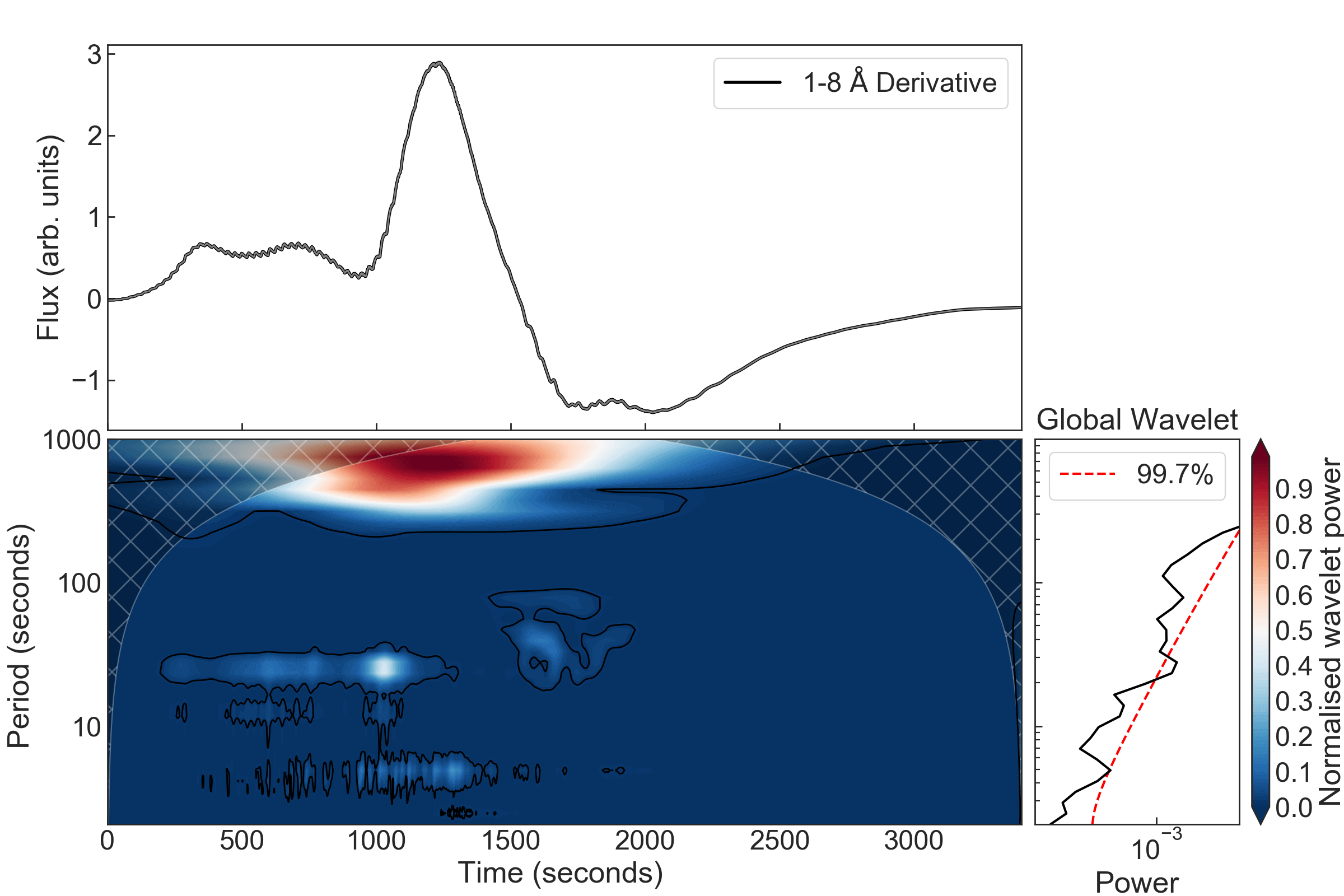}
\includegraphics[width=0.48\textwidth]{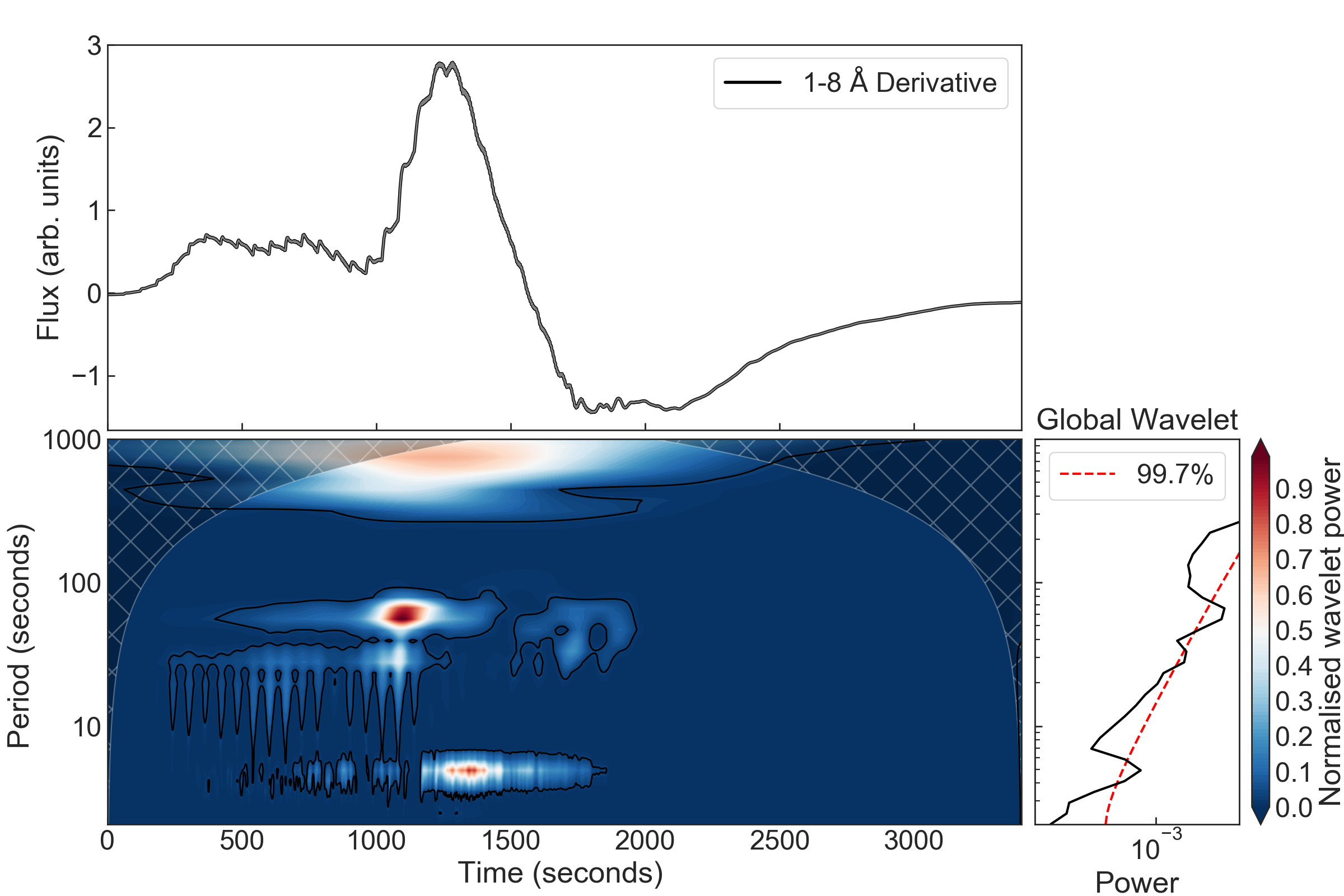}
\caption{Wavelet analysis of three synthetic GOES light curves of the 23 July 2016 M5.0 flare, using three different loop injection rates (plots similar to those in Figure \ref{fig:obs_qpp}).  From top, injection rates of 10 s, 26 s, and 60 s, respectively.  In each case, we find a significant peak in the wavelet power at the same period as we energize loops (10, 26, 60\,s), and smaller peaks at shorter periods. \label{fig:qpp_test}}
\end{figure}

We first focus briefly on the M5.0 flare that was found to have an approximately 26 second QPP.  We construct a flare model using the same abundances and heating durations, but we change the rate of loop injection.  We test three cases: 10 s, 26 s (observed), and 60 s.  In Figure \ref{fig:qpp_test}, we show the wavelet analysis of these three cases.  The figures are similar to the observed ones, showing the derivative of the 1--8\,\AA\ channel, along with the wavelet power spectrum and global wavelet spectrum.  In the first case, we find that there is a significant peak in the wavelet power at 10 s as compared to a red-noise background model, as with the observations.  In the second case, there is significant power at 26 s, with smaller peaks at shorter periods.  In the third case, there are significant peaks at 60 s and 30 s.  From this analysis, we conclude that the assumed period of loop energization will produce significant peaks in the wavelet power corresponding to the period of energization of loops.  That is, the timing of the bursty reconnection can reproduce QPP observed in GOES/XRS.

\begin{figure*}
\centering
\includegraphics[width=0.48\textwidth]{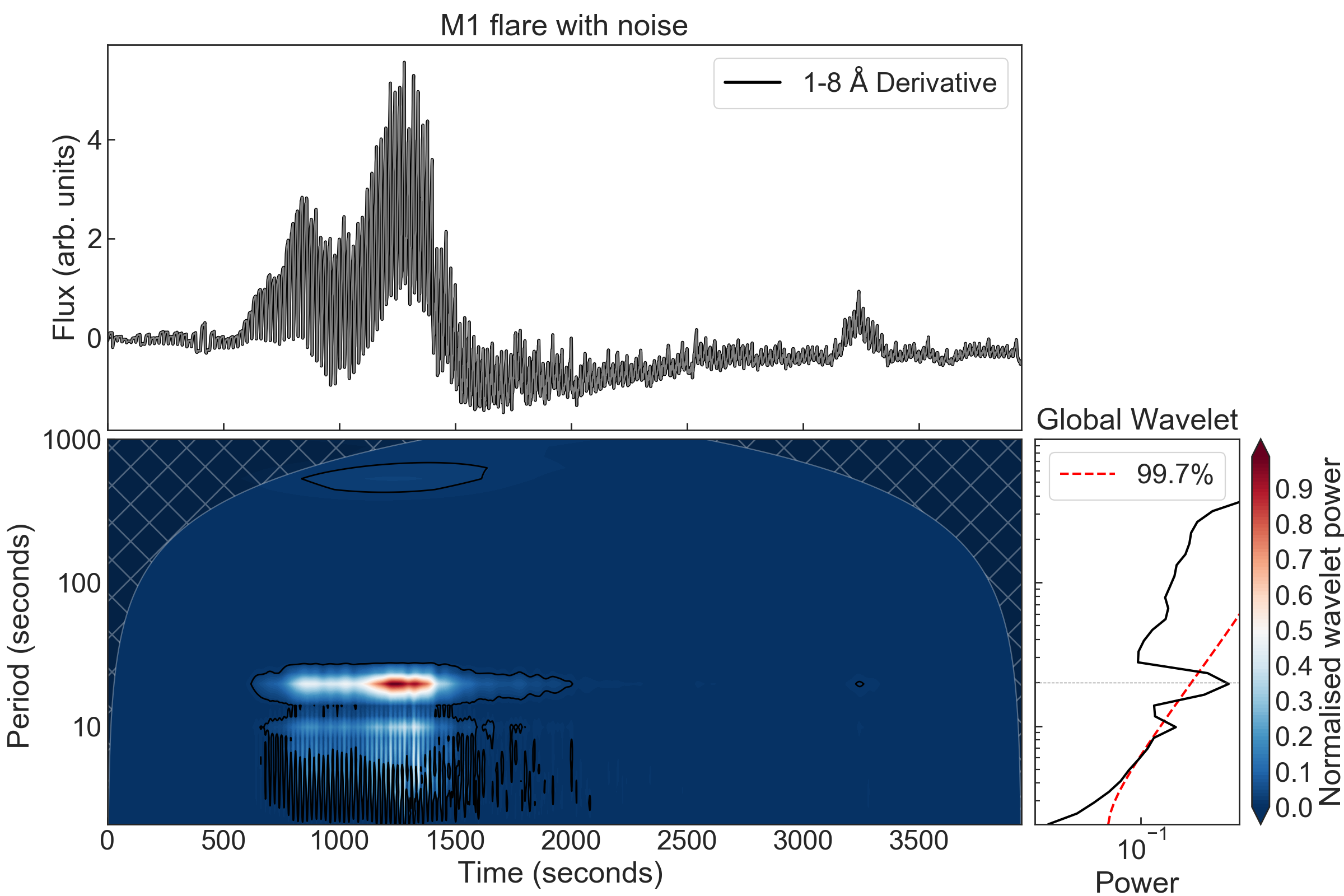}
\includegraphics[width=0.48\textwidth]{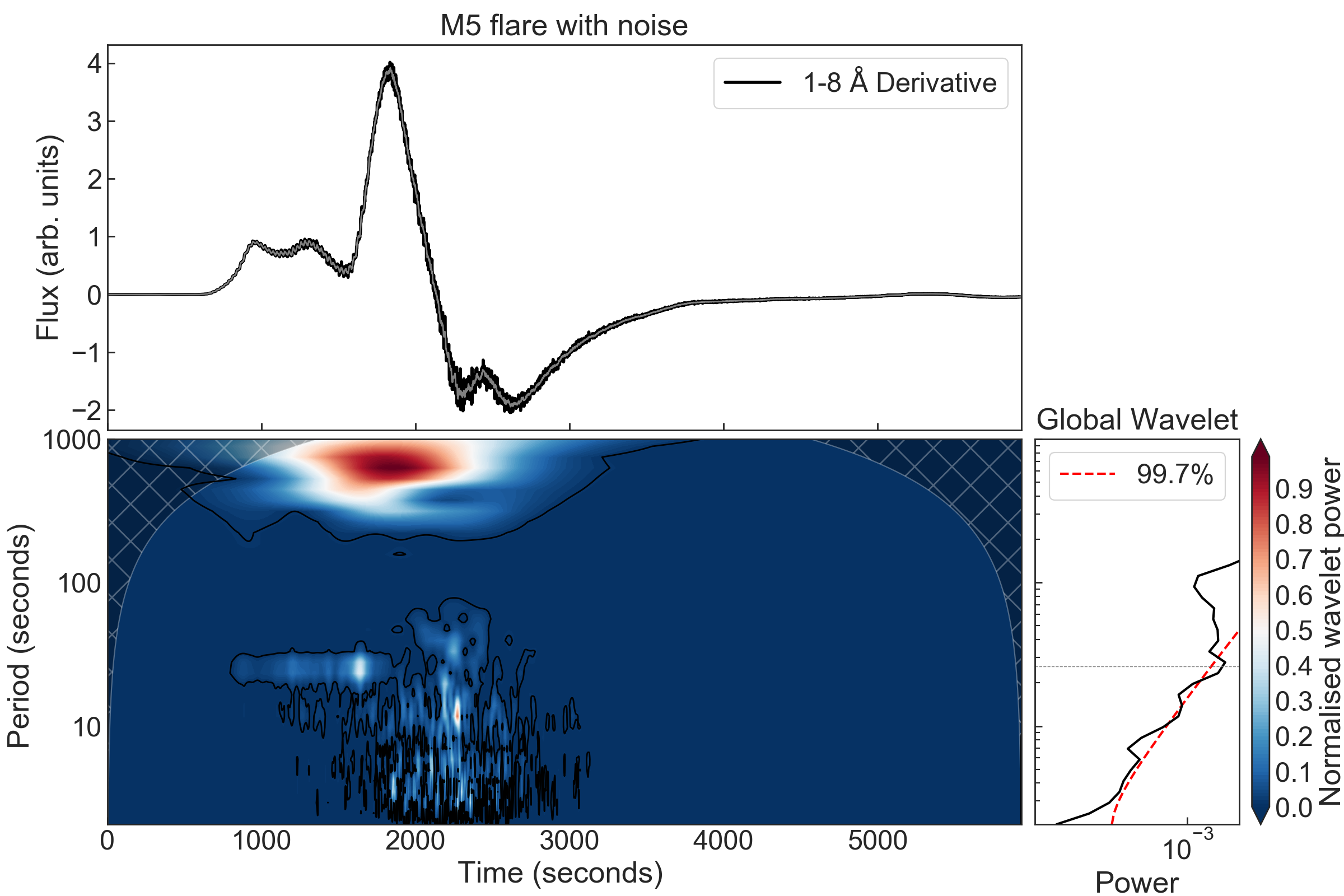}
\includegraphics[width=0.48\textwidth]{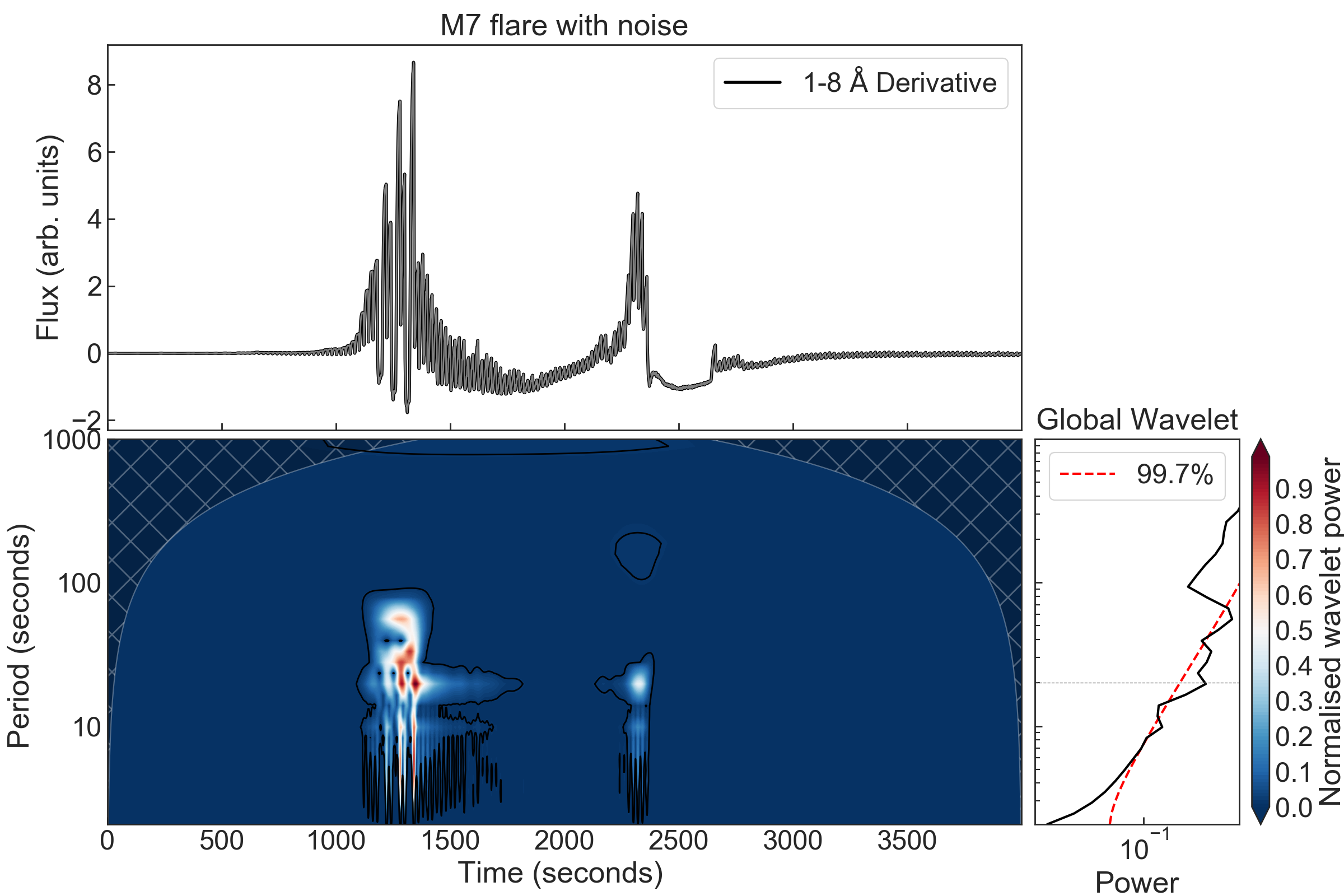}
\includegraphics[width=0.48\textwidth]{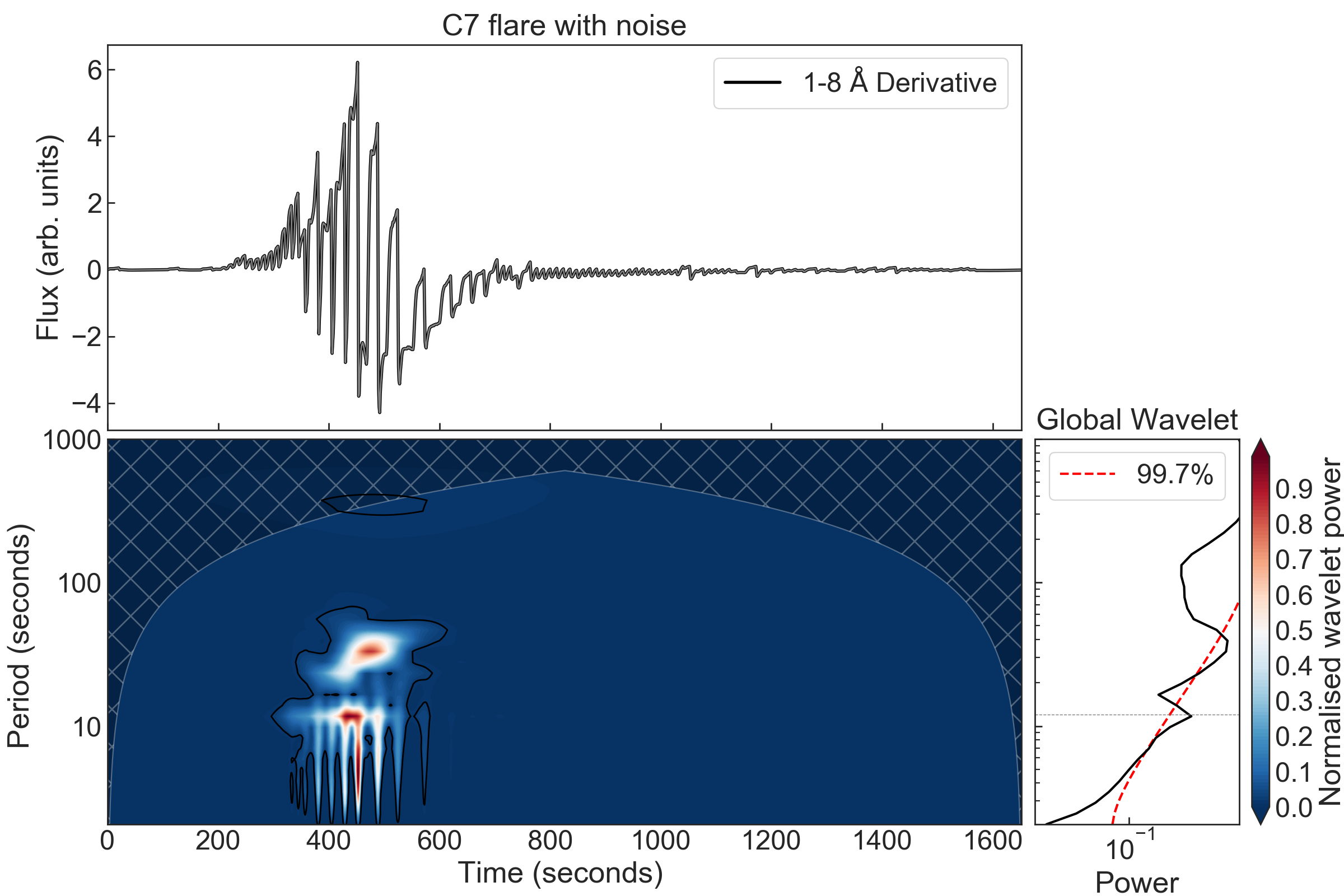}
\includegraphics[width=0.48\textwidth]{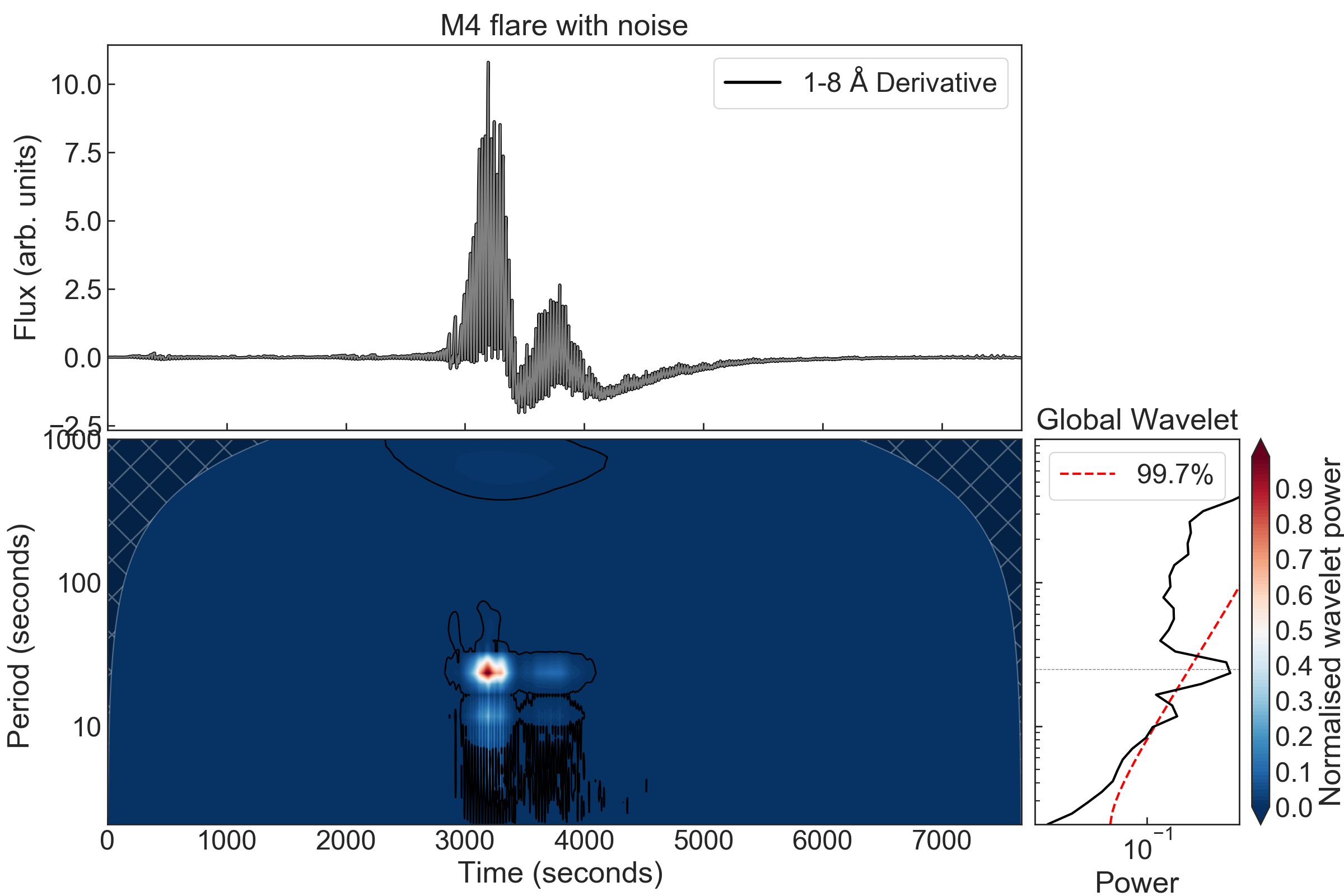}
\caption{Wavelet analysis of synthetic GOES light curves for each of the five flares, including 0.01\% Gaussian noise added to the light curves.  From top left, 21 July 2016 M1.0, 23 July 2016 M5.0, 23 July 2016 M7.7, 29 November 2016 C7.8, and the 1 April 2017 M4.4 flares.  In each case, the wavelet power shows significant peaks at the periods of loop energization (20, 26, 20, 12, and 25 s, respectively). \label{fig:qpp_noisyflares}}
\end{figure*}

In Figure \ref{fig:qpp_noisyflares}, we show the wavelet analysis for all five flares, where we have assumed that loop energization period is the same as the observed peak in QPPs.  We additionally add a Gaussian noise to the light curves of around 0.01\%, consistent with observed levels of noise in GOES/XRS light curves \citep{simoes2015}.  In each case, the wavelet power shows significant peaks at periods of 20, 26, 20, 12, and 25 s, corresponding to the assumed loop energization.  The noise does not significantly affect the measured periods.

This analysis confirms that the peaks of observed QPPs can be reproduced by assuming that the period of loop energization is equal to the observed QPPs.  While this analysis shows that they are consistent with observed QPPs, it does not definitively establish that the observed QPPs are caused by this mechanism (which is still an open question, \citealt{nakariakov2009}).  The mechanism(s) that produce QPPs require further study beyond the scope of this paper.  For our purposes, however, in the rest of this paper, we choose a loop energization period that is consistent with observed QPPs, thus constraining one of the assumed values of the flare model.  In the rest of the paper, we use the observed QPP periods for the timing of successive loop energization in each of the five flares.

\subsection{Heating Duration}
\label{subsec:duration}
\citet{warren2006} found that while the GOES light curves can be reconstructed with essentially any assumed heating duration on each loop, light curves in other channels (Yohkoh BCS in that paper) will be strongly impacted by this assumption.  In that paper, it was concluded that a 200 second heating duration is more consistent with the Yohkoh light curves than a shorter 20 s duration.  Importantly, that work only considered light curves, whereas spectral features were not examined, which could perhaps constrain the model even more stringently.  Since we do not know \textit{a priori} how long individual loops are heated, we must assume those values.  We therefore repeat the exercise using the current model and test a few heating durations to see if the AIA light curves and MinXSS-1 spectra can constrain the heating duration for individual loops.

We first assume that the plasma composition is photospheric in nature, based on the result of \citet{warren2014}, who found that the FIP bias for a set of 21 flares in SDO/EVE data was close to photospheric with little variation.  Other studies with other instruments have found larger enhancements (\textit{e.g.}\,\citealt{dennis2015}), and some have even found an inverse FIP effect where high FIP elements are enhanced \citep{doschek2015,doschek2016}, although these results are for relatively small areas.  We begin by using the abundance set derived by \citet{asplund2009}, and will examine how abundances affect these results in Section \ref{subsec:abund}.  

For each of the five flares, we have synthesized the GOES, MinXSS-1, and AIA emission for five different temporal profiles of heating.  We use fixed heating rates lasting for 20, 60, and 200 s.  We also use two cases with fixed heating rates lasting 60 s and 200 s, which then decays over another 60 s and 200 s, respectively.  Since the GOES light curves are constrained by the model to be approximately reproduced for any of these values, we then examine the MinXSS-1 and AIA emissions to determine how the heating duration affects the light curves and spectra.  

\begin{figure*}
\centering
\includegraphics[width=2.66in,height=1.6in]{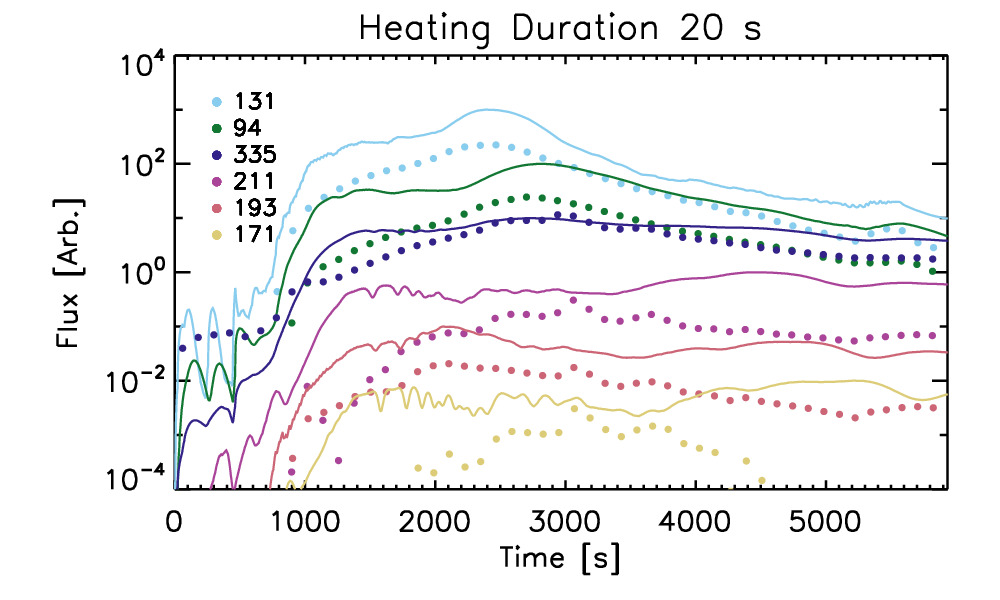}
\includegraphics[width=2.66in,height=1.6in]{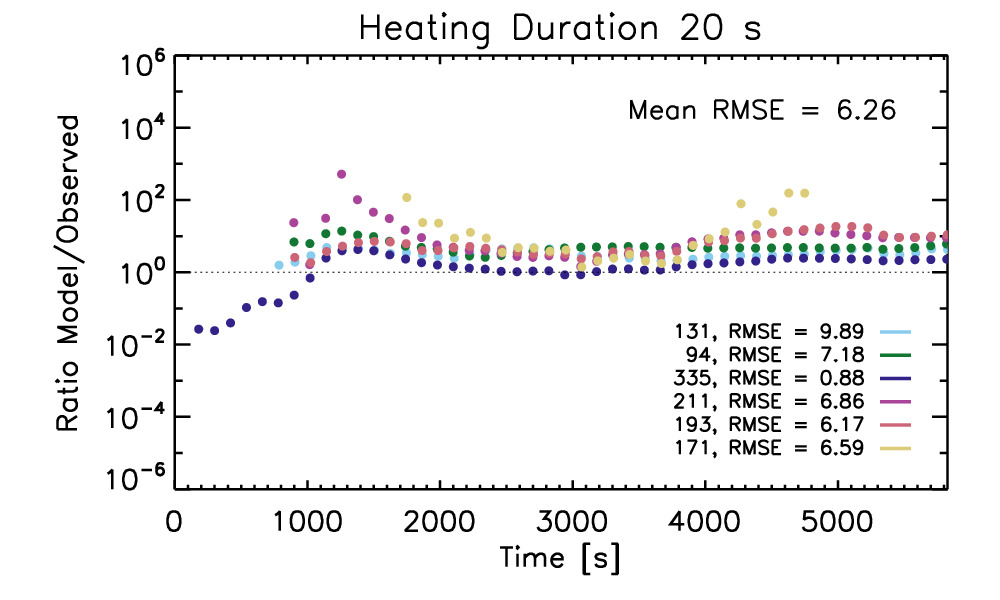}
\includegraphics[width=2.66in,height=1.6in]{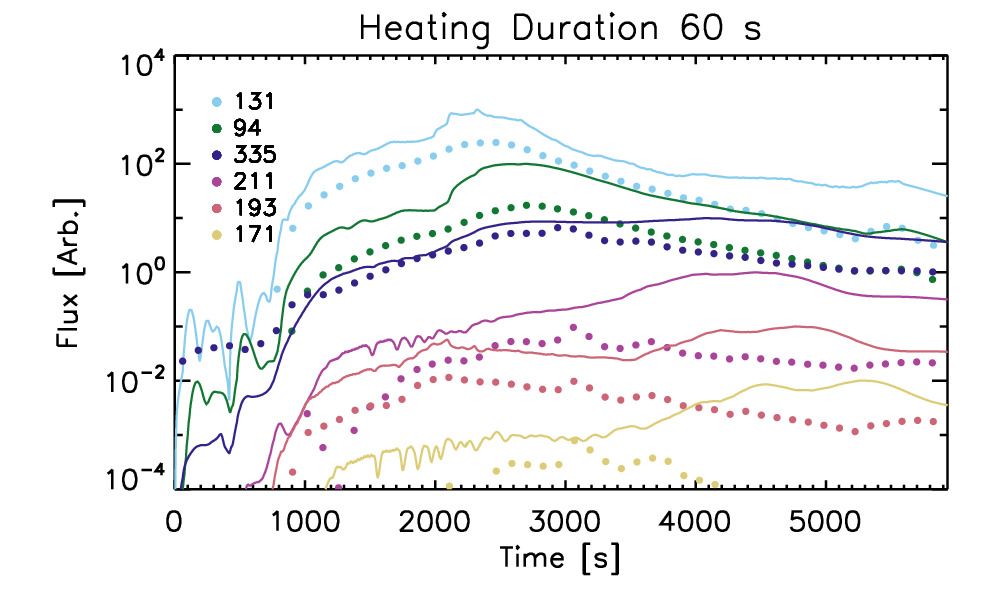}
\includegraphics[width=2.66in,height=1.6in]{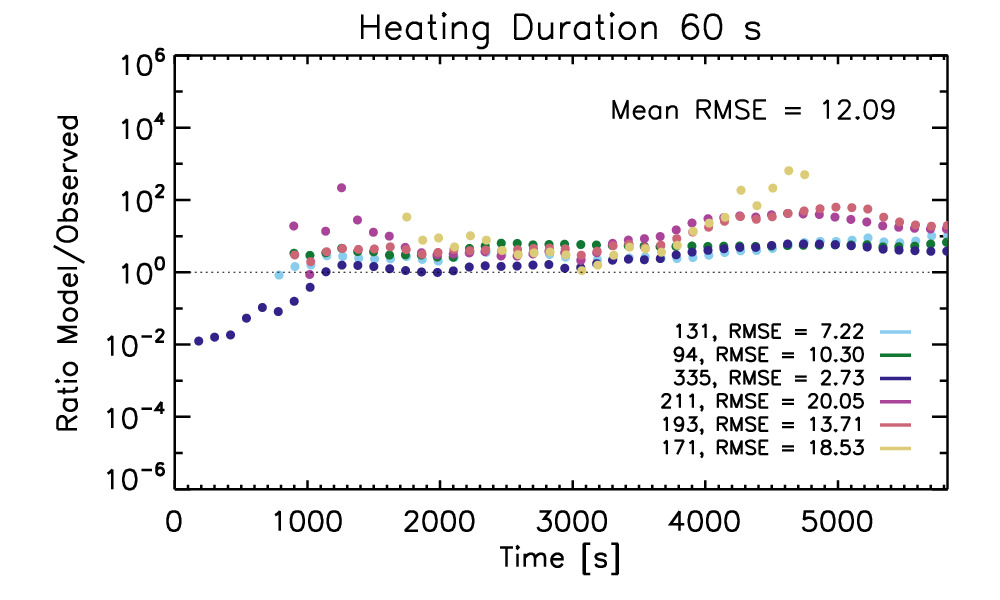}
\includegraphics[width=2.66in,height=1.6in]{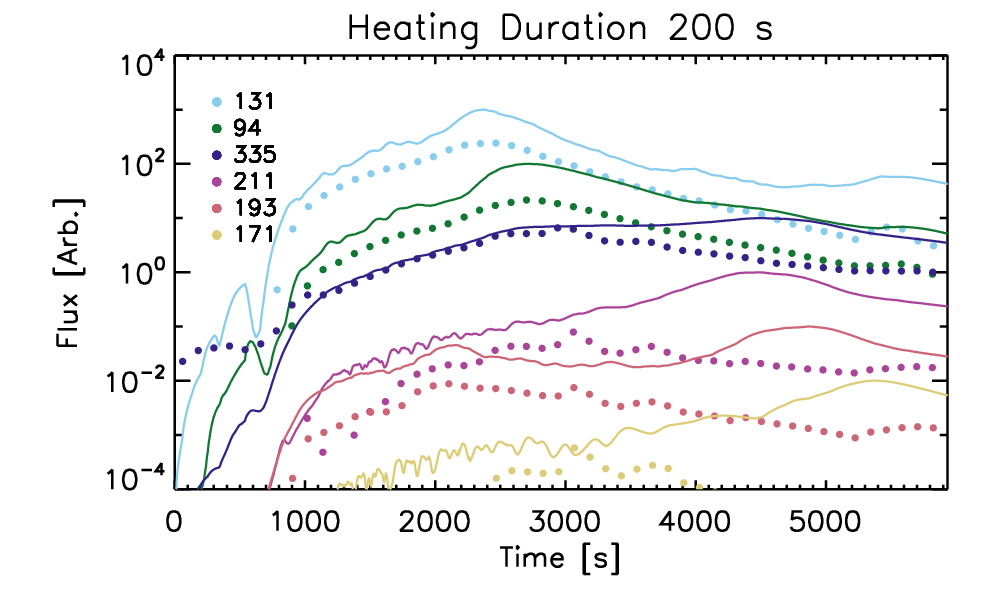}
\includegraphics[width=2.66in,height=1.6in]{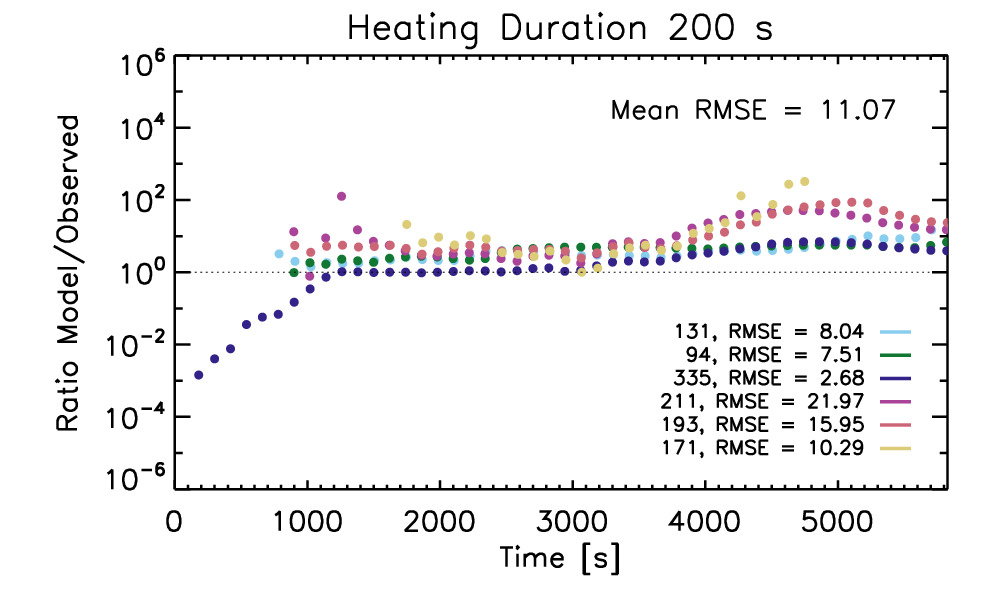}
\includegraphics[width=2.66in,height=1.6in]{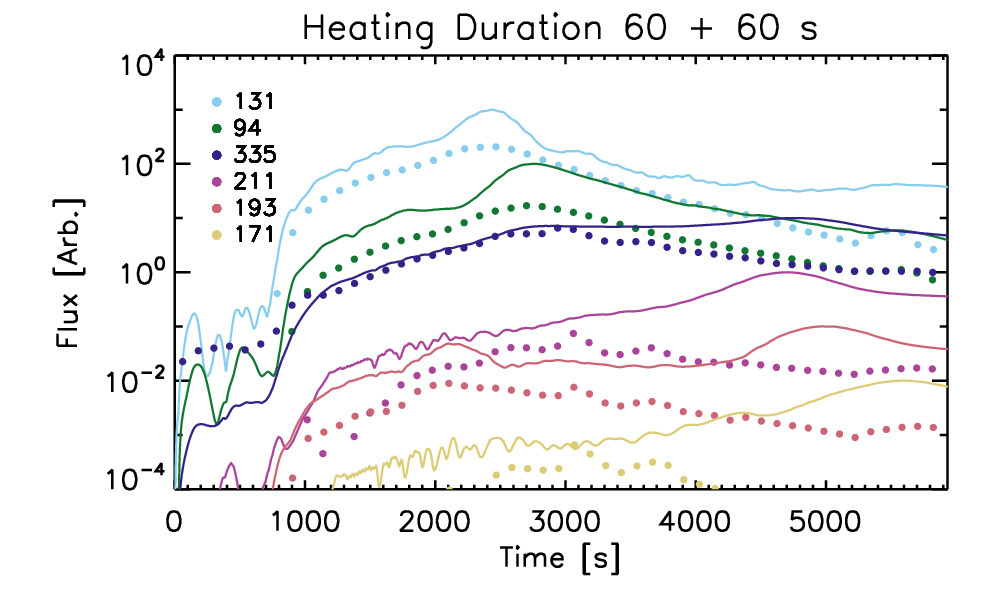}
\includegraphics[width=2.66in,height=1.6in]{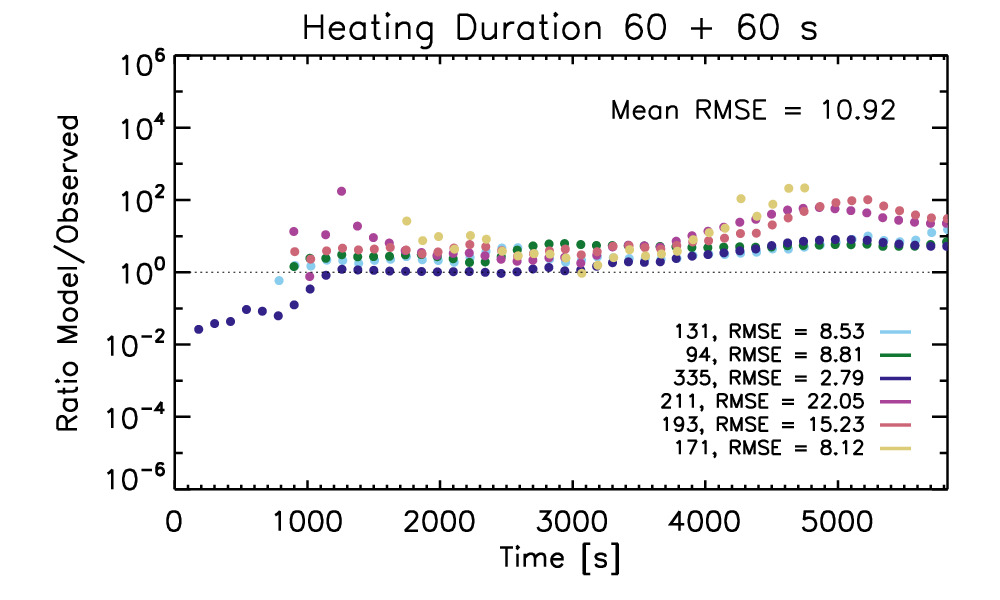}
\includegraphics[width=2.66in,height=1.6in]{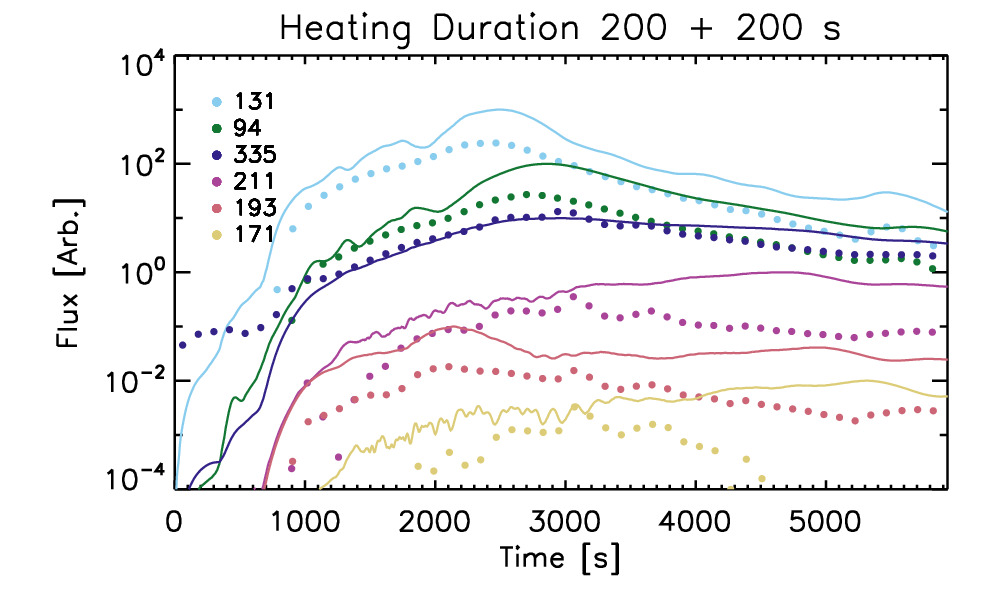}
\includegraphics[width=2.66in,height=1.6in]{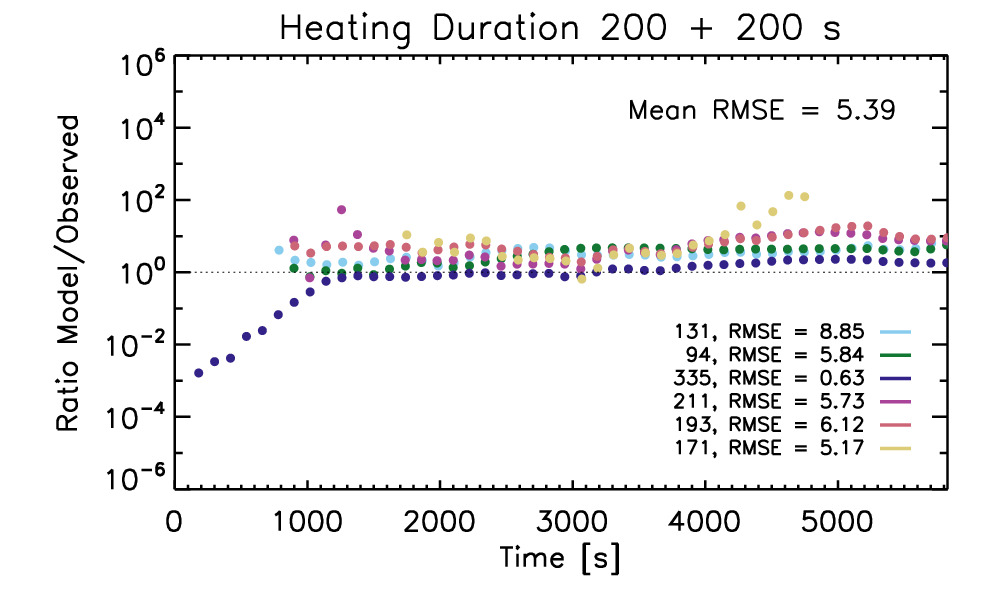}
\caption{A comparison of the synthetic and observed AIA light curves for the 2016 July 23 M5.0 flare, synthesized with various heating durations.  The left hand column shows the light curves, with each channel normalized to a different power of 10, while the right hand column shows the ratio of model to observed for each channel.  The heating durations are, respectively, fixed heating rates for 20 s, 60 s, and 200 s, as well as fixed with a long decay phase lasting 60 + 60 s and 200 + 200 s.  The RMSE for each channel is indicated, as well as the mean value of the RMSE for all channels.  The last case, 200 s with a long decay phase, produces light curves that are the most consistent with observations, although the 20 s case is comparable. \label{fig:m5_aia}  }
\end{figure*}

To begin, consider the 2016 July 23 M5.0 flare.  In Figure \ref{fig:m5_aia}, we show a comparison of the observed and modeled light curves for 6 AIA channels, as well as the ratio of model to observed in each case.  We then determine how well the data is fit by calculating a root-mean-squared-error for each channel:
\begin{align}
    \text{RMSE} = \Bigg[\frac{1}{N} \sum_{i = 1}^{N} \Big(\text{Model}(t_{i}) - \text{Observed}(t_{i})\Big)^{2}\Bigg]^{1/2}
\end{align}
\noindent This is summed over all times $t_{i}$ when there are observations in the given AIA channel.  Since the observed counts in these flares are very large and the uncertainty from the counting statistics with AIA are essentially zero, it would be difficult to accurately calculate a chi-squared value for the AIA light curves.  For all five flares, the uncertainty-to-signal ratio in the 94\,\AA\ channel was on the order of 0.1\% (estimated with the SSWIDL routine `aia\_bp\_estimate\_error'), while the ratio in the 131, 171, 193, 211, 304, and 335 channels were on the order of 0.01\%.  Finally, we normalize the RMSE in each channel to the median observed intensity of that channel so that the values in different channels can be directly compared.  

The left hand column of Figure \ref{fig:m5_aia} shows the comparison between the synthetic and observed light curves for the six AIA channels, ordered roughly from hot to cold.  The fluxes have been normalized in decreasing factors of 10.  The right hand column shows the ratio of the modeled to observed values in each case, where a ratio of 1.0 would be considered perfect agreement.  From top, the plots show the five heating durations: fixed for 20, 60, and 200 s, as well as the cases with fixed heating for 60 and 200 s, with a long decay phase lasting 60 and 200 s, respectively.

Although none of the cases reproduce the observed light curves exactly, there are a couple of points worth noting.  The first is that the hotter channels (131, 94, 335) are more closely aligned to the observations in all cases, while the cooler channels, particularly 171, are poorly reproduced.  This is because the hot emission is constrained by the GOES ratio, whereas the cooler channels are essentially unconstrained.  The second point is that the longest heating duration, with a long decay phase, does a better job at reproducing all of the light curves than shorter heating durations.  When we repeat this process for the other four flares, we find similar results.  

\begin{figure*}
\centering
\includegraphics[width=0.48\textwidth]{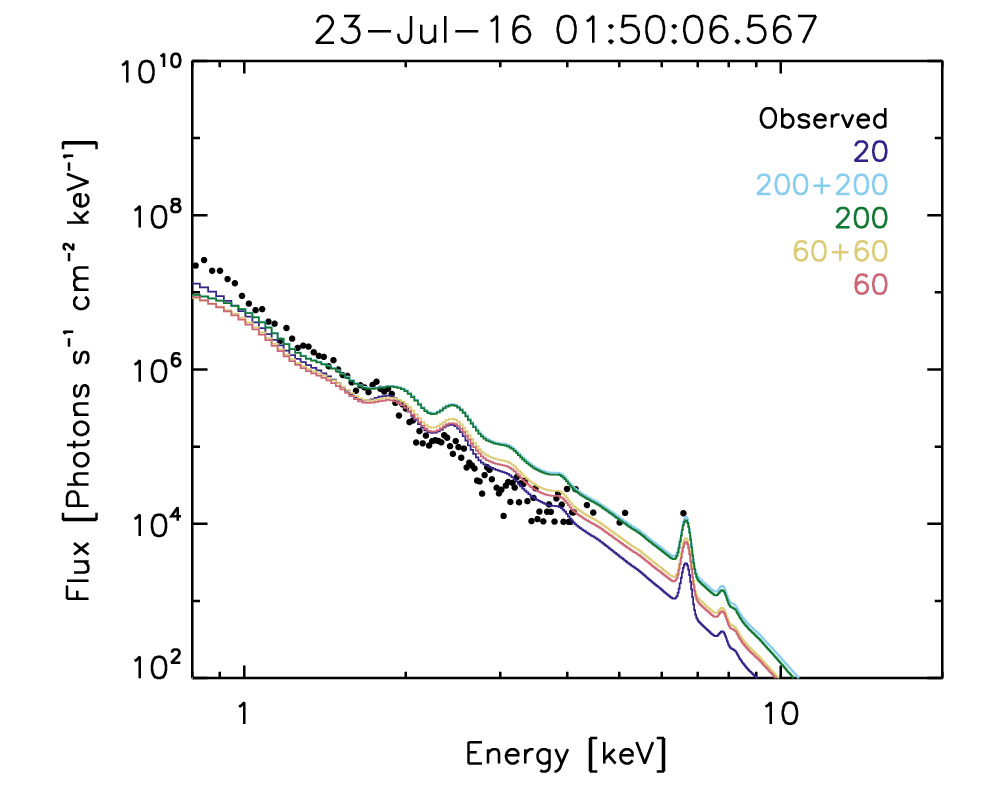}
\includegraphics[width=0.48\textwidth]{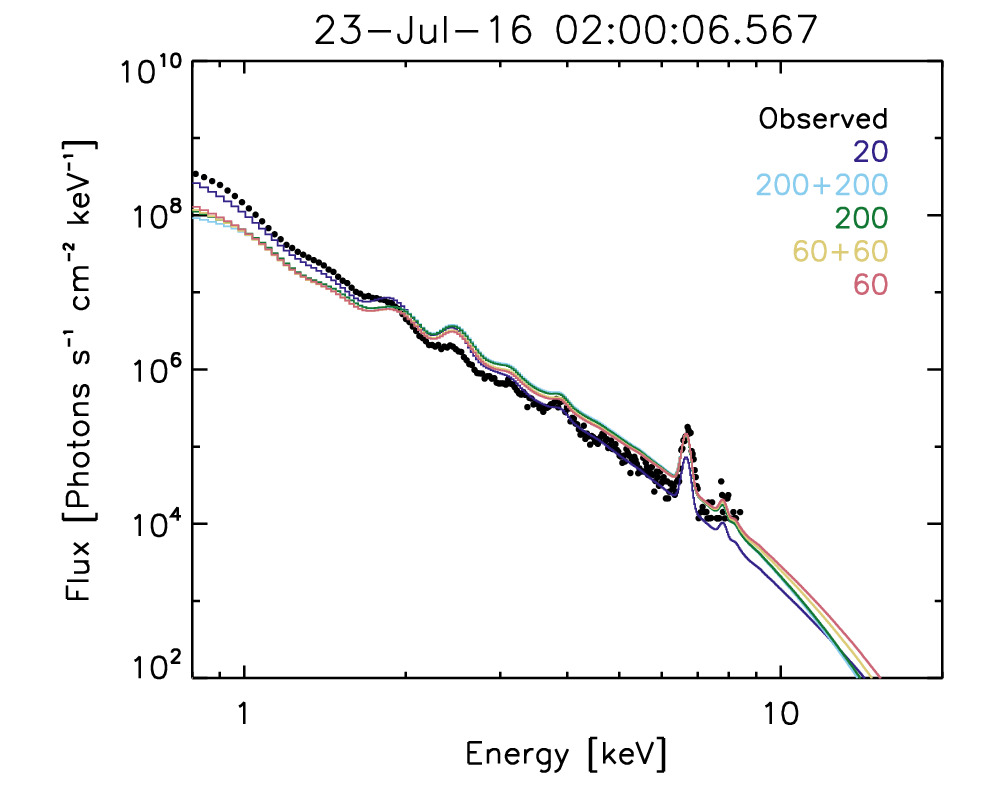}
\includegraphics[width=0.48\textwidth]{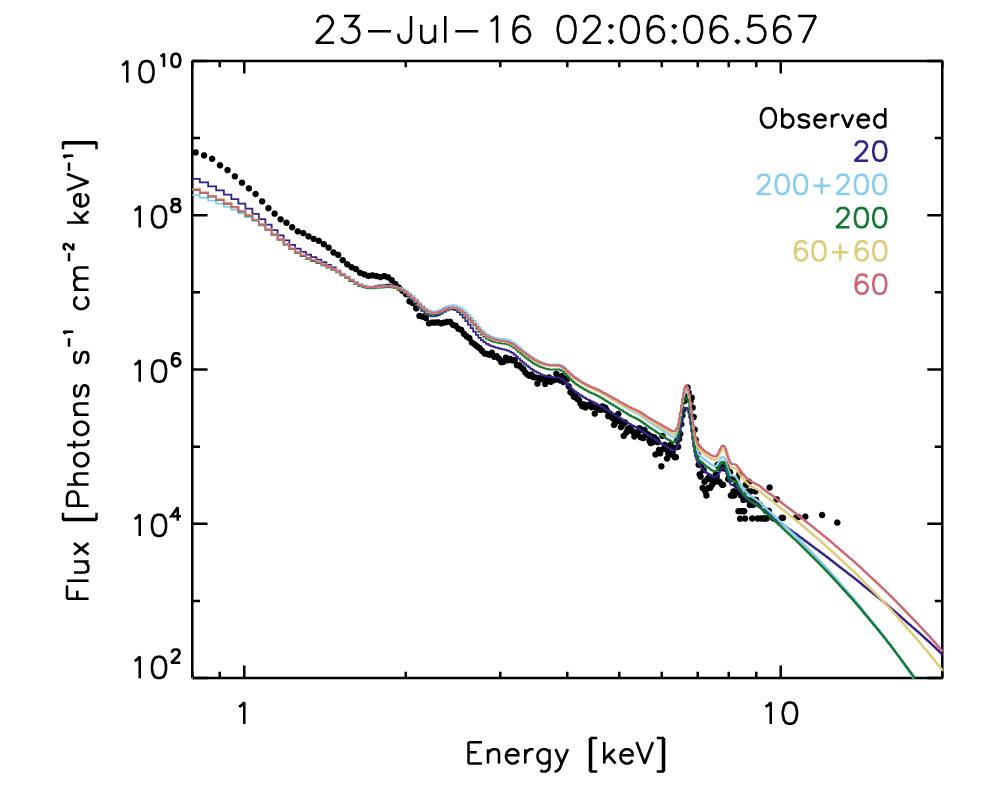}
\includegraphics[width=0.48\textwidth]{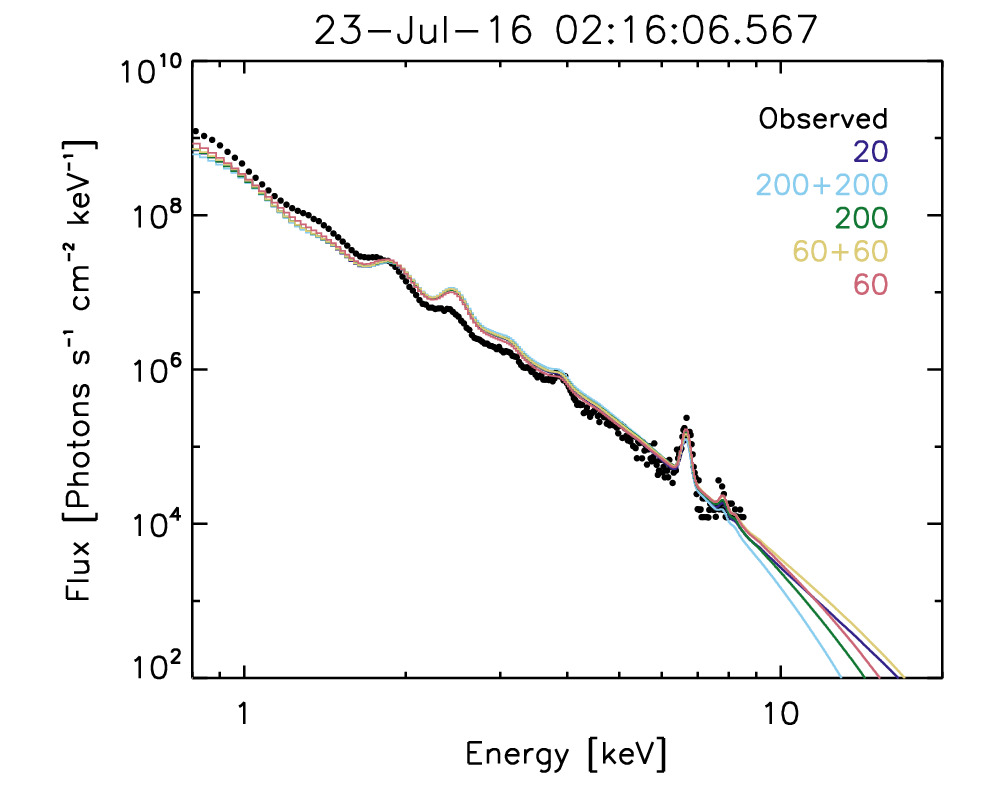}
\includegraphics[width=0.48\textwidth]{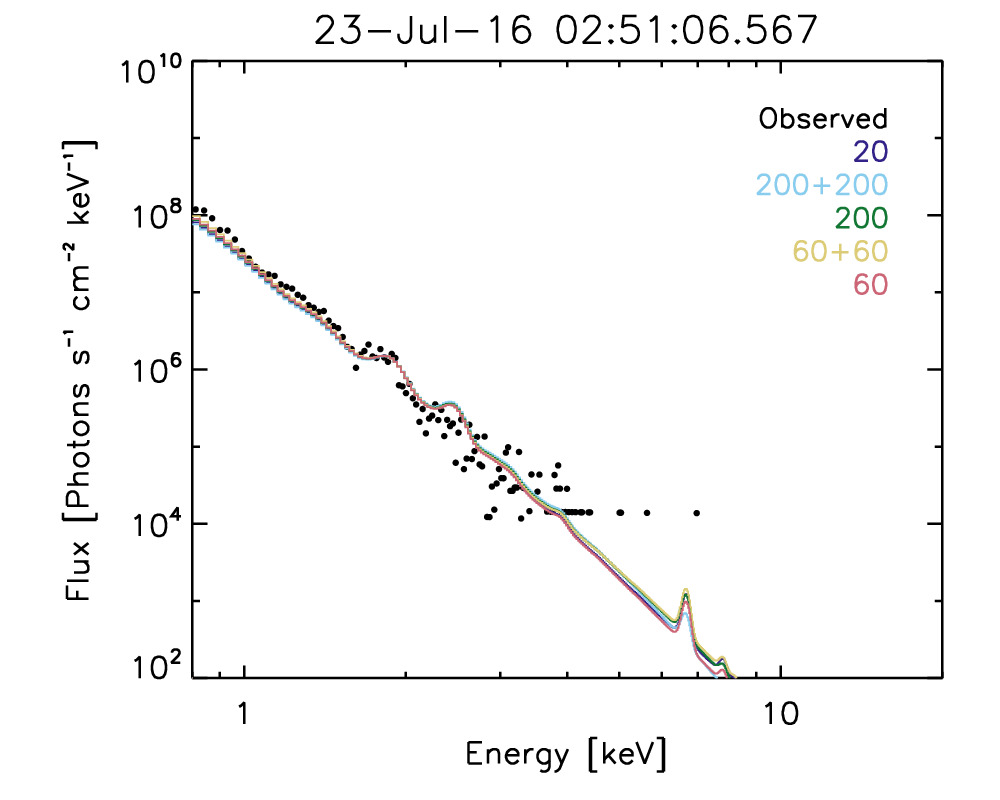}
\includegraphics[width=0.48\textwidth]{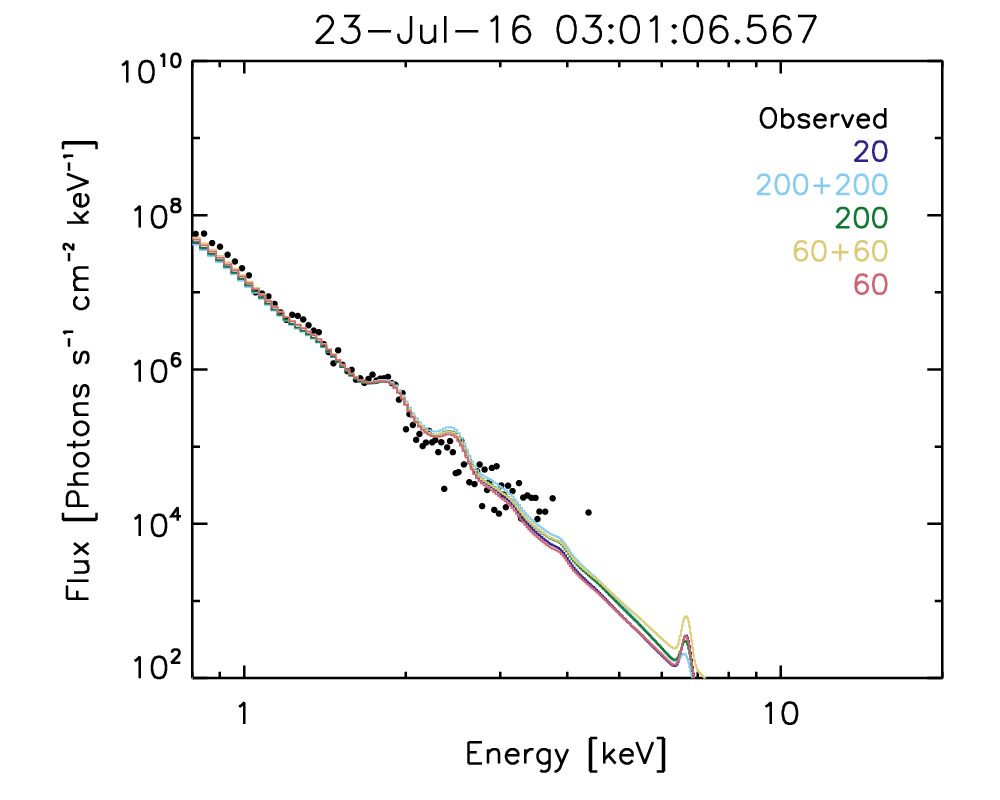}
\caption{A comparison of the synthetic (colors) and background-subtracted observed (black dots) MinXSS-X123 one-minute averaged spectra for the 23 July 2016 M5.0 flare at 6 selected times for the five different assumed heating durations as labeled.  The differences between the synthetic spectra are small, particularly near the peak of the flare (UT 02:16) when they all provide adequate fits. \label{fig:minxss_heating}}
\end{figure*}

We next turn our attention to the MinXSS-1 data.  We have similarly synthesized the MinXSS-X123 spectra with CHIANTI to compare to the observed values.  To improve the signal-to-noise, we use one-minute averaged spectra.  In Figure \ref{fig:minxss_heating}, we show an example of the 23 July 2016 M5.0 flare, comparing the synthetic (colors) and background-subtracted observed (black dots) MinXSS-X123 spectra.  We show the five different heating durations, from top, 20 second heating duration, 60 second with 60 second decay, 200 second with 200 second decay.  In all cases, we find similar levels of agreement, particularly near the peak of the flare (UT 02:16).  The soft X-ray spectrum is generally well-reproduced by this model.  At low energies, the magnitudes of the synthetic spectral intensity are consistently smaller than observed, while at higher energies there is good agreement in both the lines and continuum.  The difference for different assumed heating durations is small, which we find for all five flares, suggesting that the assumed heating duration does not strongly impact the synthetic MinXSS-1 spectra.  

\begin{figure}
    \centering
    \includegraphics[width=\linewidth]{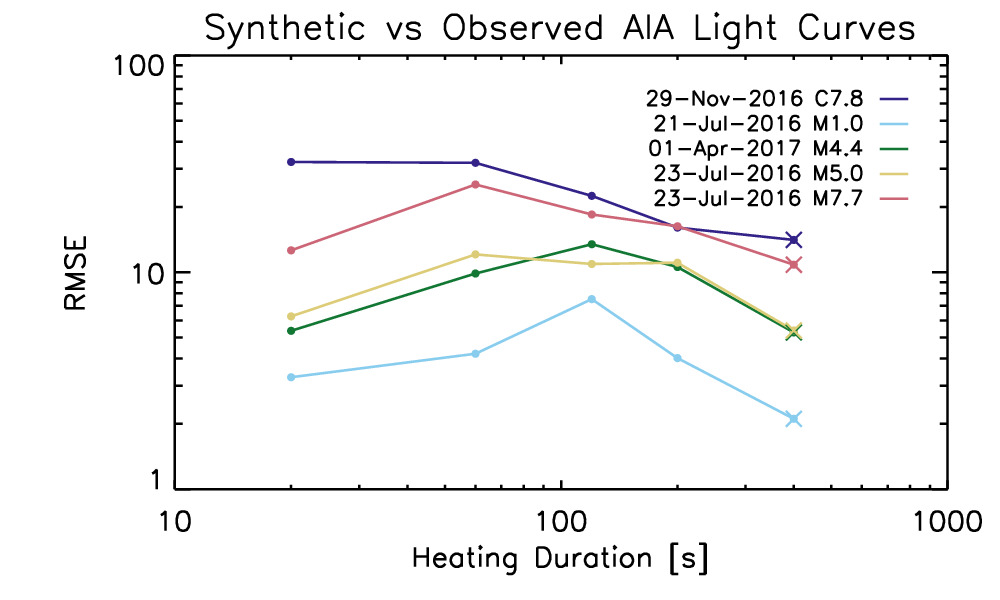}
    \includegraphics[width=\linewidth]{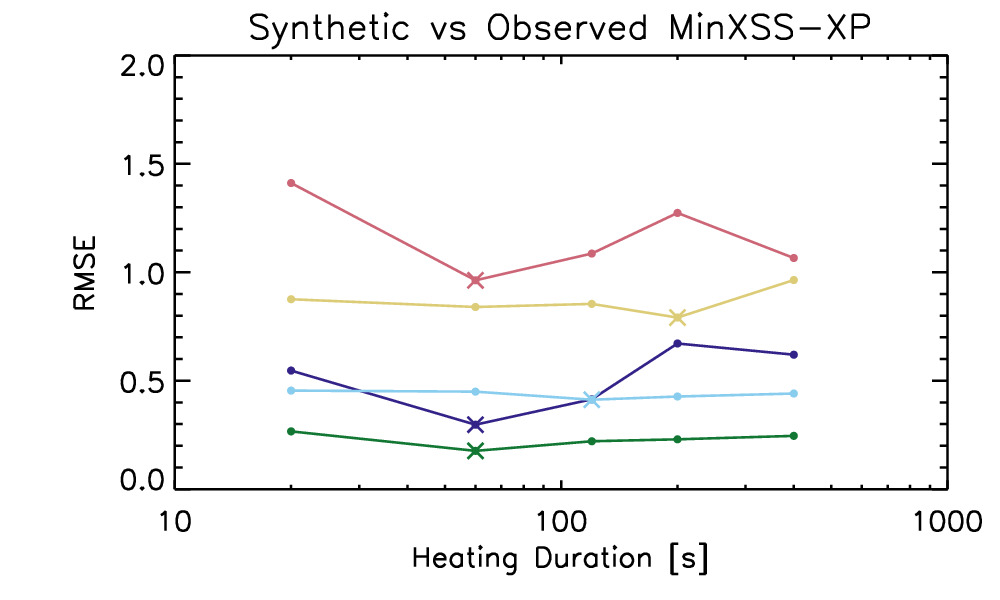}
    \includegraphics[width=\linewidth]{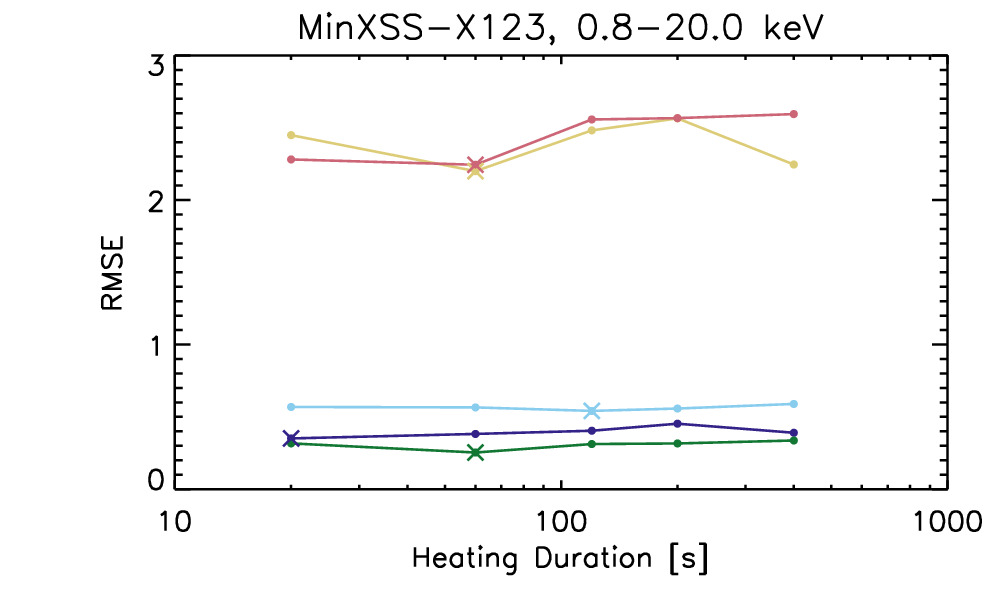}
        \caption{The RMSE for the mean of the six AIA channels (top), MinXSS-XP (middle), and MinXSS-X123 (bottom) comparing synthetic to observed light curves for each of the five flares and each of the five assumed heating durations.  The minima are marked with an asterisk.  }
    \label{fig:duration_rmse}
\end{figure}

Finally, in Figure \ref{fig:duration_rmse}, we summarize the results of this section.  We show the RMSE quantifying the differences between the synthetic to observed AIA (top), MinXSS-XP (middle) and MinXSS-X123 (bottom). For MinXSS-X123, we have computed the RMSE over the energy range 0.8--20.0\,keV.  In the case of AIA, the light curves are significantly better reproduced for all fives flares with the longest heating duration (200 s with a 200 s decay).  In the case of MinXSS-1, however, heating duration does not cause significant variation in the fits for XP or X123.  This suggests that while we do find good agreement between the observed and synthetic SXR spectra, we cannot use them as a diagnostic of the heating duration, for which AIA is much more strongly discriminating.  In each case, longer heating durations improve the comparison between the observed and synthetic AIA light curves, which is similar to the conclusion of \citet{warren2006}.  The MinXSS-1 spectra are always well-reproduced, which is likely because its energy range overlaps significantly with the GOES channels, which we have constructed to be in good agreement with observations.

\subsection{Abundances}
\label{subsec:abund}

We now turn our attention to the elemental abundances in each flare.  It is generally assumed that flares are photospheric in composition, which is broadly consistent with observations \citep{warren2014}.  However, measuring abundances is not trivial, and, since abundances influence the rate of cooling, it may be possible to constrain the abundances with this modeling.  In order to self-consistently study this, we have re-run the ebtel++ simulations with FIP enhancement factors of $f = $[1, 1.5, 2, 2.5, 3, 3.5, 4], meaning that all the low-FIP elements are enhanced by the same factor of $f$ in the radiative losses.  We then re-run the total flare simulations, similarly calculating the GOES light curves with appropriate assumed abundances (see the Appendix), and finally re-calculate the AIA light curves and MinXSS-1 spectra.  We use the longest heating duration, which produces the best fit for the AIA light curves as found in the previous section.

\begin{figure*}
\centering
\includegraphics[width=0.3\textwidth]{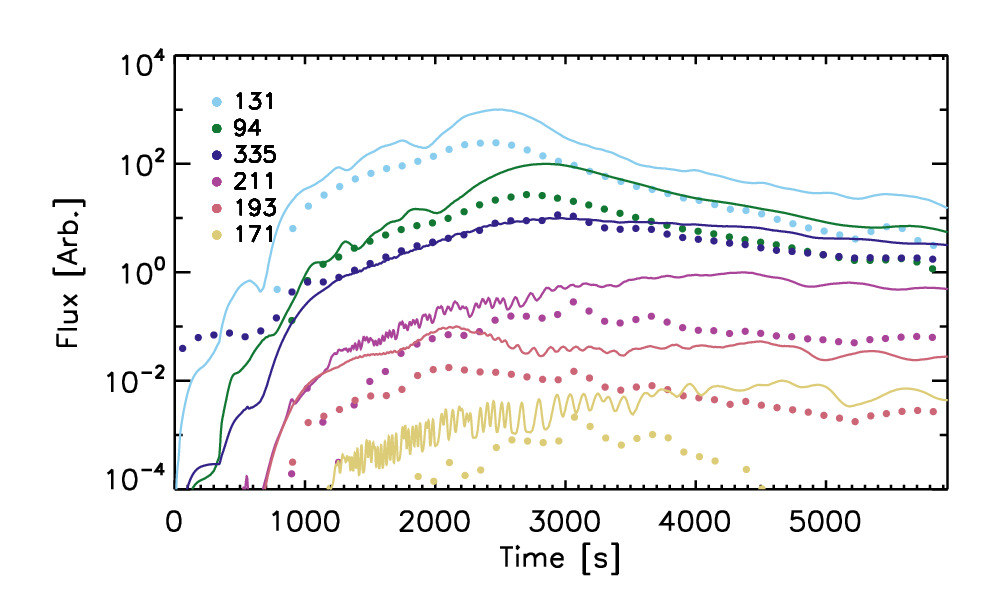}
\includegraphics[width=0.3\textwidth]{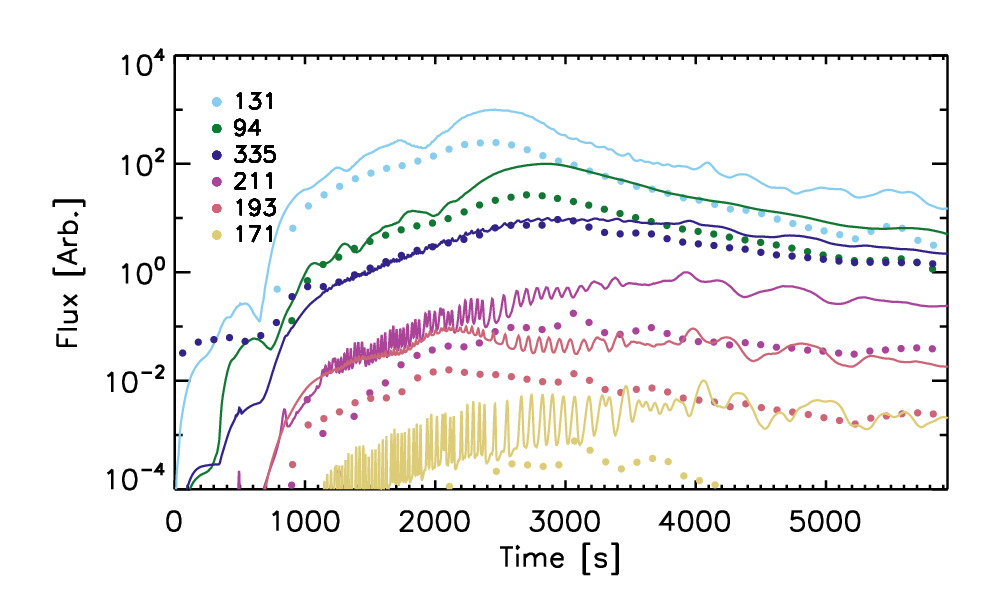}
\includegraphics[width=0.3\textwidth]{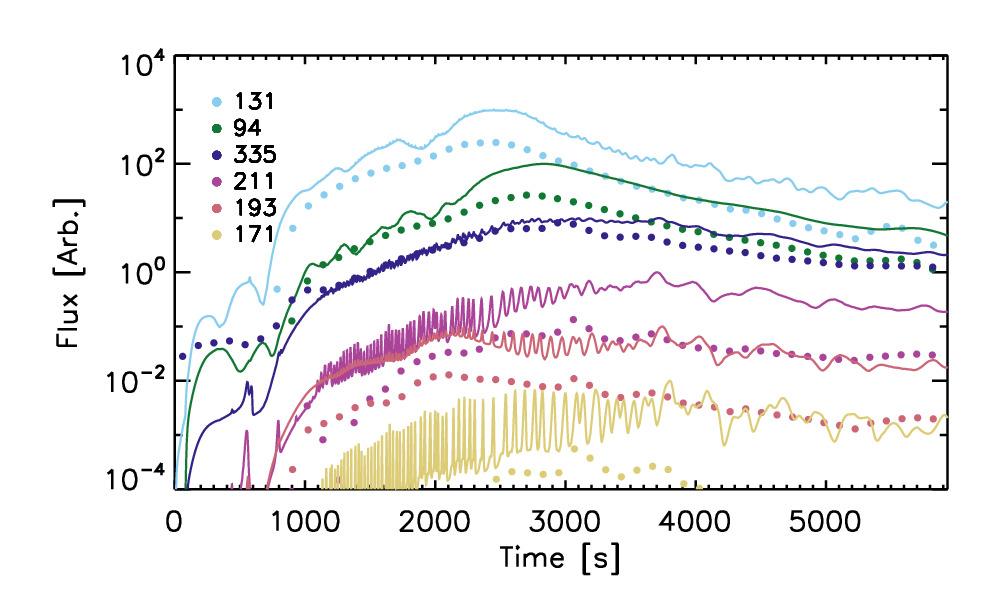}
\includegraphics[width=0.3\textwidth]{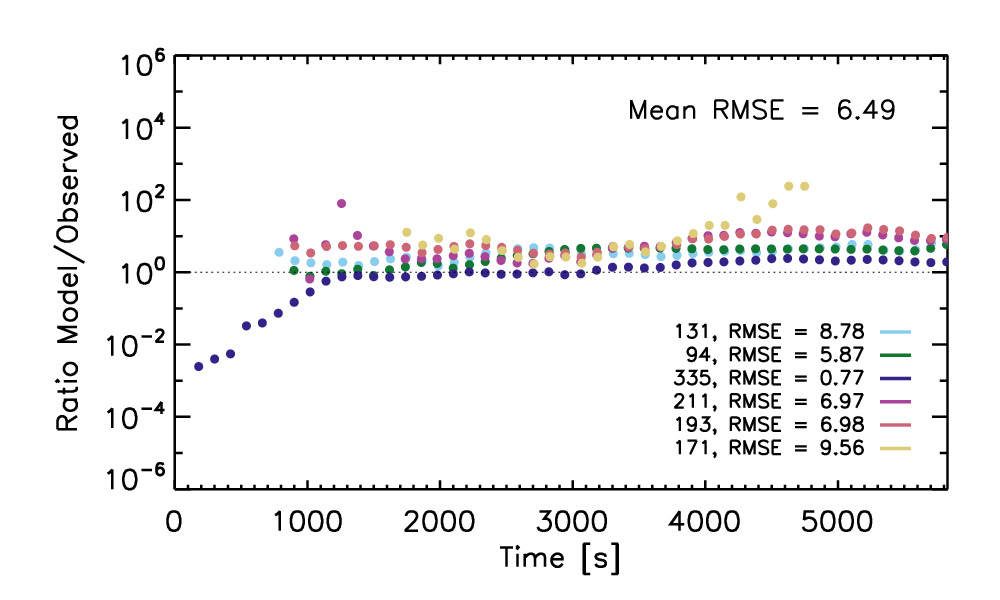}
\includegraphics[width=0.3\textwidth]{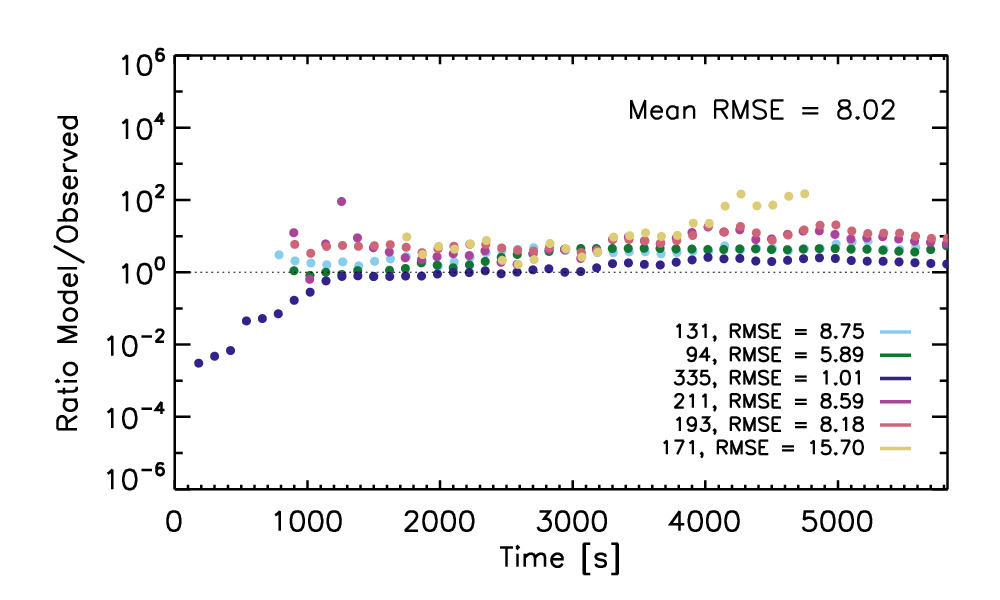}
\includegraphics[width=0.3\textwidth]{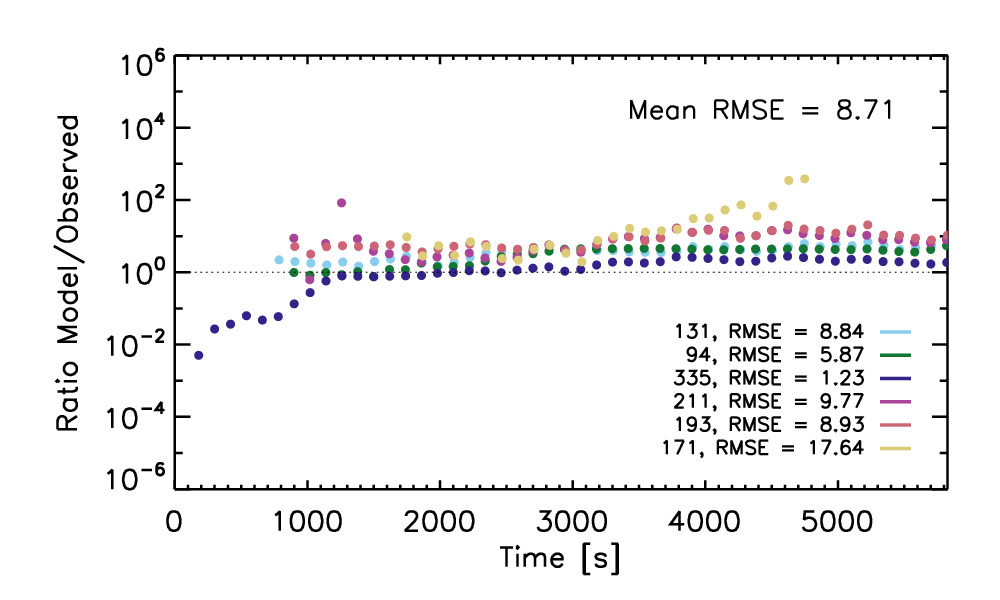}
\includegraphics[width=0.3\textwidth]{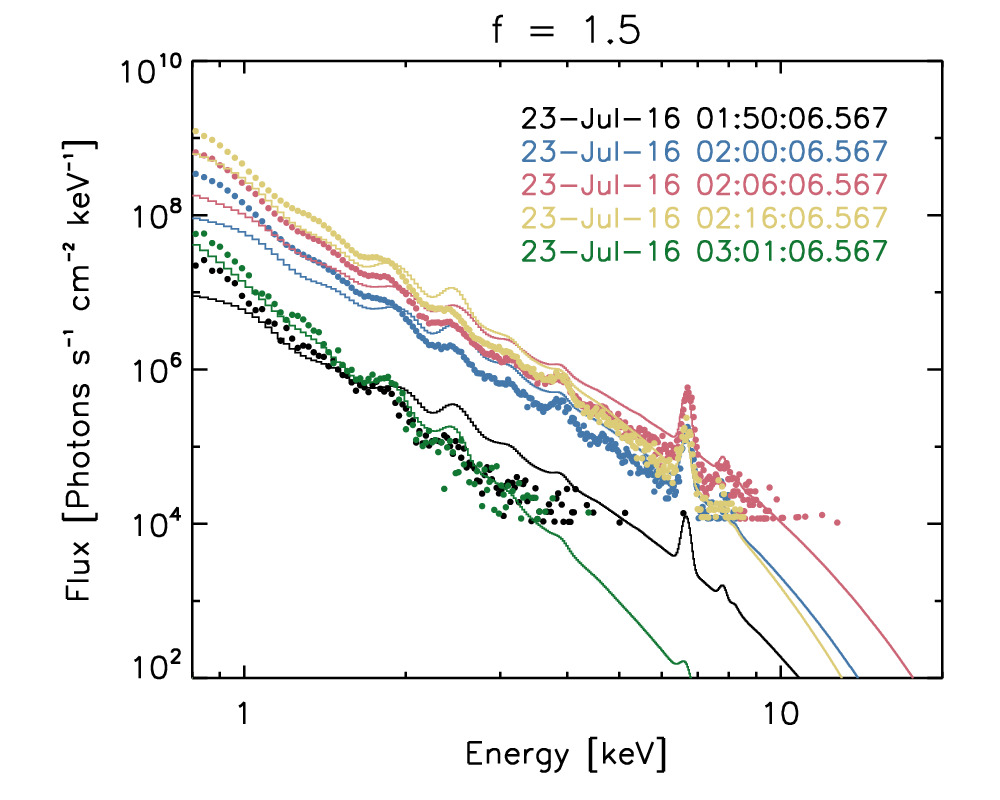}
\includegraphics[width=0.3\textwidth]{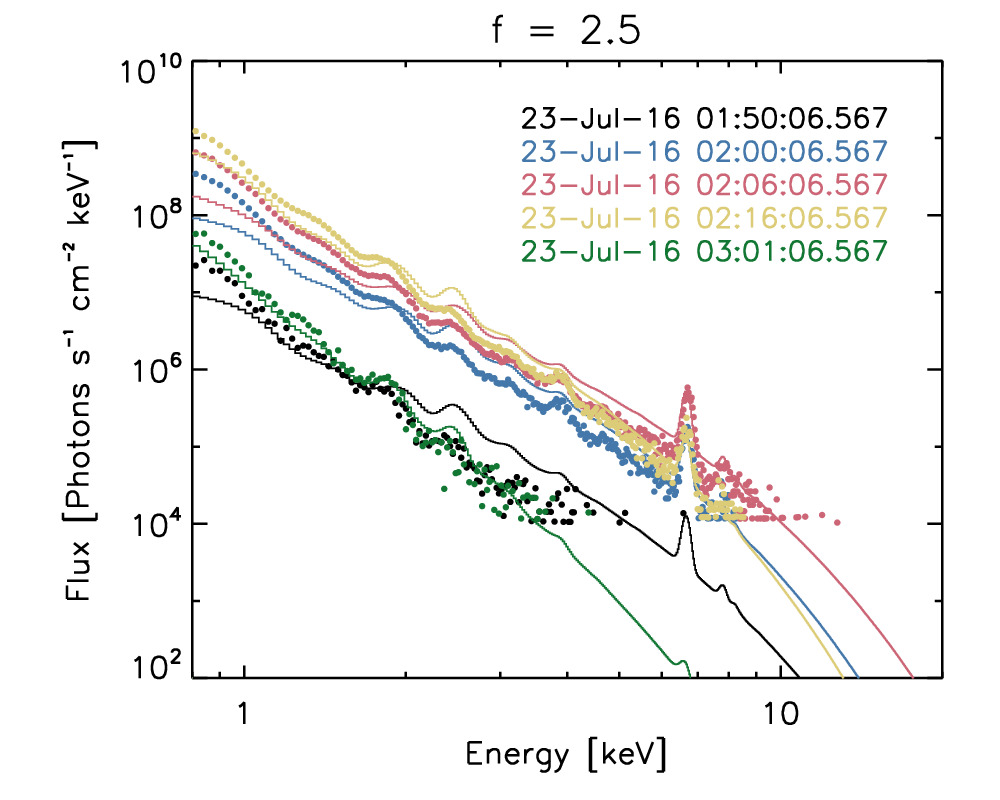}
\includegraphics[width=0.3\textwidth]{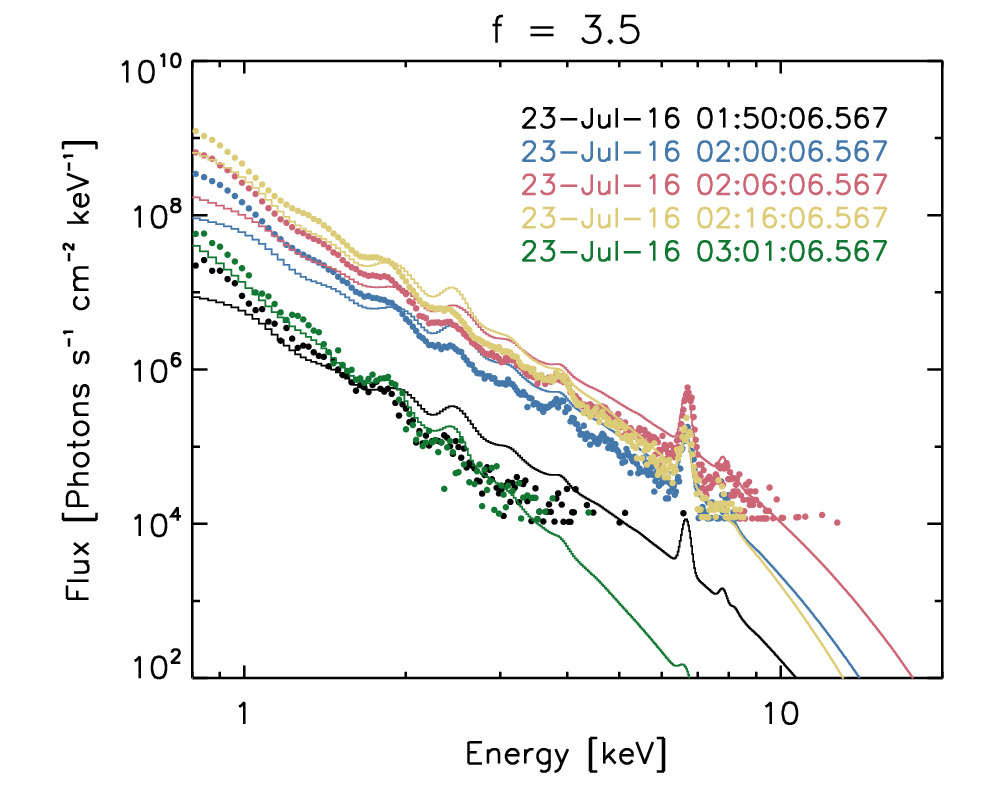}
\caption{A comparison of the synthetic and observed AIA light curves (top), their ratio (middle), as well as the MinXSS-X123 spectra (bottom) at selected times, for three different FIP enhancement factors $f$ = 1.5 (left), 2.5 (middle), and 3.5 (right) for the 23 July 2016 M5.0 flare.  The AIA light curves are marginally closer to observations for abundances close to photospheric, while the MinXSS-1 spectra show essentially no variation. \label{fig:m5_abund}} 
\end{figure*}

In Figure \ref{fig:m5_abund}, we briefly compare the effect of three separate values of FIP enhancement factors on the 23 July 2016 M5.0 flare.  The plots show the synthetic and observed AIA light curves (top), their ratio (middle), as well as the MinXSS-X123 spectra (bottom) at selected times, for three different FIP enhancement factors $f$ = 1.5 (left), 2.5 (middle), and 3.5 (right).  In this case, the AIA light curves, particularly in the cooler channels, are more consistent with the observations for abundances close to a photospheric scaling, while they diverge in the case of a more coronal scaling.  The MinXSS-X123 spectra are nearly the same in all three cases.  The MinXSS spectra are moderately well reproduced regardless of what FIP multiplier for the abundance is assumed, which is once again due to the constraints from spectrally-integrated GOES signals.

\begin{figure}
    \centering
    \includegraphics[width=\linewidth]{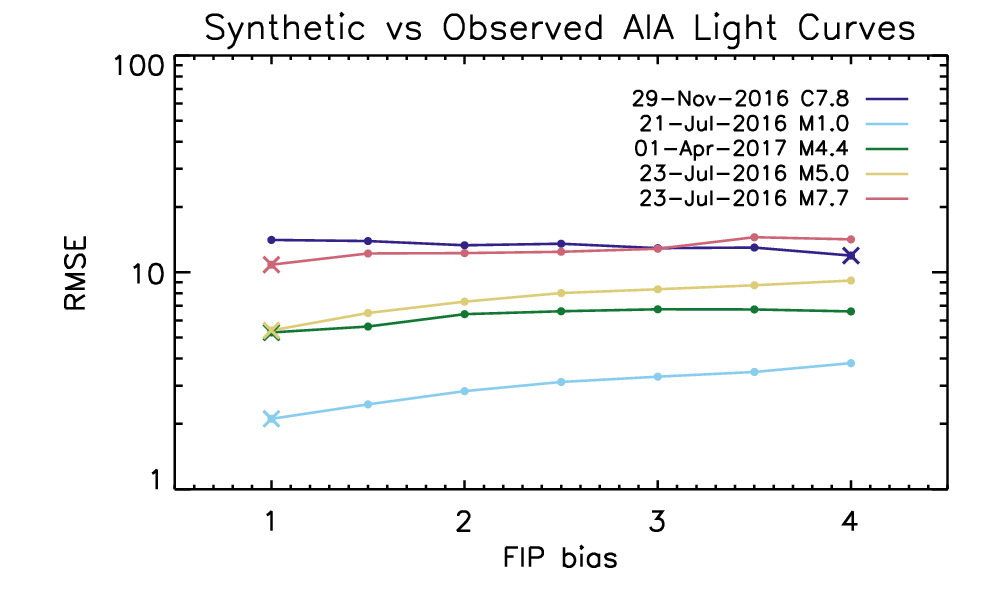}
    \includegraphics[width=\linewidth]{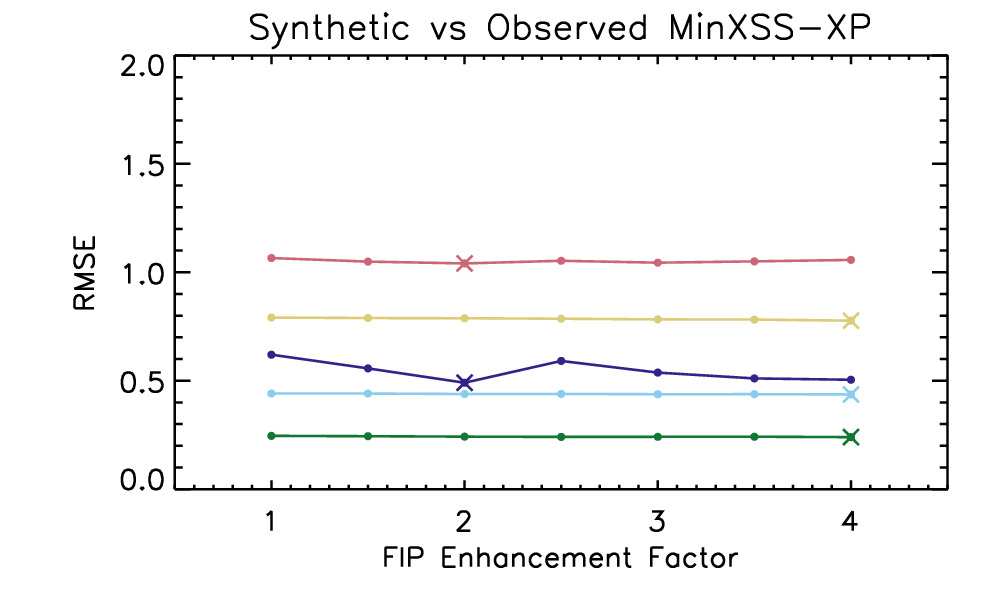}
    \includegraphics[width=\linewidth]{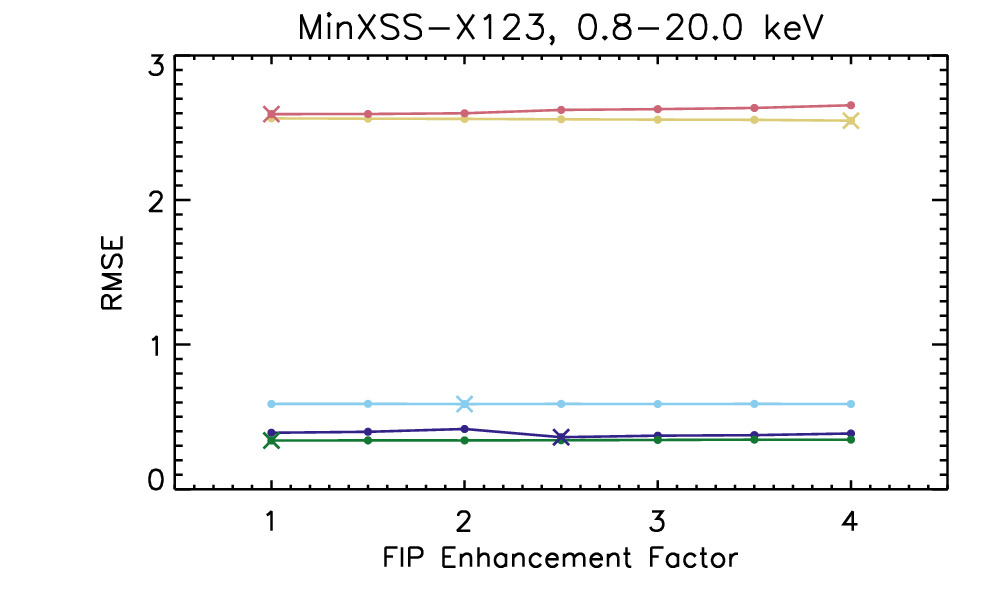}
        \caption{The RMSE for the mean of the six AIA channels (top), MinXSS-XP (middle), and MinXSS-X123 (bottom) comparing synthetic to observed light curves for each of the five flares and each of the seven assumed FIP enhancement factors $f$.  The minima are marked with an asterisk.  In four flares, the minimum RMSE are for the photospheric case $f = 1$, while the 29 November 2016 C7.8 flare is best fit with a coronal abundance ($f = 4$).  }
    \label{fig:abund_rmse}
\end{figure}

To demonstrate this, we finally show a comparison in Figure \ref{fig:abund_rmse} of the RMSE values for AIA (top), MinXSS-XP (middle), and MinXSS-X123 (bottom), for FIP enhancement factors of $f = $[1, 1.5, 2, 2.5, 3, 3.5, 4] and for each of the five flares.  With AIA, in four of the five flares, the RMSE is minimized with $f = 1$, that is, with photospheric abundances.  The 29 November 2016 C7.8 flare, however, has a minimum RMSE with a FIP enhancement factor $f = 4$, corresponding roughly to coronal.  As in the previous section, the MinXSS-XP and MinXSS-X123 show almost no variation for the various FIP enhancements, which is due to the algorithm we use to construct the flare model that strongly constrains the SXR emission.  Unlike changes in the heating duration, the variations of RMSE for the AIA light curves is small here.  We cannot make definitive conclusions about the abundances of these flares from the model.  

\section{Extrapolation}
\label{sec:extrap}

This model is capable of reproducing many aspects of the GOES, AIA, and MinXSS emission for the observed flares, but can it be used to extrapolate to larger or smaller flares that may not have such coverage?  One goal for this modeling is to extrapolate the predicted irradiance to superflares like the Carrington event \citep{carrington1859,cliver2013}, with energies exceeding $10^{32}$ erg, for which there are limited modern solar observations and GOES/XRS would saturate.  There are four events in the GOES catalogue which have saturated in the 1--8\,\AA\ channel: 02 April 2001, 28 October 2003, 04 November 2003, and 07 September 2005.  The largest of these, 04 November 2003, was estimated to have had a bolometric radiative output of $4 \times 10^{32}$ erg \citep{emslie2012}.  In this section, we develop a method to scale the modeled flares based on the total input energy $E$ in order to test whether it can reproduce observed scalings.  The synthesis of irradiance spectra for superflares will be considered in a future paper.   

For this arcade model, the total energy $E$ in a given flare can be calculated by summing the energy input over all of the loops
\begin{align}
    E = \sum_{i} H_{i} \tau_{i} V_{i}
\end{align}
\noindent where $H_{i}$ is the heat input to loop $i$ (erg\,s$^{-1}$\,cm$^{-3}$), $\tau_{i}$ is the heating duration (s), and $V_{i}$ is the volume of the thread (cm$^{-3}$).  Suppose that we introduce a scaling factor $\varphi$ for the energy, such that $E \rightarrow \varphi E$ in our extrapolated simulation.  Since the heating duration did not strongly impact our results, for simplicity we assume $\tau_{i}$ is constant from loop to loop, and therefore can pull it out of the summation.  Then, there are a number of ways we could scale each thread: we could simply increase the volume of each thread by a factor of $\varphi$, we could increase the heating rate of each thread by $\varphi$, or some permutation of the two.  Simply put, we can introduce a factor $\alpha$ such that the scaling can be generally written:
\begin{align}
    V_{i} &\rightarrow \varphi^{\alpha} V_{i} \\
    H_{i} &\rightarrow \varphi^{1-\alpha} H_{i} 
\end{align}
\noindent where $\alpha \in [0,1]$.

\begin{figure*}
\centering
\includegraphics[width=0.49\textwidth]{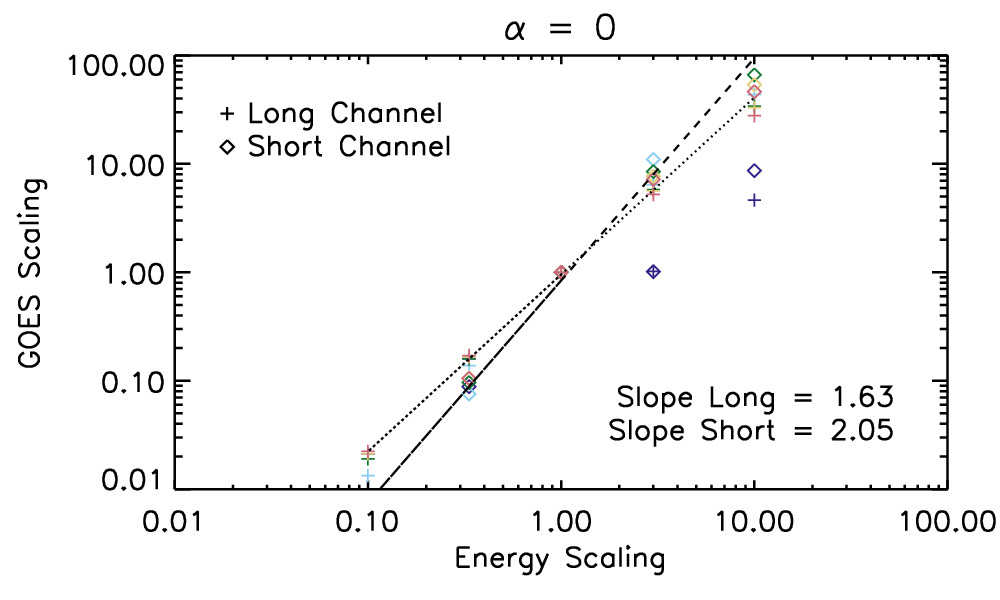}
\includegraphics[width=0.49\textwidth]{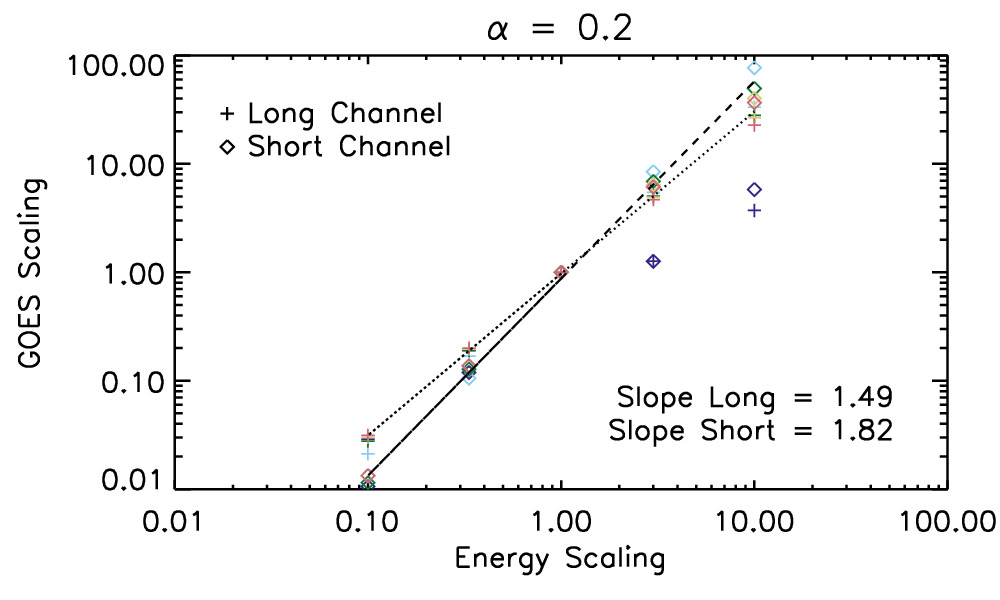}
\includegraphics[width=0.49\textwidth]{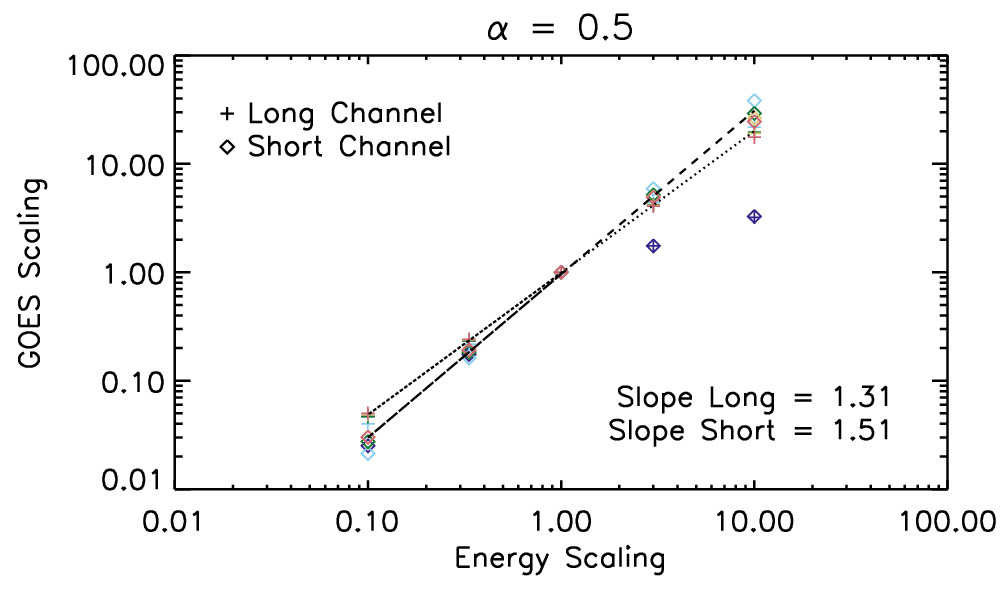}
\includegraphics[width=0.49\textwidth]{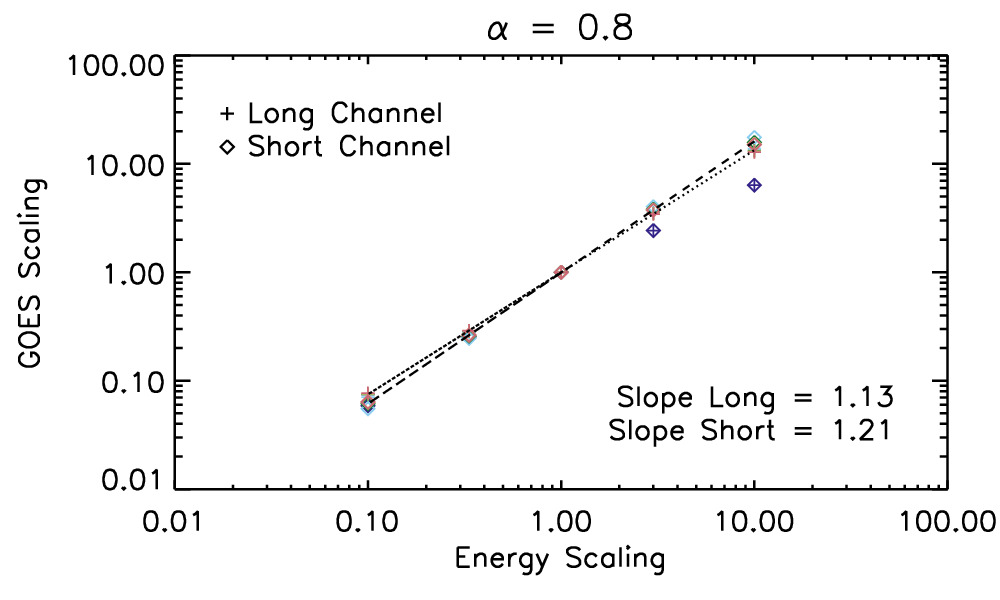}
\includegraphics[width=0.49\textwidth]{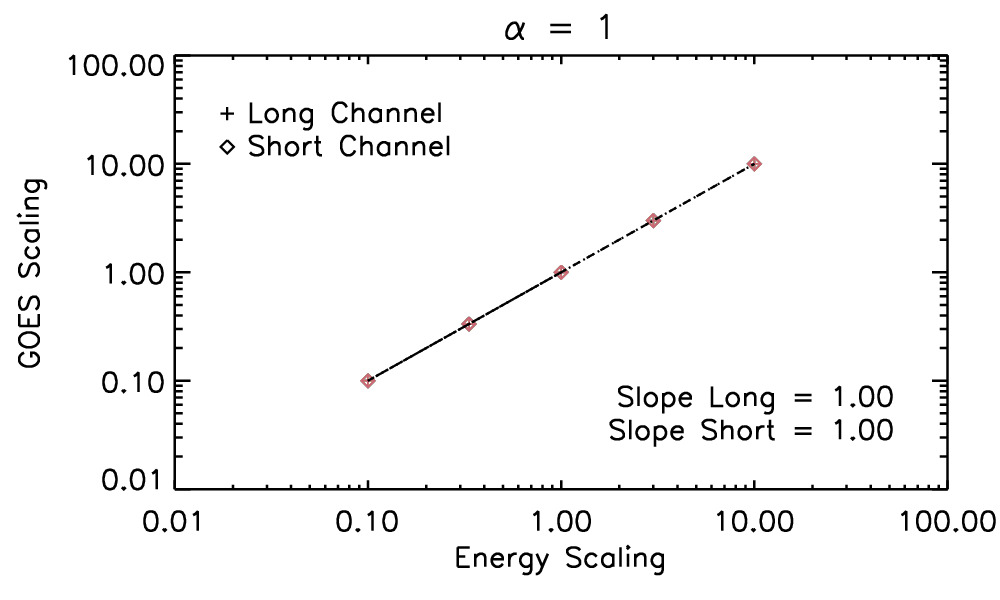}
\includegraphics[width=0.49\textwidth]{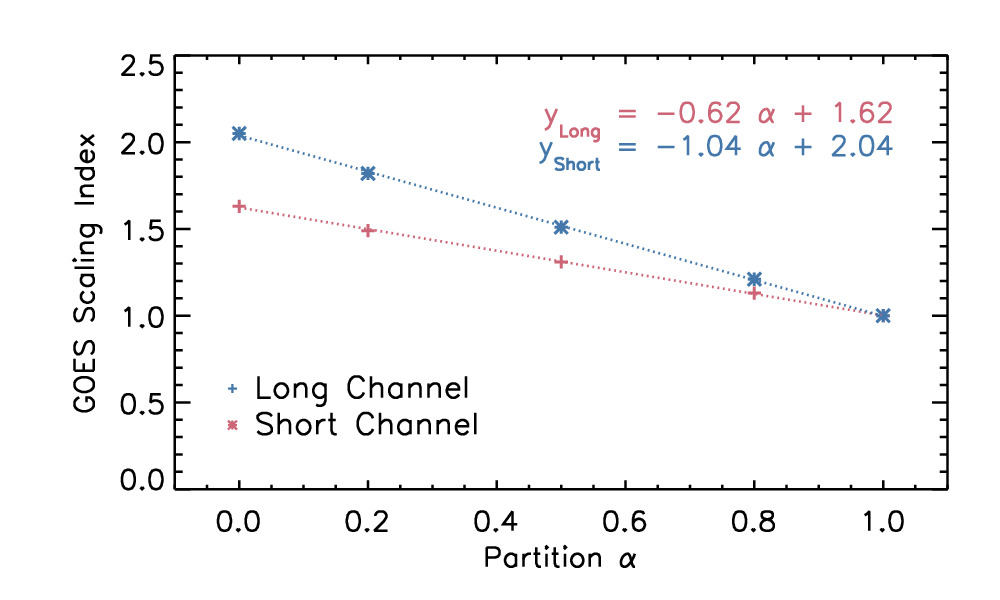}
\caption{The scaling of GOES flux with energy input, for different permutations of scaled heating rates and volumes quantified by the parameter $\alpha$ (see text).  The first five plots show how the GOES flux scales with energy input for values of $\alpha = [0, 0.2, 0.5, 0.8, 1]$.  The bottom right plot then shows how the power law fit changes as a function of $\alpha$.  \label{fig:scaling}}
\end{figure*}

We test this scaling by calculating all of the permutations with $\varphi$ = [1/10, 1/3, 1, 3, 10] and $\alpha$ = [0, 0.2, 0.5, 0.8, 1], for all five of the flares presented in Section \ref{sec:observations}.  We use the best fits for the heating duration and abundances, as well as the appropriate time-scales for successive reconnection for each flare, as found in Section \ref{sec:simulations}.  In Figure \ref{fig:scaling}, we show the results of this experiment.  The first five plots show how the GOES emission scales with the energy scaling $\varphi$ for each value of $\alpha$, labeled at the top.  Each flare is colored differently.  We also show the fitted slopes in both the long 1--8\,\AA\ channel and the short 0.5--4\,\AA\ channel in each case.  Because the C7 flare (dark blue) has a different best fit abundance, its scaling differs from the other four, particularly with small $\alpha$.  Finally, the bottom right plot shows how the slopes of this scaling depends on the parameter $\alpha$.  

As might be expected, in the case $\alpha = 1$, that is, where the volume is increased linearly with no change in the heating rate, the GOES emission scales linearly in both channels.  This is because the EM scales directly with volume.  In the other limiting case with $\alpha = 0$, where the heating rate is increased linearly, the two channels scale super-linearly with slopes $F_{1-8} \propto E^{1.63}$ and $F_{0.5-4} \propto E^{2.05}$.  Finally, in between these limiting cases, there is a linear gradient in both GOES channels for different values of $\alpha$.  Compare these results to \citet{warren2004}, who predicted slopes of 1.75 and 2.24, respectively, based on the RTV scaling laws, or \citet{reep2013} who found slopes of 1.7 and 1.6 based on hydrodynamic simulations.  The primary issue is that these two results were not based on a proper modeling of the flare arcade, focusing only on dynamics of a single loop.  

\begin{figure}
\centering
\includegraphics[width=0.49\textwidth]{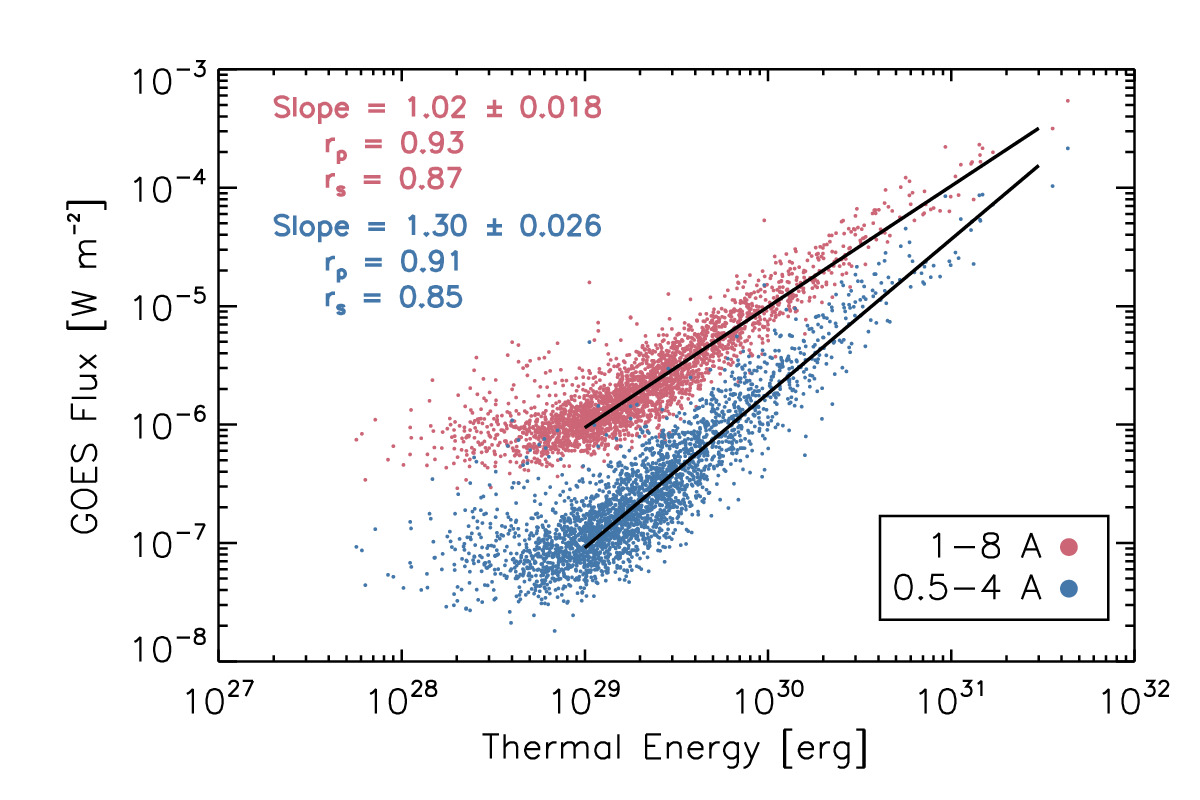}
\caption{The observational scalings of GOES flux against thermal energy $E$, taken from the data set in \citet{reep2019}.  The fits have been calculated for energies above $10^{29}$\,erg, rather than being unrestricted as in that paper.  The Pearson correlation coefficient $r_{p}$ and Spearman rank coefficient $r_{s}$ are also indicated. \label{fig:obsscaling}}
\end{figure}

How do these scalings compare to observations?  We use the data set of \citet{reep2019} to show this.  In Figure \ref{fig:obsscaling}, we show how the GOES class scales with thermal energy, where the data has been fit with a non-parametric Theil-Sen estimator \citep{sen1968,theil1992} only for flares with energy above $10^{29}$ erg.  We find slopes of $1.02$ and $1.30$ in the long and short GOES channels, respectively (slightly higher than the fits in \citealt{reep2019}, in which the energy ranges were not restricted).  For reference, both the Pearson correlation coefficient $r_{P}$ and Spearman rank coefficient $r_{S}$ are also indicated on the plots, showing the degree of correlation and monotonicity, respectively.

These observations lead us to conclude that values of $\alpha$ between approximately 0.7 and 1.0 or so are consistent with observed quantities.  Importantly, this tells us that between 70 and 100\% of the increase in energy for larger flares is due to having a larger volume, while the remaining 0 to 30\% is due to an increase in the heating rate.  The heating rate only increases marginally, which is consistent with the marginal increases in temperature with flare class (\textit{e.g.} \citealt{feldman1995,feldman1996,reep2019}).  For $\alpha = 0.8$, this means that a superflare of energy 100 times larger than an observed X-flare will have a volume larger by $(10^{0.8})^{2} \approx 40$, with an average heating rate larger by $(10^{0.2})^{2} \approx 2.5$.  This gives a method to extrapolate the simulations to larger events or superflares, which can be used to predict irradiance spectra and the resultant ionospheric response, and this will be the focus of future work.  

It should be noted, however, that this extrapolation is only based on five flares, and that they do not encompass the full range of conditions of flare-productive active regions (three of these flares occurred in the same active region).  There are many factors that could impact the extrapolation here: physical size of the active region, magnetic flux density, association with a coronal mass ejection or lack thereof, whether the events are eruptive or confined, the magnetic topology of the active regions, \textit{etc.} \citep{kazachenko2017,tschernitz2018}.  A wider range of flare energies and active region configurations could improve this extrapolation.  

One other possible implication of this scaling is that it predicts how heating rates and volumes scale down to microflares and nanoflares, assuming that they are driven by the same processes as macro-scale flares.  For example, using a value of $\alpha = 0.8$, if we assume that the total energy of an event is $10^{6}$ times smaller than a flare, then its volume will scale down by a factor of $(10^{0.8})^{6} \approx 6.3 \times 10^{4}$, while the heating rate will scale down by a factor of $\approx 16$.  This is a crucial prediction to test whether nanoflares are scaled down versions of full-sized flares (see the discussion in \textit{e.g} \citealt{hudson1991}).  

\section{Summary}

In this paper, we have developed a model of solar flares that can be used to calculate the irradiance spectrum across a wide wavelength range.  The model uses the observed GOES light curves to deduce the heating rates, volumes, and ribbon expansion as a function of time for a series of successively heated loops in a flare arcade.  The flare is then simulated using a series of hydrodynamic simulations of loops with the appropriate heating rates, and then the spectra of all the loops are summed together to create a global irradiance spectrum.  Many, though not all, of the input parameters can be constrained by the GOES observations alone, though lower temperature emission requires other constraints.  

From this model, we have found that
\begin{itemize}
    \item The arcade model developed in this paper can accurately reproduce the soft X-ray light curves and spectra (RMSE $< 3$), and to a good extent light curves from cooler emission (RMSE $< 10$).  This model can therefore be used to extrapolate predictions of intensities to wavelengths which were not observed.
    \item The primary period of quasi-periodic pulsations in GOES/XRS light curves is consistent with and can be reproduced by the successive reconnection model.  The five flares in this paper exhibited main QPP periods of 20, 26, 20, 12, and 25 s (see Table \ref{tab:obs_summary}).
    \item The duration of heating on individual loops in this model strongly impacts the light curves of cooler emission, such as that observed by SDO/AIA.  Longer heating durations improve consistency with observed AIA light curves since this factor strongly affects the cooling rate.  The soft X-rays (spectra or light curves), however, cannot be used as a strong discriminator of heating duration because of the constraints on the model.
    \item Elemental abundances can be similarly constrained through comparison to observed emission in cooler AIA channels, though it is difficult to definitively measure their values in this way.   
    \item This model can be extrapolated to extreme events by scaling the energy appropriately.  If the energy increases, for consistency with observed scaling laws, between 70 and 100\% of that energy increase must be due to an increase in volume, while 0 to 30\% of that extra energy is due to an increase in the heating rate.
\end{itemize}

Because we wished to explore the parameter space to better understand the effects of various parameters, we have used the ebtel++ 0D hydrodynamics code \citep{barnes2016} for this model due to its speed in running large numbers of simulations.  In future work, in order to produce more realistic emission measure distributions for each individual loop, we will switch to the field-aligned HYDRAD code \citep{bradshaw2013}, which additionally includes the effects of many more physical mechanisms that occur in flares that are not (nor could be) treated by ebtel++, which also can better constrain the model.  For example, heating due to an electron beam produces non-thermal HXR emission, which can be directly contrasted with observations by RHESSI to better constrain the parameters of that electron beam.  Additional constraints like this will allow us to improve comparison with the cooler emission emitted in the chromosphere, in particular.  Finally, we will use this model to calculate full irradiance spectra ranging from the X-rays through the ultraviolet, which will be used to better understand the impact of solar flares on the ionosphere.

\appendix

\addcontentsline{toc}{section}{Appendices}
\renewcommand{\thesubsection}{\Alph{subsection}}

\subsection{Abundances in the IDL routine `goes\_fluxes'}

Synthetic GOES/XRS fluxes are often calculated with the `goes\_fluxes' routine bundled in the SolarSoftWare package \citep{Freeland:1998}.  For the 0D simulations used in this work, this routine is the simplest option (see also \citealt{reepwarren2018}).  The standard version of this code (at the time of writing) calculates fluxes using CHIANTI version 7 \citep{landi2012} and only allows for two options in elemental abundances roughly corresponding to photospheric and coronal abundances.  In this work, to be truly self-consistent with abundance variations, we have modified this routine to allow it to use any specified abundance set with the most recent version of CHIANTI (9.2.1, \citealt{dere2019}).  

\begin{figure*}
    \centering
    \includegraphics[width=\textwidth]{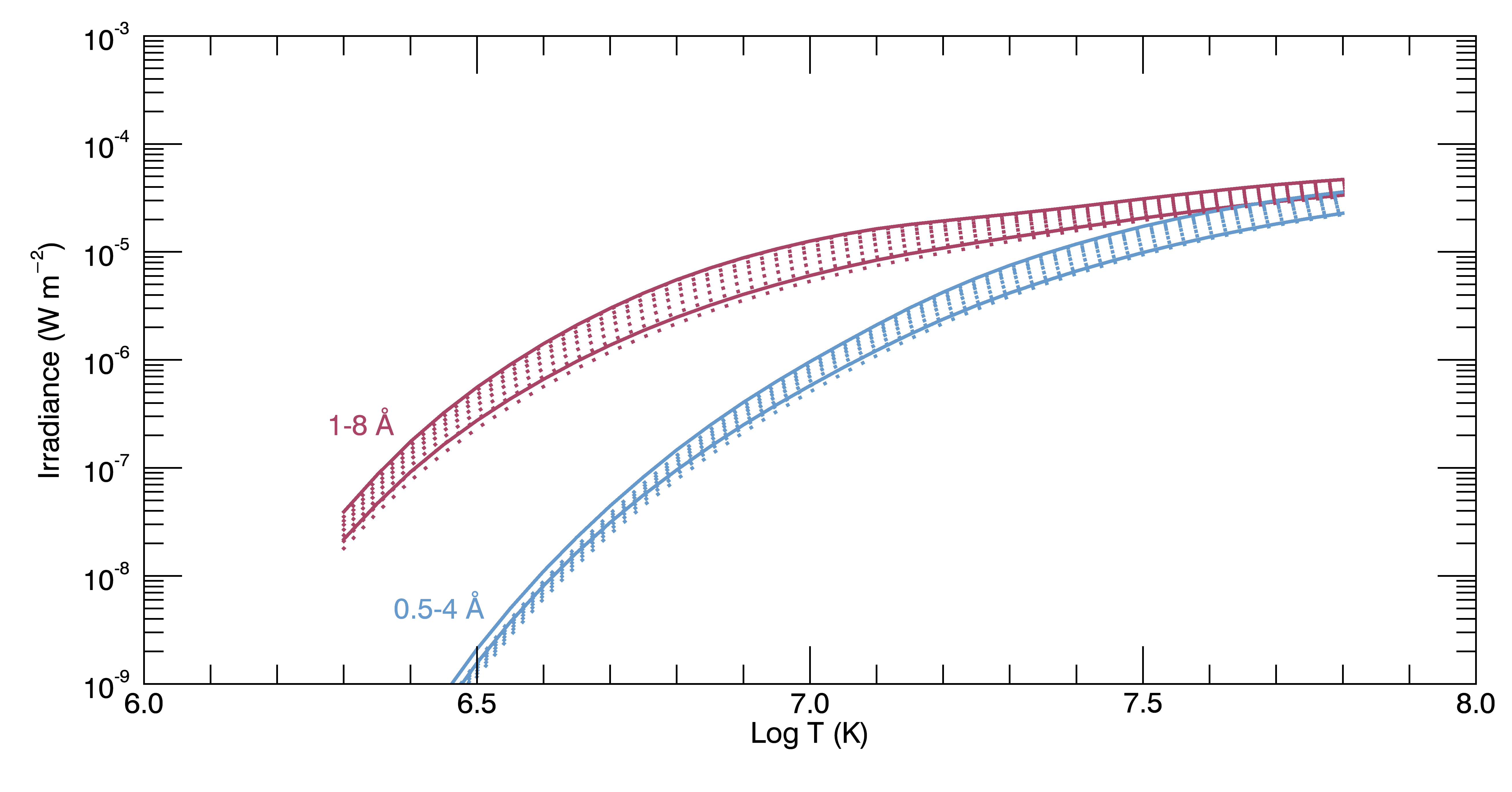}
    \caption{The dependence of synthetic GOES fluxes on temperature, for various abundance values. The red trends show the 1--8\,\AA\ channel while blue shows 0.5--4\,\AA.  The solid lines for each channel show the calculations with `goes\_fluxes' using the original coronal (top) and photospheric (bottom) abundances in that routine, calculated with CHIANTI version 7 \citep{landi2012}.  The dotted lines show the synthetic GOES fluxes with seven sets of abundances, using the photospheric abundance set of \citep{asplund2009}, modified for FIP enhancement factors $f$ = [1.0, 1.5, 2.0, 2.5, 3.0, 3.5, 4.0], where 4.0 corresponds approximately to coronal, calculated with CHIANTI version 9 \citep{dere2019} and a modified `goes\_fluxes' routine.  }
    \label{fig:goes_fluxes}
\end{figure*}

The code calculates the fluxes in both XRS channels, given a temperature and assuming a total volumetric emission measure.  In Figure \ref{fig:goes_fluxes}, we show the comparison of the original routine to the modified one.  The red lines show the synthetic fluxes in 1--8\,\AA\ channel as a function of temperature, while the blue lines show the 0.5--4\,\AA\ channel.  The solid lines for each channel show the original routine, using CHIANTI version 7, for coronal (top solid line) and photospheric (bottom solid line) abundances.  There are 7 dotted lines for each channel, showing the synthetic fluxes using the \citet{asplund2009} photospheric abundances, where the low FIP elements have been enhanced with FIP enhancement factors of $f$ = [1.0, 1.5, 2.0, 2.5, 3.0, 3.5, 4.0].  The resultant spectra are then calculated with CHIANTI version 9 \citep{dere2019} and used to synthesize the GOES emission, folding in the instrumental response appropriate for a specified GOES satellite (the plot shows GOES-15).  The bottom dotted line, $f$ = 1.0, corresponds to photospheric, while the top dotted line, $f$ = 4.0, is roughly coronal.  In each case, it is clear that the synthetic fluxes depend strongly on the abundances.  The difference between the two versions of CHIANTI mostly affects calculations at low temperature (less than 10 MK), where spectral line emission becomes increasingly important.  


\acknowledgments JWR was supported by NASA's Living With a Star program.  HPW was supported by
NASA's MinXSS project and NASA's Heliophysics Guest Investigator program.  L.A.H is supported by an
appointment to the NASA Postdoctoral Program at Goddard Space Flight Center, administered by USRA
through a contract with NASA.  CS was supported by NSF- Fisk-Vanderbilt Master’s-to-Ph.D. Bridge
Program Grant No. HRD-1547757 and the Latino Initiatives administered by the Smithsonian Latino
Center.  MinXSS-1 CubeSat mission is supported by NASA Grant NNX14AN84G. CHIANTI is a collaborative
project involving George Mason University, the University of Michigan (USA) and the University of
Cambridge (UK).


\bibliography{apj}
\bibliographystyle{aasjournal}

\end{document}